\documentclass[12pt,a4paper]{article}

\newcommand{\tsecompldate}{13\textsuperscript{th} December 2025}

\usepackage[utf8]{inputenc}

\usepackage{booktabs}
\usepackage{multirow}

\usepackage[UKenglish]{isodate}

\usepackage[a4paper,margin=2cm, marginparwidth=2cm]{geometry}

\usepackage{amsmath,amssymb,amscd,amsthm}
\DeclareMathOperator*{\argmax}{arg\,max}

\numberwithin{equation}{section}

\usepackage{graphicx}
\graphicspath{{./figures/}}

\usepackage[margin=10pt,font=small,labelfont=bf]{caption}

\usepackage{xcolor} 

\definecolor{TSEBrickRed}{RGB}{203,65,84}
\definecolor{TSEDarkMagenta}{RGB}{139,0,139} 
\definecolor{HPDarkGreen}{RGB}{1, 50, 32} 

\newcommand{\tsenote}[1]{}
\newcommand{\tsecomment}[1]{}
\newcommand{\hpcomment}[1]{}

\newcommand{\tsedef}[1]{\textsc{#1}}
\newcommand{\tseabbv}[1]{\textbf{#1}}

\usepackage[square,comma]{natbib}
\providecommand{\cite}[1]{\citep{#1}}
\renewcommand{\harvardurl}[1]{\textbf{URL:} \url{#1}}

\usepackage[colorlinks,allcolors=blue,pdfauthor={T.S. Evans, H.C.W. Price},pdftitle={Understanding Main Path Analysis}]{hyperref}

\providecommand{\href}[2]{\texttt{#2}}
\providecommand{\url}[1]{\texttt{#1}}

\newcommand{\bea}{\begin{eqnarray}}
\newcommand{\eea}{\end{eqnarray}}
\newcommand{\beq}{\begin{equation}}
\newcommand{\eeq}{\end{equation}}

\providecommand{\eqref}[1]{(\ref{#1})}
\newcommand{\equationref}[1]{equation (\ref{#1})}

\newcommand{\figref}[1]{Fig.\ \ref{#1}}
\newcommand{\tabref}[1]{Table~\ref{#1}}

\newcommand{\secref}[1]{Section~\ref{#1}}

\newcommand{\appref}[1]{Appendix~\ref{#1}}

\newcommand{\tseurl}[1]{\href{https://#1}{\texttt{#1}}}

\newcommand{\texpect}[1]{\langle #1 \rangle}

\newcommand{\tsevec}[1]{\mathbf{#1}}
\newcommand{\tsemat}[1]{{\mathbf{\textsf{#1}}}}

\newcommand{\pvec}{\tsevec{p}}

\newcommand{\svec}{\tsevec{s}}
\newcommand{\tvec}{\tsevec{t}}
\newcommand{\uvec}{\tsevec{u}}
\newcommand{\vvec}{\tsevec{v}}

\newcommand{\xvec}{\tsevec{x}}

\newcommand{\kin}{k^{(\mathrm{in})}}
\newcommand{\kout}{k^{(\mathrm{out})}}

\newcommand{\ktarget}{k^{(\mathrm{out})}_{\mathrm{target}}} 

\newcommand{\Amat}{\tsemat{A}}

\newcommand{\Gmat}{\tsemat{G}}

\newcommand{\Bcal}{\mathcal{B}}

\newcommand{\Dcal}{\mathcal{D}}
\newcommand{\Dcalint}{\mathcal{D}^\mathrm{(int)}}
\newcommand{\Dcalhc}{\mathcal{D}^\mathrm{(hc)}}
\newcommand{\Dcalrev}{\mathcal{D}^\mathrm{(rev)}}
\newcommand{\Ecal}{\mathcal{E}}
\newcommand{\Ecalrev}{\mathcal{E}^\mathrm{(rev)}}

\newcommand{\Ncal}{\mathcal{N}}

\newcommand{\Pcal}{\mathcal{P}}
\newcommand{\Pcalrev}{\mathcal{P}^\mathrm{(rev)}}

\newcommand{\Vcal}{\mathcal{V}}

\newcommand{\Vcalint}{\mathcal{V}^\mathrm{(int)}}

\newcommand{\Ibb}{\mathbb{I}}

\newcommand{\Rbb}{\mathbb{R}}
\newcommand{\Zbb}{\mathbb{Z}}

\newcommand{\vbar}{\bar{v}}

\newcommand{\Deltabar}{\bar{\Delta}}
\newcommand{\Deltafifty}{\Delta_\mathrm{50}}

\newcommand{\cG}{c^{\mathrm{(G)}}}
\newcommand{\dG}{d^{\mathrm{(G)}}}
\newcommand{\hG}{h^{\mathrm{(G)}}}
\newcommand{\HG}{H^{\mathrm{(G)}}}
\newcommand{\LG}{L^{\mathrm{(G)}}}
\newcommand{\BG}{B^{\mathrm{(G)}}}
\newcommand{\PcalG}{\mathcal{P}^{\mathrm{(G)}}}

\newcommand{\Bcalunit}{\mathcal{B}^{\mathrm{(unit)}}}
\newcommand{\cunit}{c^{\mathrm{(unit)}}}
\newcommand{\Gunit}{G^{\mathrm{(unit)}}}
\newcommand{\Lunit}{L^{\mathrm{(unit)}}}
\newcommand{\Pcalunit}{\mathcal{P}^{\mathrm{(unit)}}}

\newcommand{\cspc}{c^{\mathrm{(spc)}}}

\newcommand{\Gspc}{G^{\mathrm{(spc)}}}
\newcommand{\Gmatspc}{{\Gmat}^{\mathrm{(spc)}}}

\newcommand{\Lspc}{L^{\mathrm{(spc)}}}
\newcommand{\Pcalspc}{\mathcal{P}^{\mathrm{(spc)}}}
\newcommand{\Wspc}{W^\mathrm{(spc)}}
\newcommand{\Xspc}{X^\mathrm{(spc)}}

\newcommand{\dentropy}{d^\mathrm{(H)}}
\newcommand{\Mdentropy}{M^{\mathrm{(d)}}}

\newcommand{\Gmatentropy}{\Gmat^{\mathrm{(H)}}}

\newcommand{\cspe}{c^{\mathrm{(spe)}}}

\newcommand{\Gspe}{G^{\mathrm{(spe)}}}
\newcommand{\Gmatspe}{{\Gmat}^{\mathrm{(spe)}}}

\newcommand{\Lspe}{L^{\mathrm{(spe)}}}

\newcommand{\Pcalspe}{\mathcal{P}^{\mathrm{(spe)}}}
\newcommand{\Wspe}{W^\mathrm{(spe)}}
\newcommand{\Xspe}{X^\mathrm{(spe)}}

\newcommand{\Pcalrnd}{\mathcal{P}^{\mathrm{(rnd)}}}

\newcommand{\Pcalgpd}{\mathcal{P}^{\mathrm{(gpd)}}}

\newcommand{\Pcallp}{\mathcal{P}^{\mathrm{(lp)}}}

\newcommand{\That}{\widehat{T}}

\newcommand{\Ntilde}{\tilde{N}}

\newcommand{\Nbox}{\tilde{N}}
\newcommand{\Nint}{N^\mathrm{(int)}}

\begin{document}


\begin{center}
 {\Large\textbf{Understanding Main Path Analysis}}\tsenote{Notes such as this not present in final version}
 \\[\baselineskip]
 {\large H.C.W.\ Price\footnote{ORCID:  \href{http://orcid.org/0000-0003-0756-0652}{\texttt{0000-0003-0756-0652}} }} 
 and
 {\large \href{http://www.imperial.ac.uk/people/t.evans}{T.S.\ Evans}}\footnote{ORCID:  \href{http://orcid.org/0000-0003-3501-6486}{\texttt{0000-0003-3501-6486}} }
  \\[0.5\baselineskip]
 \href{http://complexity.org.uk/}{Centre for Complexity Science}, and \href{http://www3.imperial.ac.uk/theoreticalphysics}{Abdus Salam Centre for Theoretical Physics},
 \\
 Imperial College London, SW7 2AZ, U.K.
 \\
 \tsecompldate 
\end{center}

\begin{abstract}
Main path analysis has long been used to trace knowledge trajectories in citation networks, yet it lacks solid theoretical foundations.
To understand when and why this approach succeeds, we analyse directed acyclic graphs created from two types of artificial models and by looking at over twenty networks derived from real data. 

We show that entropy-based variants of main path analysis optimise geometric distance measures, providing its first information-theoretic and geometric basis.
Numerical results demonstrate that existing algorithms converge on near-geodesic solutions.
We also show that an approach based on longest paths on the unweighted network produce similar results, are equally well motivated yet are much simpler to implement.

However, the traditional single-path focus is unnecessarily restrictive, as many near-optimal paths highlight different key nodes.
We introduce an approach using ``baskets'' of nodes where we select a fraction of nodes with the smallest values of a measure we call ``generalised criticality''. 
Analysis of large vaccine citation networks shows that these baskets achieve comprehensive algorithmic coverage, offering a robust, simple, and computationally efficient way to identify core knowledge structures.
In practice we find that those nodes with zero unit criticality capture the information in main paths in almost all cases and capture a wider range of key nodes without unnecessarily increasing the number of nodes considered. We find no advantage in using the traditional main path methods.
\end{abstract}

\setcounter{footnote}{0}
\renewcommand{\thefootnote}{\arabic{footnote}}




\section{Introduction}\label{s:intro}

Innovation leaves its marks in many documents.  For technical innovations, academic papers, patents and documents related to formal trials and approvals all contain references to relevant earlier work in their bibliography. Main path analysis, first proposed by \citet{HD89}, is a popular bibliometric technique used to trace the most influential citation chains within a citation network in order to find the main development trajectory in a research field. Numerous studies have used it (we shall cite a small selection in our paper) as shown by the plot of the number of papers in each year shown in \appref{as:mpalit}. Further evidence of the utility of main path analysis comes from several numerical implementations (see \appref{as:software}) including the popular network analysis package \href{https://mrvar.fdv.uni-lj.si/pajek/}{\texttt{pajek}} \citep{BM98,BM03,NMB18}.

It is impossible to \emph{prove} that this quantitative method captures innovation trajectories. Influences on technological evolution include social, institutional, and contextual factors and main path analysis of a citation network cannot capture all those aspects. Nevertheless, the use of main path method is supported in two ways. 

Firstly, the intuition of many experts is that the traversal-count heuristic at the core of this algorithm is a measure of the connectivity and reachability of nodes and that these are good proxies for influence in a citation network. The traversal count of an edge is the number of paths passing through that edge given some set of initial and final nodes for these paths. 
The main path is then the path, a sequence of edges, with highest total traversal count.

Secondly, empirical studies such as  \citet{HD89,CHH93,B03d,V07a,LL08,FNV09,LL12,YKLK13,CB14a,XLLZ14,BDFK14,BGGMN15,LK15,LPC17,KJKS22} (or see \citet{LLH19} for a recent overview)\tsenote{Maybe look at the following web sites: \href{http://vlado.fmf.uni-lj.si/pub/networks/Data/cite/default.htm}{pajek}, 
	\href{http://www.garfield.library.upenn.edu/histcomp/index.html}{garfield},
	\href{https://www.sussex.ac.uk/webteam/gateway/file.php?name=sewp166&site=25}{sussex},
	\href{https://davincierlab.weebly.com/list-of-main-path-articles.html}{da Vincier Lab Main Path} web page.
} 
have found strong evidence that main path analysis captures key knowledge flows in citation networks but this is not always true \citep{F21}. The known bias against negative results in the published literature makes the empirical support for main path analysis harder to interpret.

The main aim of this paper is to provide a more rigorous testing of main path analysis. We do this by making the conjecture that the main path should be a geodesic, a straight line between two points. That is we use concepts from geometry which underlies the fact that in many contexts, such as diffusion in media or the path of objects in mechanics, the geodesic is the path of least effort and so the path most likely to be followed. To show this relationship, we first work with artificial models where the geometry is well defined and the geodesic is the straight line that we would intuitively pick out as the most direct path between two points. We later move on to compare various different paths to the paths found by main path analysis in over twenty networks derived from real data.

One of our major conclusions is that by our metrics there is little to choose between the main paths we study and a path containing the largest number of edges in these networks. However, whichever path we choose, they are all seriously flawed because they highlight a very small fraction of  nodes in the network while our analysis shows there are a much larger group of nodes with a similar level of importance.  So our main practical recommendation is that it is sufficient to look at the nodes that lie on longest paths (paths with the largest number of edges) as this is much simpler to calculate numerically and as this seems to capture most of the important nodes.

\tsecomment{An additional output is that we have produced fast open-source code that we used to calculate our results \citep{???}.}

We start in \secref{s:methods} by looking at the methods we use.  This includes a definition of main path analysis and the particular variant we use in \secref{s:mpadetails}.  We will also define a new main path variant based on the entropy of the edge traversal counts in \secref{s:spe}  followed by various other paths we use for comparisons. The methods used to compare paths are given in \secref{s:pathcomp} which includes our key tool, the criticality of a node based on these different edge weights. We then move on to look at how various paths compare to the geodesic in two simple artificial models, a lattice in \secref{s:lattice} and a random geometric DAG in \secref{s:rgdag}. Our methods are then applied to over twenty networks derived from real data in \secref{s:realworldnetworks}.
We end with a discussion in \secref{s:discuss}.

\tsecomment{Perhaps discuss some alternative approaches. **Alternative methods**: Techniques like *intermediacy*, *betweenness centrality*, and *influence difference analysis* offer complementary views.}


\section{Generic Methods}\label{s:methods}

\subsection{Directed acyclic graphs}\label{s:DAG}

A \tsedef{directed acyclic graph} (\tseabbv{DAG})  is a network with an overall sense of order and direction, for example see \figref{f:traversalcountex} below. When following a path in a DAG you are always moving ``forwards'' in some sense. The most common source for this sense of direction is time. For instance a citation network is a good example where nodes are documents and there is a link from the bibliography of one document to older documents listed in the bibliography. 

One special property of DAGs is that we can always define the \tsedef{longest path} between any two connected vertices and it is a useful tool for DAGs \citep{KW59,BB91,ES91,RW09,CE16,CE17,ECV20,J22b,HPEO24,HPEO24a,HPEO25}. Unlike most networks which contain cycles, the acyclic property of a DAG means that if a path exists between two vertices, then the longest such path does not generally visit a large fraction of the network making such longest unit length paths a useful tool for DAGs.

\subsubsection{Formal definitions}\label{s:DAGformal}

A DAG $\Dcal$ is a network $\Dcal = (\Vcal, \Ecal)$ with a set $\Vcal$ of $N=|\Vcal|$ nodes and a set $\Ecal$ of $E=|\Ecal|$ directed edges. An edge $(u,v)$ is a pair of nodes in which the edge is directed from node $u$ to node $v$. We will never have more than one edge between each pair of nodes. We also use an \tsedef{adjacency matrix} $\Amat$ to represent the connectivity where $A_{vu}=1$ if $(u,v)$ is an edge from $u$ to $v$, otherwise $A_{vu}=0$. Later we will assign values to edges and these will be the entries in a matrix $\Gmat$ where $G_{vu}$ is the weight of the edge $(u,v)$, and is zero if there is no such edge.

A \tsedef{path} $\Pcal$ in a DAG is a \emph{sequence of nodes}, denoted as $\Pcal = (v_0,v_1,\ldots,v_L)$, in which consecutive pairs of nodes are always edges and the order of the sequence is compatible with the direction of the edges. That is for any $j=1,2,\ldots,L$ we have that $(v_{j-1},v_j)\in\Ecal$. We call this set of edges $\Ecal(\Pcal)= \{(v_{j-1},v_j)|j=1,2,\ldots,L\}$ the edges in the path $\Pcal$. 
The \tsedef{unit length} of a path is $L=|\Ecal(\Pcal)|=|\Pcal|-1$, the number of edges in the path which is also one less than the number of nodes in the path. 
A notation we will also use is that $\Pcal(i,f)$ is the \emph{set} of \emph{all} paths which start an initial node $i$ and end at the final node $f$. 


The order in a DAG also means that our DAGs must have a beginning and an end. The network starts from \tsedef{source} nodes which have no incoming edges while \tsedef{sink} nodes are nodes with no outgoing edges. We will simplify our analysis by always working with an \tsedef{interval} which is a DAG with a single source node $s$ and a single sink node $t$, denoted as $\Dcal(s,t)$. Every node lies in such interval DAGs lies on at least one path from $s$ to $t$.

If the edges in the DAG have weights then we can define another path `length' for each type of weight. The \tsedef{weight length}, or simply the \tsedef{weight},  $\LG(\Pcal)$ of a path $\Pcal= (v_0,v_1,\ldots,v_L)$ is the sum of the edge weights $G_{vu}$ of the edges $(u,v)$ from node $u$ to node $v$ in the path. 
That is 
\beq
	\LG (\Pcal) 
	= 
	\LG \big( (v_0,v_1,\ldots,v_L) \big) 
	= 
	\sum_{i=1}^L  G_{v_i v_{i-1}} \, .
	\label{e:LGdef}
\eeq 
We will use several different ways to measure the length of a path. The simplest and standard way is to count one for every edge in the path, the unit length $\Lunit$ where $\Gmat = \Amat$,
\beq
\Lunit (\Pcal) 
= 
\Lunit \big( (v_0,v_1,\ldots,v_L) \big) 
= 
\sum_{i=1}^L  A_{v_i v_{i-1}} 
=
L
\, .
\label{e:Lunitdef}
\eeq 
Below we will define two more weight lengths $\LG$ based on different matrices of weighted edge values $\Gmat$. So it is important to be clear which length we are using.  If we do refer to path length without any qualification, then it is the unit path length $L=\Lunit$ as this is always defined on any network.

We have noted above that another property of a DAG not shared by most networks is that the longest path by unit length, i.e.\ counting the number of edges on a path, between any two nodes is a very useful concept. We will have both unweighted DAGs, $\Gmat=\Amat$ which we will refer to as a \tsedef{unit weight}, and weighted graphs, using values to be defined below. Whatever edge weights we use, we can always find the path with the largest path weight, a generalisation of the ``longest path'' which uses unit weights. Formally, the definition of a longest path $\PcalG$ using weight $\Gmat$ in a DAG is
\beq
  \PcalG 
	= \argmax_{\Pcal \in \Pcal(s,t)} \LG(\Pcal)  \, .
	\label{e:lpGdef}
\eeq
Note that there can be multiple longest paths and we will indicate how we have chosen just one path if this occurs.

In general, DAGs have no particular symmetry except for an order reversal property. That is we can always flip the direction of every edge in any DAG $\Dcal= (\Vcal,\Ecal)$, to create a related DAG $\Dcalrev = (\Vcal,\Ecalrev)$ where $\Ecalrev = \{ (v,u) | (u,v) \in \Ecal \}$.

\subsection{Main path analysis}\label{s:mpa}

\subsubsection{Overview}\label{s:mpaoverview}

\begin{figure}[htb]
	\begin{center}
		\includegraphics[width=0.75\textwidth]{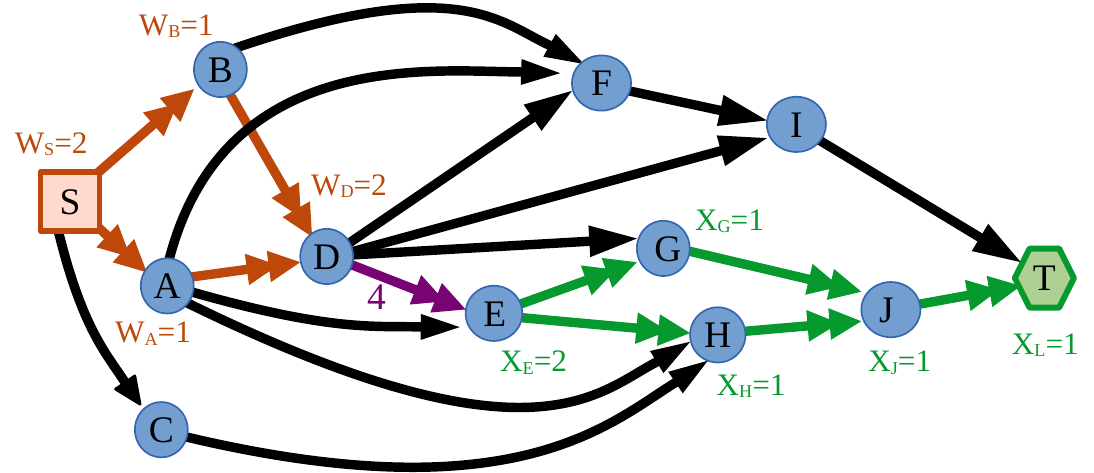}
	\end{center}
	\caption{Diagram illustrating the traversal count edge weight used in main path analysis on an interval DAG. The solid red square is the only source node $S$  while the solid green hexagon indicates the only sink node $T$. Each distinct path from $S$ to $T$ is counted once in SPC main path analysis. We illustrate the traversal count of the edge $(D,E)$ shown in purple with a double headed arrow which is $\Gspc_{ED}=4$. This comes from two parts. There are just two paths flowing to $D$ from initial vertices, the paths $(S,A,D)$ and $(S,B,D)$ shown in red with double headed arrows. These paths are counted by the $W_v$ node values shown so $W_D=2$. In a similar way we count paths from node $E$ to the only sink node $T$. In this case there are two such paths, $(E,H,J,T)$ and $(E,G,J,T)$ shown with green double headed arrows, giving $X_E=2$. These paths to the sink node $T$ are counted by the $X_v$ values shown by nodes. This then gives the traversal count of $(D,E)$ to be $G_{ED}=W_D X_E=4$. These two SPC main paths are also longest paths (i.e.\ by unit weight) but there are two additional longest paths found by swapping node $B$ for $A$ in the two SPC main paths.
	}
	\label{f:traversalcountex}
\end{figure}

In main path analysis, the aim is to identify a particular path, the \tsedef{main path}, in a DAG (directed acyclic graph). In main path analysis we first give each edge $(u,v)$ a weight $G_{vu}$, which measures the importance of an edge (an edge centrality measure). These are almost always based on some sort of  \tsedef{traversal count} which counts the number of paths\footnote{It is worth contrasting these Main Path traversal counts with the standard network centrality measure edge betweenness (for example see section 11.2 of \citet{C21}). For edge betweenness only shortest paths are used and the initial and final node sets are the set of all nodes $\Vcal$. The path set used for Main Path analysis is closer to what appears in centrality measures such as eigenvector, Katz, PageRank and current based measures of centrality (for example see section 11.4 of \citet{C21}) where counting many paths weighted in different ways gives the value assigned to edges.} 
passing through that edge, as illustrated in \figref{f:traversalcountex}. 
Different variants of Main Path Analysis make different choices for the initial and final node set \citep{LLH19,K22}. 

In our case, we will work only with interval DAGs and we limit ourselves to the simple case where we look at all paths from the single source node $s$ to the single sink node $t$, the \tsedef{Search Path Count} (SPC) approach, available in the social network analysis package \href{http://mrvar.fdv.uni-lj.si/pajek/}{\texttt{pajek}}\footnote{For version \texttt{5.16} look under \emph{Network - Acyclic Network - Create Weighted Network + Vector - Traversal Weights - SPC/SPLC/SPNP}.} and which is one of the ``most popular'' variants according to \citet{K22} (see \appref{as:variants} for a discussion on some other main path variants).

The main path is then the path $\PcalG$ with the largest weight $\LG$ as defined in \eqref{e:LGdef} and \eqref{e:lpGdef}.
Finding such longest paths in terms of edge weights (here the traversal counts) is equivalent to finding the critical path in a scheduling problem \citep{KW59,LTW63,LTW63a}.

\tsenote{Understanding the processes implicit in the traversal count used for main path analysis is essential if we are to understand how main path analysis captures the innovation story. However the traversal count is defined, main path methods are all designed to encode the idea that the higher the value of the traversal count is for a particular edge, the more important an edge is.}

\begin{figure}[htb]
	\begin{center}
		\includegraphics[width=0.75\textwidth]{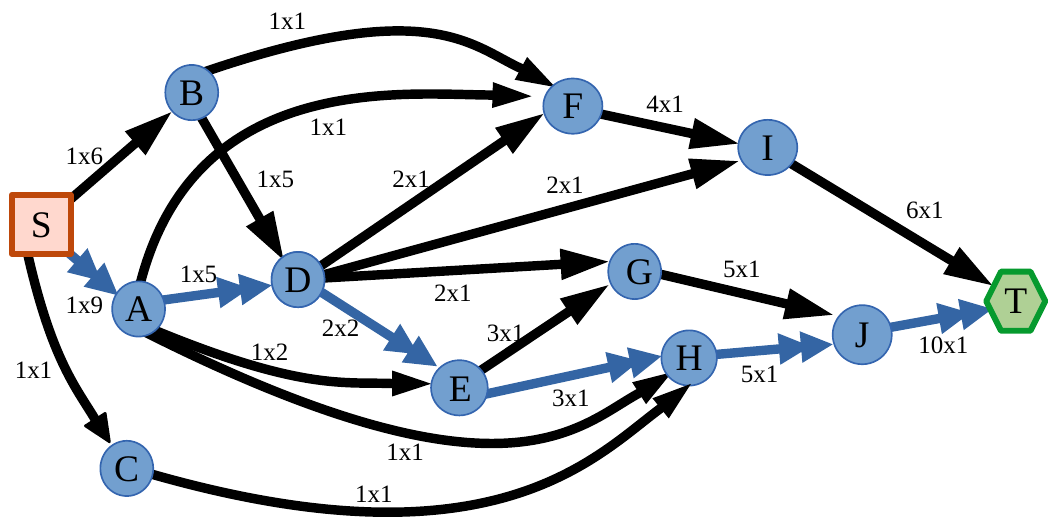}
	\end{center}
	\caption{An example of an SPC main path on an interval DAG. The solid red square indicates the source node $S$ and all paths in the main path analysis start from this. The solid green hexagon is the sink node $T$. The labels on an edge from node $u$ to node $v$ show the $W_u \times X_v$ values where we count every path from the initial nodes to final node with equal weight. The products of these numbers give the edge weight for each edge, the traversal counts $\Gspc_{uv}$. The blue edges with double arrow heads indicate that the path $(S,A,D,E,H,J,T)$ which is one of two possible SPC main paths in this example (we can swap nodes $G$ and $H$ in this main path to find the other main path). 
	}
	\label{f:mpaex}\textit{}
\end{figure}

\subsubsection{Search Path Count}\label{s:mpadetails}

We will start by working with the set of all paths between the source and the sink in which every path carries equal weight.
In this case, the number of paths $\Omega_{vu}$ starting from $u$ and ending at $v$ is
\beq
	\Omega_{vu}
	=
	\left[ \sum_{m=0}^\infty \Amat^m \right]_{vu} \, ,
	\label{e:Omegadef}
\eeq
since our adjacency matrix $A_{vu}$ is equal to one (zero) if there is an edge (no edge) from node $u$ to node $v$.

The number of such paths passing through and edge $(u,v)$ from node $u$ to node $v$ is the traversal count, the edge weight, $\Gspc_{vu}$ and we will call this the \tsedef{search path count}. 
These search path counts are given by\footnote{The central factor of $A_{vu}$ is to ensure that we only have a non-zero traversal count when $u$ and $v$ are connected by an edge.}
\begin{subequations}
	\bea
		\Gspc_{vu}
		&=&
		\Xspc_v
		\, 
		\Amat_{vu}
		\,
		\Wspc_u \, ,
		\label{e:Gspcdef}
		\\
		\Wspc_u
		&=&
		\Omega_{us} \, ,
		\label{e:Wspcdef}
		\\
	\Xspc_v
	&=&
	\Omega_{tv} \, .
	\label{e:Xspcdef}
	\eea
\end{subequations}
The form for the weight $\LG$ of a path $\Pcal$ is the sum of these search path counts $\Gspc_{vu}$ for the edges $\Ecal(\Pcal)$ in the path, so
\bea
	\Lspc(\Pcal)
	&=&
	\sum_{(u,v) \in \Ecal(\Pcal)}
	\Gspc_{vu} 
	\label{e:Mspcdef}
	=
	\sum_{(u,v) \in \Ecal(\Pcal)}
	\Xspc_v \Wspc_u 
	\, .
	\label{e:Mspcdef2}
\eea 
Finally, the SPC main path is the path $\Pcalspc$ that has the largest SPC weight $\Lspc$, formally
\beq
	\Pcalspc 
	= \argmax_{\Pcal} \Lspc(\Pcal) \, .	
	\label{e:spcdef2}
\eeq

\subsubsection{Search Path Entropy (SPE)}\label{s:spe}

When talking about the flow of innovations and information in a network, it makes sense to draw on the tools of information theory \citep{C06c,M19b}. One of the simplest is the entropy of a system. 

In the language of statistical physics, we treat each path as a microstate. Since all paths are equally weighted here, the formula for entropy is simply the logarithm of the number of such microstates (Boltzmann's formula). Then the number of microstates to which an edge from node $u$ to node $v$ contributes is the number of different paths from source $s$ to sink $t$ that pass through the edge $(u,v)$. i.e.\ it is the search path count of the edge $\Gspc_{vu}$ of \eqref{e:Gspcdef}.
We want to use the \emph{entropy} associated with these paths on this edge to define the SPE edge weight $\Gspe_{vu}$ of an edge $(u,v)$. Since all paths are treated equally, this is simply the logarithm of the number of microstates associated with edge $(u,v)$, that is
\begin{subequations}
\bea
	\Gspe_{vu} 
	&=& 
	\ln(\Gspe)
	=
	\Wspe_{u}+\Xspe_{v} 
	\label{e:Gspedef}
	\\
	\Wspe_u
		&=&
		\ln(\Omega_{us} )
		\, ,
		\label{e:Wspedef}	
	\\
	\Xspe_v
	&=&
	\ln(\Omega_{tv} ) \, .
	\label{e:Xspedef}
	\eea
\end{subequations}	
Note that when counting paths we allow for a trivial path of length zero which exists for any node $v$ and which is the sequence of just that one node itself, $\Pcal=\{v\}$. In this way the source node $s$ has zero in-entropy $\Wspe_s =0$ and the sink node $t$ has zero out-entropy $\Xspe_t = 0$.

So the SPE edge weights are the sum of the entropy of the incoming paths from the source node $s$ and the entropy associated with the outgoing paths to the sink node $t$. So rather than counting the paths passing through the edge $(u,v)$ as used in the SPC main path method, we are suggesting we could use the entropy, the information content not the number, of the paths to define these new edge SPE weights. 

Since entropy is additive, it is natural to define the \tsedef{search path entropy} (SPE) weight $\Lspe(\Pcal)$ for a path $\Pcal = (v_i )_{i=0}^L$ by adding these entropy edge weights $\Gspe_{vu}$. That is we define the entropy $\Lspe(\Pcal)$ of a path $\Pcal$ to be 
\bea
	\Lspe(\Pcal)
	&=&
	\sum_{v \in \Pcal}
	\left(\Wspe_v + \Xspe_v \right) \, .
	\label{e:Lspedef}
\eea 
Finally, we define the SPE main path as we did for the SPC case to be the path $\Pcalspe$ running from the source node $s$ to the target node $t$ has the largest weight $\Lspe(\Pcal)$, so
\beq
	\Pcalspe = \argmax_{\Pcal(s,t) } \Lspe(\Pcal) \, .
	\label{e:spedef2}
\eeq

\subsection{Other paths}\label{s:pathother}

\subsubsection{Geodesics}\label{s:geodesic}

Many of the toy models we use to study the properties of paths in a DAG are embedded in a continuous real space of $D$-dimensions. That is every point $\pvec$ in the space can be described by $D$ real coordinates $p_i$ and there are always other points arbitrarily close to point $\pvec$. The nodes of the DAGs formed in some of our models will be a finite number of individual points in this space.

We will use standard (Euclidean) geometry on this space and this enables us to define what is known as a \tsedef{geodesic}, here the shortest curve $\Gamma(\svec,\tvec)$ between two points $\svec$ and $\tvec$. 
The geodesic plays an important role in many applications. Typically, a geodesic is the path of least resistance between two points. For instance the geodesic is the path followed by a particle moving between two points if no other forces are acting on it and it is also the shortest curve as measured by the Euclidean instance between two points and so also the quickest route for a free particle at any given speed. 

For that reason, a key part of our analysis is to compare the paths we form on a DAG embedded in a space to the path of the geodesic of the embedding space. Our central conjecture is that paths in that DAG which are close to the geodesic represent the best route for information to flow through the DAG. 

In a social context, it is difficult to prove our conjecture rigorously. Our goal here is to provide strong evidence for our conjecture, enough to justify the use of certain paths when analysing the properties of DAGs obtained from real data.

However, in a mathematical sense, it is possible to prove rigorously a statement that is closely related to our conjecture. That is it has been shown that the unit length of the longest paths (paths with the longest unit length) in DAGs embedded in what are called ``Box spaces'' scale in very precise manner \citep{BB91}. This enabled \citet{BG91} to show that in simple models of discrete space-time the longest path (by unit length) in a DAG was the best approximation to the geodesic of Minkowski space-time, a result supported by numerical work \citep{ITR06,RW09}.

\subsubsection{Greedy Perpendicular Distance (GPD) Path}\label{s:gpdpath}

In order to compare the various paths in our models to the geodesic, we often use the perpendicular distance $\Delta(v)$ of each point $v$ in a path to the geodesic. 
In terms of paths, this enables us to define another useful path which we will call the \tsedef{Greedy Perpendicular Distance path} (GPD path) $\Pcalgpd(s,t)$ running from node $s$ to node $t$ in a DAG. That is we define this path using a greedy algorithm in which we grow our path $\Pcal$ of length $L$ by adding the successor node $w$ of the last node $v_L$ which has the smallest perpendicular distance $\Delta(w)$. This greedy perpendicular distance path is only available when we have a geodesic so it is only useful for our toy models and not in real data.

\subsubsection{Greedy Degree Path}\label{s:gdpath}

The drawback of using an algorithm based on the properties of the geodesic is that in a network derived from data there is no inherent embedding space and so no geodesic. For practical work, it is often better to use properties intrinsic to any network, based on the network topology alone, when constructing paths. So the \tsedef{Greedy Degree path} is an example of an intrinsic path on a network which is easy to construct. \tsenote{Where has such a path been used before?}

To construct such a path, we just repeat the same greedy algorithm as outlined for the Greedy Perpendicular Distance path in \secref{s:gpdpath} but we replace the selection criteria so that the next node added to the greedy path is one which has the highest degree of all possible successor nodes. Note that as degree is integer and the degree of most nodes in our toy models and in data come from a limited range of values, ties will be common so there are many different Greedy Degree paths in any one network.

\subsubsection{Random Paths}\label{s:randompath}

A \tsedef{random path} $\Pcalrnd(s,t)= (v_0=s, v_1, \ldots, v_L=t)$ starting from node $s$ and finishing at node $t$ is one of the simplest paths we can consider. There are many ways we could choose one path at random from the set of all possible paths, but, in the context of this work, the phrase random path will refer to a path chosen in a very specific way using a greedy algorithm. We start from node $s$ and this is the first node in our path $\Pcalrnd$ so $v_0=s$. We now iterate. Suppose we currently have a path $\Pcalrnd$ of length $n$. We take the last node, $v_n$, in our current random path $\Pcalrnd$ and we choose the next node in the path $v_{n+1}$ uniformly at random from the set of successor nodes $\Ncal^+(v_n) = \{ w | (v_n,w) \in \Ecal\}$. We add $v_{n+1}$ to the end of $\Pcalrnd$. If this $v_{n+1}$ is the target node $t$ then we have found our random path and we stop, otherwise we repeat the iteration.

\tsecomment{This means we are choosing a given path $\Pcal$ with probability $\Pi(\Pcal) = \prod_{e = (u,v) \in \Ecal(\Pcal)} A_{vu}/\kout_u$. That is we are using the (node-based) diffusion matrix rather than adjacency matrix of SPC.}

Note that the path created is largely based on local properties (the existence and number of neighbouring successor nodes. The global properties of the network only play a role when the random path hits the boundary of the interval, usually for a short section at the end of the path. So this iterative construction does not sample the set of all possible paths from $s$ to $t$ uniformly.

\subsection{Comparing paths}\label{s:pathcomp}

One of the objectives of this paper is to compare and contrast the SPC path found using main path analysis with alternative paths e.g.\ SPE, longest (unit length) path, random paths, geodesics, and so on. Our conjecture is that the SPC path, the SPE path, the longest (unit length) path and the geodesic between two nodes are all ``similar'' whereas other paths, such as a random path, between the same two nodes are quite different. To support this conjecture we will need to compare paths and to do this we will use two types of measure.

The first type of measure we use is our generalised length, simply the weight $\LG(\Pcal)$ of a path $\Pcal$. We have defined three edge weights $\Amat$, $\Gmatspc$ and $\Gmatspe$. So for any path $\Pcal$ we can define three path weights, three different ways to measure the length of any path: unit length (number of edges) $\Lunit$, SPC length $\Lspc$ and SPE length $\Lspe$. We can simply compare the values of one of these lengths for different types of path in a given DAG.  

The Euclidean space geodesic does not have a path weight since most of its points are not in the DAG. So to compare the geodesic against paths in the DAGs we must use measures based on the geometry of the continuous Euclidean space. Again this is only possible in toy models where such a space and such a geodesic exists. While there are many measures we could use, when comparing some path $\Pcal(s,t)$ to the geodesic $\Gamma(s,t)$ (both running from node $s$ to node $t$) we will focus on the values of the perpendicular distances $\{ \Delta(v_i) \}$ of the points $v_i \in \Pcal$ on the path to the geodesic. 
We may write the perpendicular distance  in terms of vector algebra as
\bea
\Delta(v)
&=& 
\left( 
|\vvec-\svec|^2 -  \frac{ \big( (\vvec-\svec).(\tvec-\svec) \big)^2}{|\tvec-\svec|^2} \right)^{1/2}
\label{e:perdistdef}
\eea
To get a useful statistic, we focus on the mean $\Deltabar$ and the median $\Deltafifty$ of the perpendicular distances from the geodesic $\Delta(v_i)$ of points  $v_i \in \Pcal$ on a path.

\subsection{Baskets of Nodes}\label{s:basket}

One of the key messages of our work will be that looking at a single path in a DAG which is optimal by some criteria is not very effective.  Reporting the nodes on a single path gives a false sense of success that those nodes are the most important node to the exclusion of \emph{all} others.

When paths are chosen by optimising the path weights which are small integers, it is highly likely that there are many ties. That is there is likely to be a large number of longest and shortest paths by unit length. This path degeneracy problem is avoided when the path weight being optimised is a very large integer, e.g.\ SPC weight, or a floating point number, e.g.\ SPE. Without symmetry, we will not find many ties so there will be very few paths with the same optimal weight. This will be the case when looking at the SPC and SPE main paths. However, in such cases we will see that there are \emph{many} paths with path weights which are \emph{very} close to the optimal.  With real data, it is unlikely the SPC and SPE measures are so good that they guarantee we find the perfect set of important nodes, those on the single optimal path, the SPC or SPE main path. Far more likely is that there are other, slightly sub-optimal paths, that actually find other important nodes, perhaps more important nodes just because of the inherent limitations of using any one measure of importance.
Either way, it seems that relying on a single optimal path from any of these algorithms makes analysis simple but it probably makes it unreliable.

So here we look at nodes in the DAG but which are not always on any of the optimal paths we defined between the source and sink. To do this we will assume that we have already defined the edge weights in $\Gmat$, using unit weights from the adjacency matrix $\Amat$, the SPC weights defined in $\Gmatspc$ of \eqref{e:Gspcdef}, or the SPE weights of $\Gmatentropy$ in \eqref{e:Gspedef}. Then, by analogy with the unweighted case, we define two quantities for \emph{any} node $v$.  First the \tsedef{generalised height} $\hG(v)$ of the node $v$ is the weight of the largest path from the source node $s$ to $v$:
\beq
\hG (v) 
= 
\max_{\Pcal \in \Pcal(s,v)} \LG(\Pcal) \, .
\label{e:hGdef}
\eeq
In the same way we define the \tsedef{generalised depth} $\dG(v)$ of a node $v$ as the largest weight of any path from $v$ to the sink node $t$
\beq
\dG (v) 
= 
\max_{\Pcal \in \Pcal(v,t)} \LG(\Pcal) \, .
\label{e:dGdef}
\eeq
The  \tsedef{generalised height} $\HG(\Dcal)$ of the DAG $\Dcal$ is simply the largest such generalised height (or generalised depth) of any node
\beq
\HG (\Dcal) 
= 
\max_{\Pcal \in \Pcal(s,t)} \LG(\Pcal) \, .
\label{e:HGdef}
\eeq
All of these quantities for the case of unit weights are well known concepts for any DAG. 

Finally, a \tsedef{generalised criticality} $\cG$ of a node $v$ is the difference between the generalised height of the DAG and the sum of the generalised height and depth of node $v$, that is 
\bea
	\cG (v) 
	&=& 
	\HG - \hG(v)-\dG(v) \geq 0 \, .
	\label{e:cGdef}
\eea
It is sometimes useful to work in terms of \tsedef{relative criticality} $\cG(v)/\max_\Dcal{\cG(v)}$ where we normalise using the largest criticality found in the DAG $\Dcal$.\tsenote{I prefer to use $\cG/\HG$.}

Again, the definition of the longest (unit weight) path, SPC main path and SPE main path means that as these are all paths with maximal lengths by their appropriate weighted length $\LG$, we are guaranteed that only nodes only that lie on a maximal path for weight $G$ will have zero criticality associated with that weight, $\cG=0$. Nodes that lie on paths that have lengths $\LG$ that are close to, but a little less, than nodes on the longest paths will still have low values of criticality. So this makes generalised criticality $\cG$ a useful measure for finding such nodes which may well be as important as those on whatever type of maximal path we have chosen to work with.

What we propose is that instead of using a single path to determine important nodes we instead examine a \tsedef{basket of nodes} $\Bcal^\mathrm{(G)}$ for any given weight $G$. Each basket is a set of important nodes that lie on optimal paths or on paths that are close to optimal. So these baskets contain nodes that lie on any of the optimal paths but they will pick up the near misses too. We define our basket $\Bcal^\mathrm{(G)}(f)$ with \tsedef{budget} $f$ to be a set of nodes containing the fraction $f$ of the nodes in the DAG that have the lowest generalised criticality $\cG$ of \eqref{e:cGdef}. 
That is
\beq
	\Bcal^\mathrm{(G)}(f) = 
	\left\{ 
		v \mid v \in \Vcal , \;
		\cG_v \leq \cG_r  
	\right\}
	\, ,
	\quad
	r = \lfloor f \cdot |V| \rfloor
	\, ,
	\label{e:Bdef}
\eeq	
where $\cG_r$ is the $r$-th smallest value among $\{\cG(v) \mid v \in V\}$.

To understand how effective these baskets of nodes might be, we will compare the nodes in a basket $\Bcal$ to the nodes on a path $\Pcal$. The baskets generally contain more nodes than a single path so what we ask is what fraction of nodes $\phi$ from a given path $\Pcal$ are contained in a given basket $\Bcal$. We call this fraction $\phi(\Pcal,\Bcal)$ the \tsedef{basket coverage of a path} and it is simply
\beq
	\phi(\Pcal,\Bcal) = \frac{|\Pcal \cap \Bcal|}{|\Pcal|} \, .
	\label{e:phidef}
\eeq

\subsection{Numerical and Computational Implementation}\label{s:numerics}

Since we are applying our ideas to a variety of formal models as well as to networks from real data, we decided to develop our own open source code \tsenote{\citep{???} BETTER PUT CODE ON GITHUB} and we did not rely on existing main path or general bibliometric packages (for a summary of alternatives, see \appref{as:software}). This enabled us to handle large networks of the type that are common in current work; some of our examples have nearly one million nodes. This enables us to highlight some serious numerical issues that can effect both speed and accuracy.

Implementation of the random geometric DAG model of \secref{s:rgdag} scales as the square of the number of nodes when trying to assign edges in the model.  However, with compiled code we were able to reach networks of $N \sim 10^5$ without difficulty. \tsecomment{HP to update this para}

The calculation of the longest unit path and the main paths are in principle straightforward and fast \citep{B03d}. This is because of the lack of cycles in a DAG which means that standard iterative algorithms can be used to find the node values $\Wspc_v$, $\Xspc_v$, $\Wspe_v$ and $\Xspe_v$ relatively fast, scaling in proportion to the number of nodes plus the number of edges. For instance, we have that 
\beq
	\Wspc_v
	=
	\sum_{u \in \Ncal^-(v) } \Wspc_u \, ,
	\quad
	\Xspc_v
	=
	\sum_{u \in \Ncal^+(v) } \Xspc_u \, ,
\eeq
where $\Ncal^+(v)$ is the number of successor (child) nodes of $v$ while $\Ncal^-(v)$ is the number of predecessor (parent) nodes of $v$. We also have that $\Wspc_s=1$ and $\Xspc_t=1$.  Once these node values are known, they give the edge weights needed immediately. A second pass using similar standard algorithms, so again scaling in proportion to the number of nodes plus edges, will find the longest path using any of these edges weights.  

However, there is a severe problem with the SPC and SPE main path algorithms that does not lead to a massive slow down in the execution times of the code but does have a major impact on development times. The problem is that the search-path counts $\Wspc_v$ and $\Xspc_v$ grow factorially with size in any normal network as can be seen explicitly in the exact forms for $\Omega_{vu}$ given for the lattice model below in \secref{s:lattice}.  
For large networks, these multiplicities reach extreme values: vaccine citation networks routinely exceed $10^{40}$, random geometric DAGs with $N=250,002$ nodes produce SPC weights with 422 digits (see \figref{af:spc250k} in \appref{s:numerics}), and networks with $N=100,000$ nodes yield SPC weights with up to 271 digits.
IEEE 754 double precision represents integers exactly only up to $2^{53} \approx 9\times 10^{15}$. So in almost all programming languages one can not use the generic integer type to calculate the the search-path counts $\Wspc_v$ and $\Xspc_v$. Without those, we can not find the logarithms of these values needed to give  $\Wspe_v$ and $\Xspe_v$. One interesting exception is \texttt{python} which at the time of writing worked with arbitrary precision integers by default. However, our initial implementation in \texttt{python} proved too slow for some of our examples.

Libraries allowing compiled languages to deal with arbitrary sized integers are standard and well tested. Adding these to our code still required extra development time when moving from our initial \texttt{python} implementation. Of course many of the calculations, say for SPE, or when producing statistics for analysis, work in standard floating point variables and ensuring that conversions to these inherent types from library based arbitrary integers also leads to additional development time.

We used arbitrary-precision integer arithmetic when calculating $\Wspc_v$ and $\Xspc_u$ which we do by recasting the sums in \eqref{e:Wspcdef} and \eqref{e:Xspcdef} in an iterative form 
\beq
	\Wspc_v = \sum_{u} A_{vu} \Wspc_u \, ,
	\qquad
	\Xspc_u = \sum_{v} \Xspc_v A_{vu}  \, ,
	\label{e:WXspciter}
\eeq
where $\Amat$ is the adjacency matrix. The initial values are $\Wspc_s=1$ for source node $s$ and $\Xspc_t=1$ for sink node $t$. By evaluating the $\Wspc_v$ in topological order and the $\Xspc_u$ in reverse topological order, we can find these quantities in $O(N+E)$ time.

The search-path entropy (SPE) variant operates in logarithmic space.
SPE weights are derived from SPC path counts using logarithmic transformation: for each edge $(u,v)$, we compute $\ln(\Wspc_u)+ \ln(\Xspc_v)$ and the logarithmic dampening means standard floating point numbers are sufficient and avoid numerical overflow whilst preserving the multiplicative path-count structure essential for identifying high-connectivity routes. 

However, even with double-precision arithmetic, we found there were still numerical problems in distinguishing between the weights of some paths.\tsenote{``at high-symmetry points''??? What does that mean?}
For example, in analysis of our DAG embedded on a three-dimensional cubic lattice (see \secref{s:latticenumerics}), floating-point computations can report up to 58 SPE-critical nodes, but exact arithmetic reveals exactly 54 critical nodes and 2 degenerate optimal paths. While a cubic lattice is highly symmetric, it is not clear that this might not happen with certain motifs in real-world networks. Simple structures, such as symmetric diamonds or near-parallel ladders (see \appref{as:motifs_precision} for some examples and further discussion), can cause the selected path to change due to rounding errors.

This numerical instability provided further motivation for us to move away from the traditional focus on a single `optimal' path and instead motivates our basket-based approach of \secref{s:basket}.

Validation of our numerical results includes comparisons of our numerical results with exact analytical results for DAGs embedded on small lattices where exhaustive enumeration is feasible.

Reproducibility is ensured via deterministic random number generation with seeds determined by network size, connection parameters, and spatial dimension, so identical parameters yield identical networks while differing parameters produce statistically independent realizations.

\section{Hypercubic Lattice Model}\label{s:lattice}

\subsection{Model Definition}\label{s:latticemodel}

The first model we shall use to study Main Path Analysis is the set of DAGs defined on intervals of directed hypercubic lattices in $D$ dimensions. The nodes of the DAG are at points given by integer coordinates, so $v_i \in \Zbb$ ($i \in \{ 1,2,\ldots,D\}$) is the $i$-th coordinate of node $v \in \Vcal$. Note the lattice spacing is one in this model and this is the unit of distance we will use for this model. The directed edges $(u,v)$ are from one node $u$ to a nearest neighbour $v$ where the neighbour has the same coordinates except in one direction, say $j$, where the coordinate of $v$ is one more than the corresponding coordinate of $u$. That is the $j$-th edge leaving node $u$ is to a node at coordinates $v_i = u_i + \delta_{ij}$ with $i,j \in \{ 1,2,\ldots,D\}$ and $\delta_{ij}$ is one if $i=j$ and is zero otherwise. 

This model is extremely unrealistic. It has a lot of structure and symmetry that is not seen in any real complex network. In particular, the network length of \emph{any} path between any two nodes is always the same. However, these are also the properties that allow us to make some analytic calculations. Despite the lack of realism, this hypercubic model illustrates several key features of main path analysis.

We shall further simplify the model to search for the main path in an \tsedef{interval} $\Dcalint(s,t)$ of a hypercubic lattice. This subgraph of the infinite lattice. The interval has nodes and edges that lie on at least one path running from a single source node $s$ to a single sink node $t$ where all sink coordinates are larger than the corresponding source node coordinate, $t_i>s_i$. The node set $\Vcalint$ of this DAG is any node $v$ where the coordinates satisfy $s_i \leq v_i \leq t_i$. The edge set is all lattice edges between nodes in the interval\footnote{So the interval is an induced subgraph, one induced by the node set $\Vcalint$.}. The interval is itself a hypercuboid with the length of each side $L_i$ in lattice units given by $L_i=t_i-s_i$. The total number of nodes $N=|\Vcalint|$ in this interval DAG $\Dcalint(s,t)$ is simply $N= \prod_{i=1}^D (L_i+1)$.

Our hypercuboid intervals are embedded in a section of Euclidean space\footnote{In many ways this Euclidean space is not the most appropriate way to embed our DAG as Euclidean space has no sense of direction. A better approach would be to use a space-time where the time-like direction gives a sense of order to the continuous space that we would match to the order in the DAG. In such cases, the diagonal would still be a geodesic, but now, in the sense used in space-times, it is Lorentzian space-time . The interpretation of geodesics in such space-times is the same; for instance, free particles in special and general relativity follow geodesics, again the line of least resistance.} 
$\Ibb(s,t)$ which is the closed set of all points $x$ whose $i$-th coordinate lies on or between $s_i$ and $t_i$, so $\Ibb(s,t) = \{x | s_i \leq x_i \leq t_i  , \;  x_i \in \Rbb\}$. The nodes in the corresponding interval $\Vcalint(s,t)$ represent a discrete approximation to the hypercuboid piece of continuous space $\Ibb(s,t)$. 

The most important path in the continuous space  $\Ibb(s,t)$ is the straight line from the source node to the target node, the diagonal across our hyper-cuboid section, the line given by $\Gamma(s,t) = \{ (1-\lambda)s + \lambda t| \lambda \in [0,1]\}$, is the geodesic between the two points $s$ and $t$. This is the shortest path between $s$ and $t$ in this continuous space when defining the distance $d(x,y)$ between two points $x$ and $y$ by the Euclidean metric, namely $d_E(x,y) = (\sum_i (y_i-x_i)^2)^{1/2}$.

In the context of main path analysis, the interpretation we want to suggest is that the diagonal $\Gamma(s,t)$
represents the ``easiest'' path from $s$ to $t$ and so this is the ``main path'' in the continuous space interval $\Ibb(s,t)$ that ``information'' will use when travelling from $s$ to $t$ in this model. 
So if we are restricted to work on the points in the interval DAG $\Dcalint(s,t)$, on the discrete points of the interval $\Vcalint(s,t)$, what we are looking for are paths in the discrete space, such as found by main path analysis, that best approximate the geodesic, the diagonal $\Gamma(s,t)$ of our continuous space hypercuboid interval $\Ibb(s,t)$. 

\begin{figure}[htb]
  \begin{center}
  \includegraphics[width=0.45\textwidth]{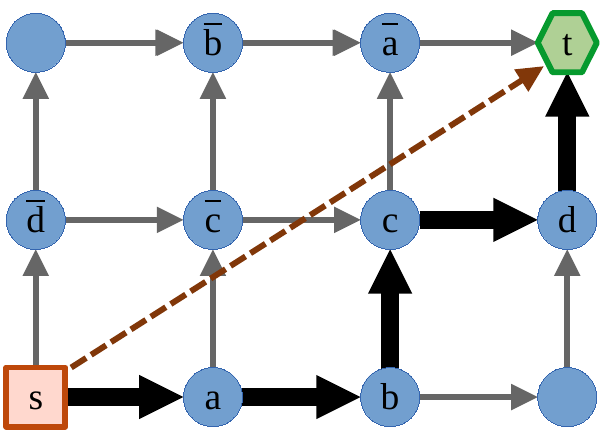}
  \end{center}
  \caption{Illustration of a hypercuboid lattice interval $\Dcalint(s,t)$ in two dimensions, $D=2$, i.e.\ on a rectangle section of a square lattice $(L_1,L_2) = (3,2)$ with a single source node $s=(0,0)$ and a single target node $t=(3,2)$. Edges are between nearest neighbours and are directed either right or up. All the nodes in this interval DAG $\Dcalint(s,t)$ lie on a path running from $s$ to $t$. All paths between any two nodes have the same network length, i.e.\ they contain the same number of edges.  Here all paths from the source to the target have length $\sum_{i=1,2} (t_1-s_i) = 5$, measured in terms of number of edges or in terms of the Euclidean length of each edge. The path shown in thick black arrows is the sequence of nodes $(s,a,b,c,d,t)$ which can be represented as a sequence of directions, here $(R,R,U,R,U)$ with $R$ for right and $U$ for up. The diagonal shown as a red dashed arrow is the shortest path, the geodesic  $\Gamma(s,t)$, between the source and target nodes as measured by the Euclidean distance in the continuous space $\Ibb(s,t)$. 
  Note that there is an order-reversal symmetry in which the direction of the arrows is reversed (equivalent here to a rotation by $180^\circ$). This symmetry links node $a$ to node $\bar{a}$ etc. It also links the path shown to a second path $(s,\bar{d},\bar{c},\bar{b},\bar{a},t)$ which is created using a sequence of moves which is the reverse of the original path, i.e.\ $(U,R, U,R,R)$.
  }
  \label{f:latticeex}
\end{figure}

A simple example of our approach is shown in \figref{f:latticeex} where we show the diagonal (red dashed line), which is defined in the continuous space, and a path on the lattice, shown in thick black lines. We will be aiming to measure the similarity of any given lattice path to this diagonal in a variety of ways. We will be creating specific types of path on the lattice, including the main path, and gathering up results. The conjecture is that if the main path is to be a good representation of the flow of innovation in a citation network then surely it must be a good approximation to the diagonal in this case. It is worth noting that the network length of any path on these intervals is always the same so in this lattice model we cannot study the shortest path or longest paths, i.e.\ paths defined as having the shortest or longest network length.

\begin{figure}[htb]
  \begin{center}
  \raisebox{11pt}{\includegraphics[width=0.45\textwidth]{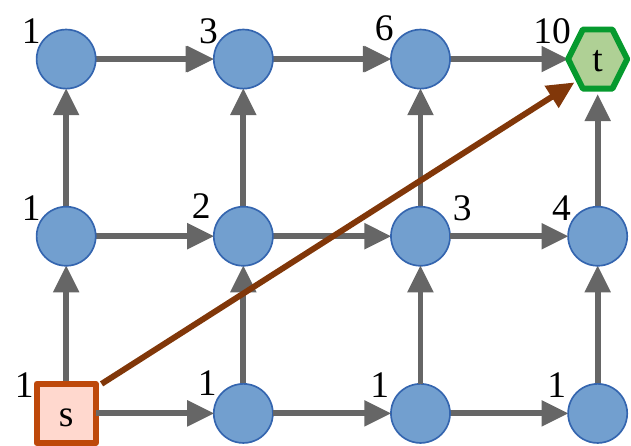}}
  \hspace*{0.05\textwidth}
  \includegraphics[width=0.45\textwidth]{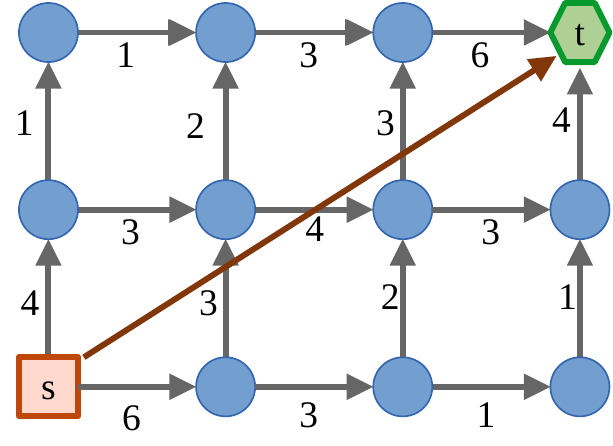}
  \end{center}
  \caption{Diagram illustrating the traversal count edge weight used in main path analysis on a square lattice. The paths all run from a single source $s$ and to a single target node $t$, so this is an example of an SPC main path analysis. On the left, we show the number of paths $\Wspc_u$ of \eqref{e:Wspcdef} reaching a vertex from $s$ showing how these values are a section of Pascal's triangle. On the right, we show the SPC weight $\Gspc_{vu}$ \eqref{e:Gspcdef} of each edge $(u,v)$. 
  The order-reversal symmetry of this directed lattice is evident from the fact that the edge labels in the right-hand network do not change under a $180^\circ$ rotation.
  }
  \label{f:traversalcountsqlatex}
\end{figure}

Finally, there are features seen in the numerical results that reflect some of the symmetry of this model. In this model we have an \tsedef{order-reversal symmetry}\footnote{More formally this is a representation of a $Z_2$ symmetry group. However, the implications of this simple symmetry group are very easy to find without using the language of discrete groups and their representations.} $\That$ where we reverse the direction of all the edges in our hypercubic DAGs to create a second DAG $\Dcalrev$. If the nodes are unlabelled, this transformation gives us exactly the same DAG. 
For our illustrations of DAGs in two dimensions, this corresponds to a symmetry under a $180^\circ$ rotation and a reversal of the direction on the edges. 
However, when we include node labels, we see that every node is paired up with a second node by this symmetry. Specifically, node $v$ with coordinate $v_i$ is paired with a node $\vbar=\That(v)$ with coordinate $\vbar_i = t_i-v_i+s_i$ where in the original DAG $\Dcal$ the source node has coordinates $s_i$ and the sink node has coordinates $t_i$. These pairs of nodes $v$ and $\vbar=\That(v)$ linked by order reversal symmetry are always distinct except when the length of the sides of the hypercube are all even when there is one node at the centre mapped onto itself, a case we avoid in all our illustrations. 
An example of this is shown in \figref{af:pd} in \appref{as:pppdeltarandom}.

What does this order-reversal symmetry mean for paths? Every path $\Pcal = (v_i | i \in \{0,1,2,\ldots,L\} )$ defined on the original DAG $\Dcal$ is matched by this symmetry to a second path $\Pcal' = \That(\Pcal) = (v_i^\prime)$ where $v_i^\prime = \vbar_{L-i} = \That(v_{L-i})$. Now there are two possible cases. Either the two paths are identical, or the order-reversal symmetry matches two distinct paths. The nature of the symmetry means that the properties of paths linked by symmetry are identical. 

For instance, if we applied these ideas in the continuous space, we would see the geodesic is symmetric, i.e.\ the order reversal symmetry does not create a second geodesic (and it is known that every geodesic between any two points in Euclidean space is unique). Most paths on the discrete lattice are related by this reversal symmetry to a distinct path. The example of the path shown in \figref{f:latticeex} is typical. 

\subsection{Analytic Results}\label{s:latticeanalysis}

There are two main goals in this section. First, the calculations illustrate how the entropic viewpoint motivates the SPE main path measures. Secondly, we can show explicitly in what sense the paths with the largest SPE weights are guaranteed to have lengths similar to the geodesic in simple interval DAGs.

One of the special properties of this hypercubic lattice DAG model is that all paths from the source node $s$ at the origin (so $s_i=0$) to the target node $t$ have the same network length which is equal to $L= \sum_i (t_i-s_i)$. Every path is made of $(t_i-s_i)$ steps along an edge in direction $i$. In fact each path can be represented as a sequence of directions $i$ where each direction $i$ appears $(t_i-s_i)$ times. The set of all possible paths can be represented as the permutations of the same set of $D$ directions. Thus the number of paths from any node $u$ to any node $v$ in this hyper-cubic lattice is $\Omega_{vu}$ where
\beq 
 \Omega_{vu} = \frac{\big(\sum_i(v_i-u_i)\big)!}{\prod_i\big((v_i-u_i)!\big)}
 \label{e:Omegasqlat}
\eeq
For example, consider the paths from node $a$ to node $c$ in \figref{f:latticeex}. These paths from $a$ to $c$  must contain one step to the right to ensure $c_1=a_1+1$ and one step up so we get to ensure $c_2=a_2+1$. These steps can be in any order as we will still get a legal path. So here we have the path right then up $(R,U)$ or the only alternative is up first then right $(U,R)$. So the set of all paths from node $a$ to node $c$ is the set of all permutations of the sequence $(R,U)$ and using \eqref{e:Omegasqlat} gives us $\Omega_{ca}=2$ for the two paths we have here.

These permutations counts have some simple properties and in particular if $L_i=(v_i-u_i)$ then $L=\sum_i L_i$ is the network length of the path from $u$ to $v$. Then we can shown that for large $L$ and provided no one side of the rectangular interval is small ($1 \ll L_i \lesssim L$) is we have (using Stirling's formula) 
\beq 
 \lim_{L \to \infty} \ln \big(\Omega_{vu} \big) 
 = 
 L \ln(L)- \sum_{i=1}^D L_i \ln(L_i) 
 \quad \text{if} \quad 
 \frac{L_i}{L} \sim O(1) \, .
 \label{e:Omegasqlat2}
\eeq
The largest value for the second term comes when $L_i \approx L/D$ for all $i$ (by symmetry and the method of Lagrange multipliers for constrained optimization) in which case we find 
\beq 
 \lim_{L \to \infty} \Omega_{vu} =  D^L \, .
 \label{e:Omegasqlat3}
\eeq
This is the expected answer as at each of the $L$ steps we have $D$ directions we can move in. The exception is at the boundaries of the interval $I(u,v)$, but these are a vanishingly small fraction $O(1/L)$ of the total number of nodes in the interval.
This also shows that the number of paths between two vertices grows extremely fast as the path gets longer\footnote{The only exception is when the path is taken to infinite length in one direction only, that is, if $L_i$ is finite for all but one direction. However, such cases were excluded by the assumption that $L_i/L \sim O(1)$.}.

From this, we see that the SPC uses the weights
\beq
 \Wspc_u
 =
 \Omega_{us}
 \, , \qquad
 \Xspc_v
 =
  \Omega_{tv} \, .
  \label{e:WXsqlat}
\eeq

For the entropy version of the edge weights we need
\begin{subequations}	
\bea
\Wspe_u
 =
 \ln\left(\Omega_{us}(s,u) \right)
 &\approx& 
 \sum_{i=1}^D  - (u_{i}-s_{i}) \ln\left(\frac{(u_{i}-s_{i}) }{|(u-s) |_1}\right) \, ,  \label{e:lnWsqlat}
 \\
 \Xspe_v
 =
 \ln\left( \Omega_{tv} \right)
 &\approx&
  \sum_{i=1}^D  -(t_{i}-v_{i}) \ln\left(\frac{(t_{i}-v_{i}) }{|(t-v) |_1}\right) \, ,
  \label{e:lnXsqlat}
  \\
  && \qquad \qquad \text{with} \;\; |\xvec|_1 =\sum_{i=1}^D x_i 
  \, .
\eea
\end{subequations}

This then gives a path weight using the search path entropy weights of 
\bea
 \Lspe(\Pcal) 
 &\approx&
 \sum_{v \in \Pcal(s,t)}
 \sum_{i=1}^D 
 \left(
 -(v_i-s_i) \ln \left( \frac{(v_i-s_i)}{|\vvec-\svec|_1} \right)
 -(t_i-v_i) \ln \left( \frac{(t_i-v_i)}{|\tvec-\vvec|_1} \right)
 \right)
 \label{e:Mentropysqlat}
 \, .
\eea 

At this point we see that our approximate expression for $\Lspe(\Pcal)$ on the hypercubic lattice is given in terms of an entropy ``distance'' measure $\dentropy$ where
\bea
 \dentropy(u,v) 
 &=& 
 -\sum_{i=1}^D  (v_i-u_i) \ln \left( \frac{(v_i-u_i)}{|\vvec-\uvec|_1} \right) 
 = 
 -|\vvec-\uvec|_1\sum_{i=1}^D  p_i \ln \left( p_i \right) 
 \, ,
  \label{e:dentropydef}
 \\
 && 
 \quad  \text{where} \quad 
 v_i \geq u_i \, , 
 \quad
 p_i = \frac{(v_i-u_i)}{|\vvec-\uvec|_1} \, ,
 \quad \forall i \in \{1,2,\ldots,D\} \, .
\eea
Note that $ 0 \leq \dentropy(u,v) \leq |\vvec-\uvec|_1 \ln(D)$.
This entropy ``distance'' measure can be defined in any real space $\Rbb^D$.
In the specific context of our lattice model, we find that
\bea
\Lspe(\Pcal)
&\approx&
\Mdentropy(\Pcal)
=
\sum_{v \in \Pcal}
\left(
\dentropy(s,v) + \dentropy(v,t)
\right) 
\, .
\label{e:Mdentropy}
\eea 

This entropy ``distance'' measure $\dentropy$ is a ``Lorentzian distance'', that is it satisfies (i) $\dentropy(u,u)=0$ (if we take the limit $v \to u$ in $\dentropy(u,v)$ carefully), (ii) $\dentropy(u,v) \geq 0$ provided there is a path from $u$ to $v$, and (iii) a reverse-triangle identity
\beq
 \dentropy(u,w) \geq \dentropy(u,v) + \dentropy(v,w) 
 \, ,
 \quad  \text{where} \quad 
 u_i \leq  v_i \leq w_i  \;\; \forall i \in \{1,2,\ldots,D\} \, .
 \label{e:dentropyrevtri}
\eeq
Equality in \eqref{e:dentropyrevtri} only occurs when point $v$ lies on the ``diagonal'' in a Euclidean sense, that is when $v_i = (1-\lambda) u_i + \lambda w_i$ with $0 \leq \lambda \leq 1$. In fact, this diagonal is also the geodesic in terms of this Lorentzian distance, though the geodesic is now the path with the \emph{longest} entropy distance between two points, reinforcing the central role of the diagonal in our hypercube interval DAGs even when we are not measuring distances with a Riemannian metric. 

We then see that we have that the upper bound on $\Mdentropy(\Pcal)$ in \eqref{e:Mdentropy} comes when every node on the path $\Pcal$ is on the geodesic. So we deduce that
\tsecomment{(CHECK THIS BOUND IN LATTICE? ALSO CAN WE USE IT FOR SPE IN PPP MODEL.)}
\bea
 \Lspe(\Pcal)
 &\approx&
 \Mdentropy(\Pcal)
 \;\; \leq \;\; 
  (L+1) \dentropy(s,t)
  \;\; \leq \;\;
  (L+1) |\tvec-\svec|_1 \ln(D)
 \label{e:Mdentropyrtri}
\eea 
where the number of nodes in the path $\Pcal$ is $|\Pcal|=(L+1)$ which is one more than the number of edges, the length $L$ of the path  $\Pcal$. 

Our DAGs are always an interval, that is with a single source node $s$ and a single sink node $t$ and where all nodes in the DAG lie on a path from $s$ to $t$. This is the case with both of the toy models we use in numerical work below. In such cases we have a single geodesic $\Gamma(s,t)$ from $s$ to $t$ given by the points $\Gamma(s,t) = \{ \lambda \tvec + (1- \lambda) \svec | \lambda \in [0,1] \}$.
This is a geodesic in a Lorentzian sense in that the straight line\footnote{Alternatively, if we work with a Euclidean metric, this straight line from $s$ to $t$ is the usual geodesic for that metric.} 
has the \emph{longest} length between two points when we use the Lorenzian distance measure $\dentropy$. 

The corollary is that when we consider a node $v$ lying on some path $\Pcal$ from $s$ to $t$, the `further' node $v$ is from this geodesic, then the smaller the contribution from $v$ is to the value of $\Mdentropy(\Pcal)$ in \eqref{e:Mdentropy}. This follows from \eqref{e:dentropyrevtri}. Since  $\Mdentropy(\Pcal) \approx \Lspe(\Pcal)$
we expect that finding the path with the largest search path entropy weight is equivalent to finding a path which has the closest SPE weight of the geodesic $\Lspe(\Gamma(s,t))$.

\subsection{Numerical Analysis of Lattice Paths}\label{s:latticenumerics}

The symmetries of an interval $\Dcalint(s,t)$ of a hypercubic lattice mean that if the lengths of the sides $L_i=(t_i - s_i)$ all share a common denominator, say $M$ so $L_i/M \in \Zbb^+ \; \forall \; i$. Then the diagonal, the geodesic, will pass through $(M-1)$ lattice points between $s$ and $t$, namely at $v_i = s_i + (m/M)(t_i - s_i)$ where $0<m<M$. In such a situation we will see periodic behaviour in the paths with the main paths of interest passing through these lattice points on the diagonal and then repeating their shape after that. So in much of our analysis, we choose the lengths of the sides of our intervals sections to have no common factors to avoid such periodic effects. At the same time, we do not want to work with one side much shorter than another so we also choose intervals where $|L_i-L_j|$ are small.

We note that even relatively small lattices generate very large numbers for some of the weights. This means that the numerical code has to be extremely carefully crafted to get the correct answer for SPC and SPE paths, which we know exactly based on our analytical expressions. 
\tsecomment{We discuss one example, the three-dimensional lattice $(15,17,19)$, in more detail in \appref{as:latcritexample}.} 
While this is bound to occur here because of the high symmetry of the hypercubic lattice, it remains a problem in general DAGs because of certain motifs in the network, see \appref{as:motifs_precision}.

\subsubsection{Path Weights}\label{s:latpathweight}

The simplest way to compare discrete paths between the same nodes in a DAG is to measure the weight of each path (the path length) using the two weights used for our main paths, the SPC weight and the SPE weight. 
In \tabref{t:lattpathweights} we have shown the results for various paths on an interval of a three-dimensional cubic lattice. 
The two main paths and the greedy perpendicular distance path are very similar in all their path weights but the small differences show that these three paths find very similar paths with small differences. Yet they must all be very close to the geodesic given that the greedy perpendicular distance is optimised to do that.
\tsenote{I THINK THEY CAN READ THE TABLE. ``the SPC main path achieves the highest weight (7.185 × $10^{22}$), followed by the SPE main path (7.183 × $10^{22}$) and the greedy perpendicular distance algorithm (7.181 × $10^{22}$).''} 
\tsenote{The slight differences in absolute values of the SPC weights show that , where only two degenerate optimal paths exist among the twenty trillion or so possible paths. However, under SPE weighting, all three paths achieve identical weights (2.466 × $10^{3}$), indicating they all lie within the broader SPE critical region that admits 24 degenerate optimal paths.} 
The random path shows significantly different weights under both measures, providing a baseline for comparison.

\begin{table}[htb!]
  \centering
  \begin{tabular}{l|| c | c }
    Path &  \textbf{SPC} $\Lspc$ (×$10^{22}$) & \textbf{SPE} $\Lspe$ (×$10^{3}$) \\
    \hline
    & & \\[\dimexpr-\normalbaselineskip+2pt] 
    SPC main path      & 7.185 & 2.466 \\
    SPE main path      & 7.183 & 2.466 \\
    Greedy Perp.Dist.  & 7.181 & 2.466 \\
    Random             & 2.964 & 2.320 \\
  \end{tabular}
  \caption{Path weights for paths from $(0,0,0)$ to $(15,17,19)$ on a cubic lattice. All paths have length (unit weight) of $15+17+19=51$ steps. SPC values are shown in units of $10^{22}$ and SPE values in units of $10^{3}$. Under SPC weighting, the three algorithms find slightly different paths with weights 7.185, 7.183, and 7.181 respectively. However, under SPE weighting, all three paths achieve the identical weight of 2.466, indicating they all lie within the SPE critical region. 
  	Note that the SPE weights $\Lspe$ are consistent with the upper bound from \eqref{e:Mdentropyrtri}, here $51 \times 52 \ln(3) \approx 2.915 \times 10^{3}$.
  	The random path shows significantly different weights, providing a baseline. All values are computational results shown to four significant figures.}
  \label{t:lattpathweights}
\end{table}

\subsubsection{Perpendicular Distance to Geodesic}\label{s:latPerDistGeo}

\begin{figure}[h!]

	\centering
	\includegraphics[width=0.75\linewidth]{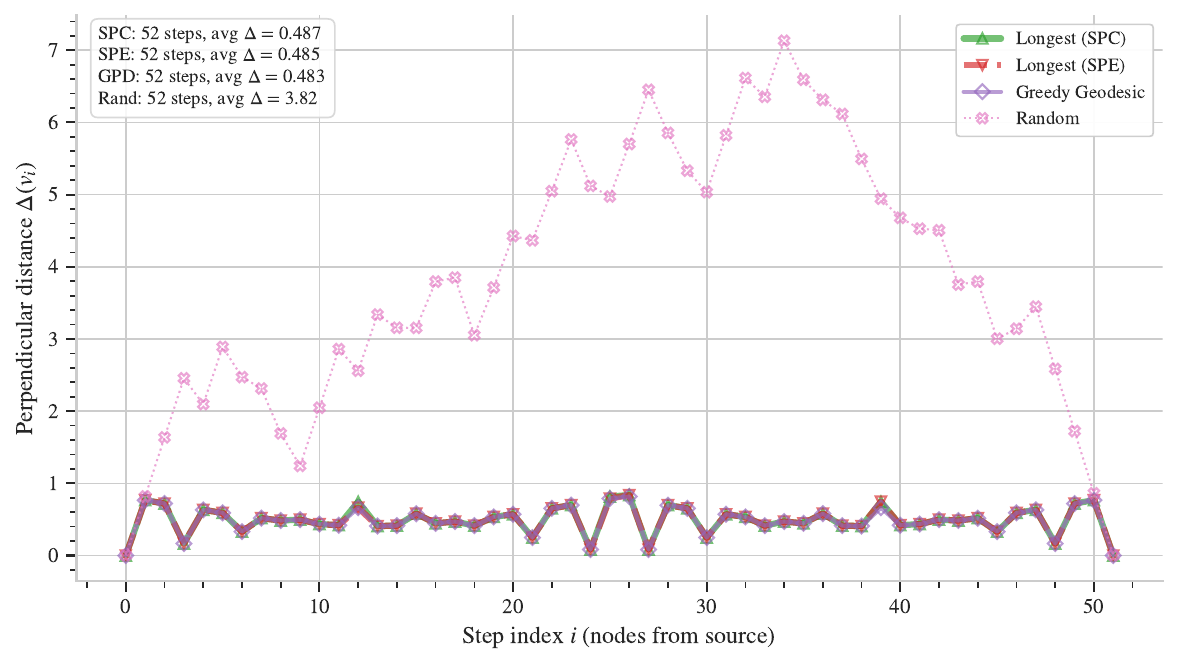}
	\caption{The perpendicular distance $\Delta (v_i)$ of \eqref{e:perdistdef} for nodes $v_i$ in various paths from $\svec=(0,0,0)$ to $\tvec=(15,17,19)$ on a cubic lattice. Each point represents a node $v_i$ at step-index $i$; lines aid visualisation. The SPC and SPE main path paths (``longest (SPC)'' and ``longest SPE'' respectively) and the greedy perpendicular distance path (``GPD'', and ``greedy (PD)'') remain within one lattice spacing of the geodesic and are almost identical (points 12 and 39 show a very small difference). A random path on the other hand wanders a significant distance from the geodesic until it hits the boundary (here around 34 steps). 
	\tsecomment{Need to change ``Greedy Geodesic'' on right. Maybe ``Greedy (GPD)''?}
	\\
	\tsecomment{Make the keys consistent across graphs and tables? I suggest we follow the standard path-name abbreviations introduced above. So ``Greedy geodesic'' becomes ``Greedy (GPD)''. In any case we are using two different labels in the two keys on the same graph. Maybe if space is a critical issue we can also use ``SPC'', ``SPE'', ``GPD'' and ``random'' and then define these in the caption. }
	}
	\label{fig:distancetodiagonal}
\end{figure}

The results for the perpendicular distances of nodes to geodesic for the same paths used in \tabref{t:lattpathweights} are shown in  \figref{fig:distancetodiagonal}.   Although the results in \figref{fig:distancetodiagonal} are for one lattice, they illustrate features we see in all lattices. 

\begin{table}[htb!]
	\centering
	\begin{tabular}{l|| c | c | c}
		\hline\hline
		Path               & Mean $\Delta$   & Median $\Delta$ & Max $\Delta$ \\
		                   & (lattice units) & (lattice units) & (lattice units) \\
		\hline
		& & & \\[\dimexpr-\normalbaselineskip+2pt] 
		Longest (SPC)      & 0.4870 & 0.4910 & 0.8380 \\
		Longest (SPE)      & 0.4850 & 0.4910 & 0.8380 \\
		Greedy  (GPD)      & 0.4830 & 0.4910 & 0.8170 \\
		Random             & 3.820  & 3.774  & 7.132 \\
		\hline\hline
	\end{tabular}
  \caption{Perpendicular distance statistics $\Delta(v_i)$ for paths from $(0,0,0)$ to $(15,17,19)$ on the same cubic lattice as \tabref{t:lattpathweights}. Means, medians, and maxima are in units of the lattice spacing $a=1$. The greedy perpendicular distance (GPD) path attains the lowest mean ($0.4830$), while SPC and SPE main paths are indistinguishable at this precision (both $\approx 0.49$ lattice units from the geodesic). The random path provides a baseline for comparison.}
  \label{t:perpendiculardistances}
\end{table}

The random path is provided to provide a simple reference point. It behaves as unconstrained random walk and the deviation of the path from the geodesic will on average,grow as $\sqrt{\ell}$ for the $\ell$-th node on the random path. This behaviour stops when the random path hits the boundary of the interval which is defined by the sink node (a global constraint), at which point the random path is pushed towards the target node. 

The remaining paths are almost, but not exactly, identical in terms of the perpendicular distance to the geodesic, $\Delta$. The summary statistics in \tabref{t:perpendiculardistances} back this up. A close examination of  \figref{fig:distancetodiagonal} (confirmed by examining the data) shows that points 12 and 39 on the paths have slightly different values of $\Delta(v_i)$. The same figure on close inspection also shows that the point with largest $\Delta$ is point 26, which from the table also has a $2\%$ difference between GPD and the SPC/SPE paths. Yet, they are all optimal paths and their SPE and SPC path weights are identical to four significant figures.  What is happening is that on the highly symmetric lattice, there can be several distinct paths all of which are optimal by one measure, such as SPC, SPE weight or equally good under the GPD algorithm.  However, other properties of these distinct but optimal paths, such as $\Delta$, will almost certainly be different between different examples of optimal paths. Since the set of optimal paths are likely to follow very similar trajectories, the differences in these other measures such as $\Delta$ are likely to be small, exactly as we see here.

While this degeneracy is likely to be high on the hypercubic lattice, this $(15,17,19)$ example of \figref{fig:distancetodiagonal} highlights an important issue. Namely, main path algorithms choose a single path, regardless of whether there are other paths with SPC or SPE path weights which are equal, or even very close. We will return to this issue in more realistic cases.

A more practical message from the example of \figref{fig:distancetodiagonal} is that there is little to choose between the SPC and SPE main paths and the GPD path. They all maintain average distances from the geodesic of less than $0.49$ lattice units, with the greedy algorithm achieving the best result by a very small margin. 
This geometric proximity, combined with identical path weights, demonstrates that for cubic lattices, these different optimisation criteria give paths, main paths, which are all very similar and which are all very close to the geodesic.

\subsection{Criticality of Points}\label{s:lattice_criticality}

We have seen that some of the variability in the properties of paths comes because there is not always one path that optimises any given criterion. Even if degenerate optimal paths are not common in citation networks, it is very possible that a key paper of interest may lie on a main path that has a path weight that is just slightly less than the maximum value. That is, relying on one path even if defined using the best algorithm possible is no guarantee that it will not miss something. Quantitative measures of social concepts are never perfect. 

To address such uncertainties, we can use the node criticality $\cG$ of \eqref{e:cGdef} to select important nodes. Here we look at the criticality of nodes on the hypercuboid intervals of a hypercubic lattice.

\begin{figure}[htb]
	\centering
	\begin{minipage}[b]{0.49\textwidth}
		\centering
		\includegraphics[width=\textwidth]{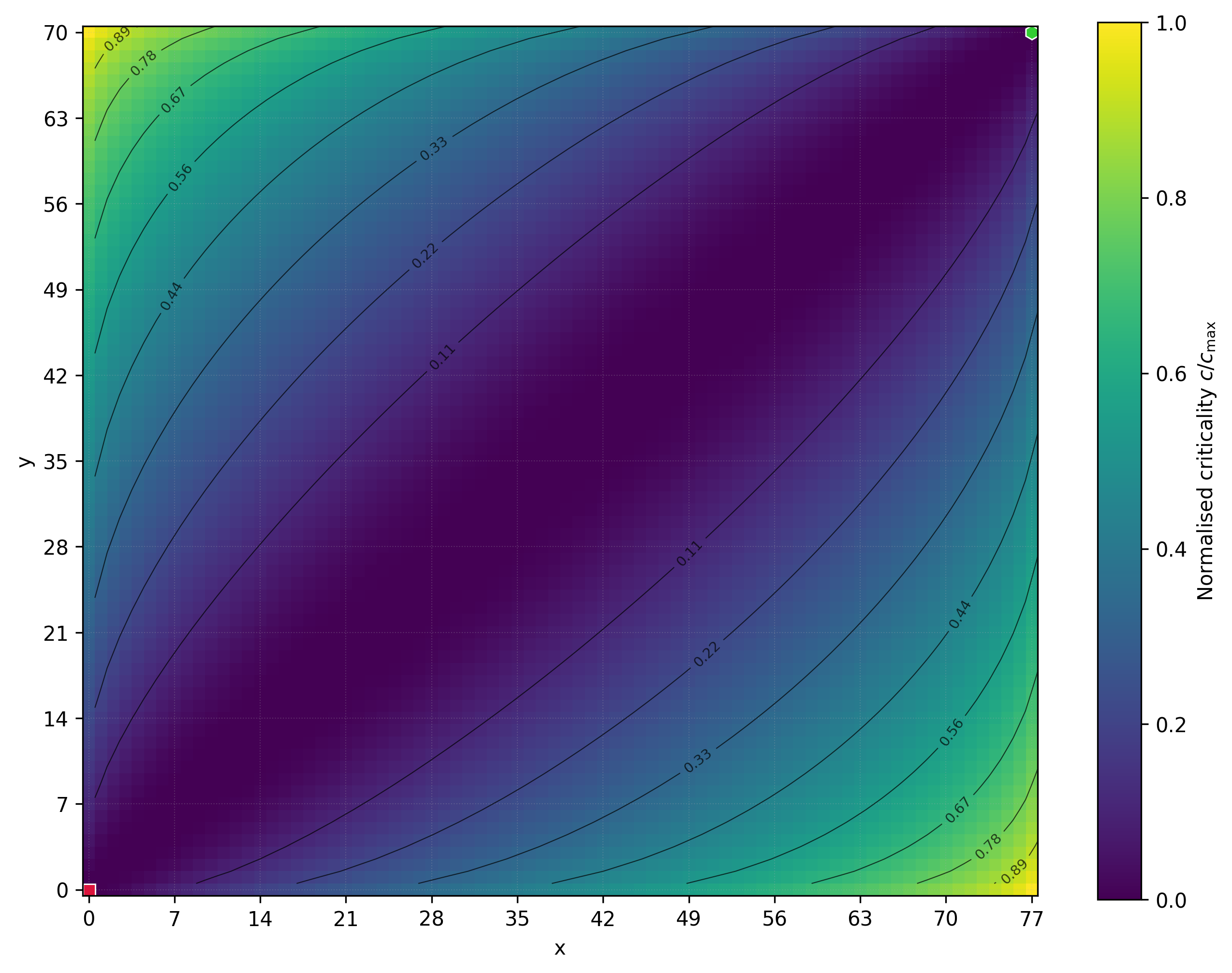}
		\caption*{(a) SPC}
	\end{minipage}
	\hfill
	\begin{minipage}[b]{0.49\textwidth}
		\centering
		\includegraphics[width=\textwidth]{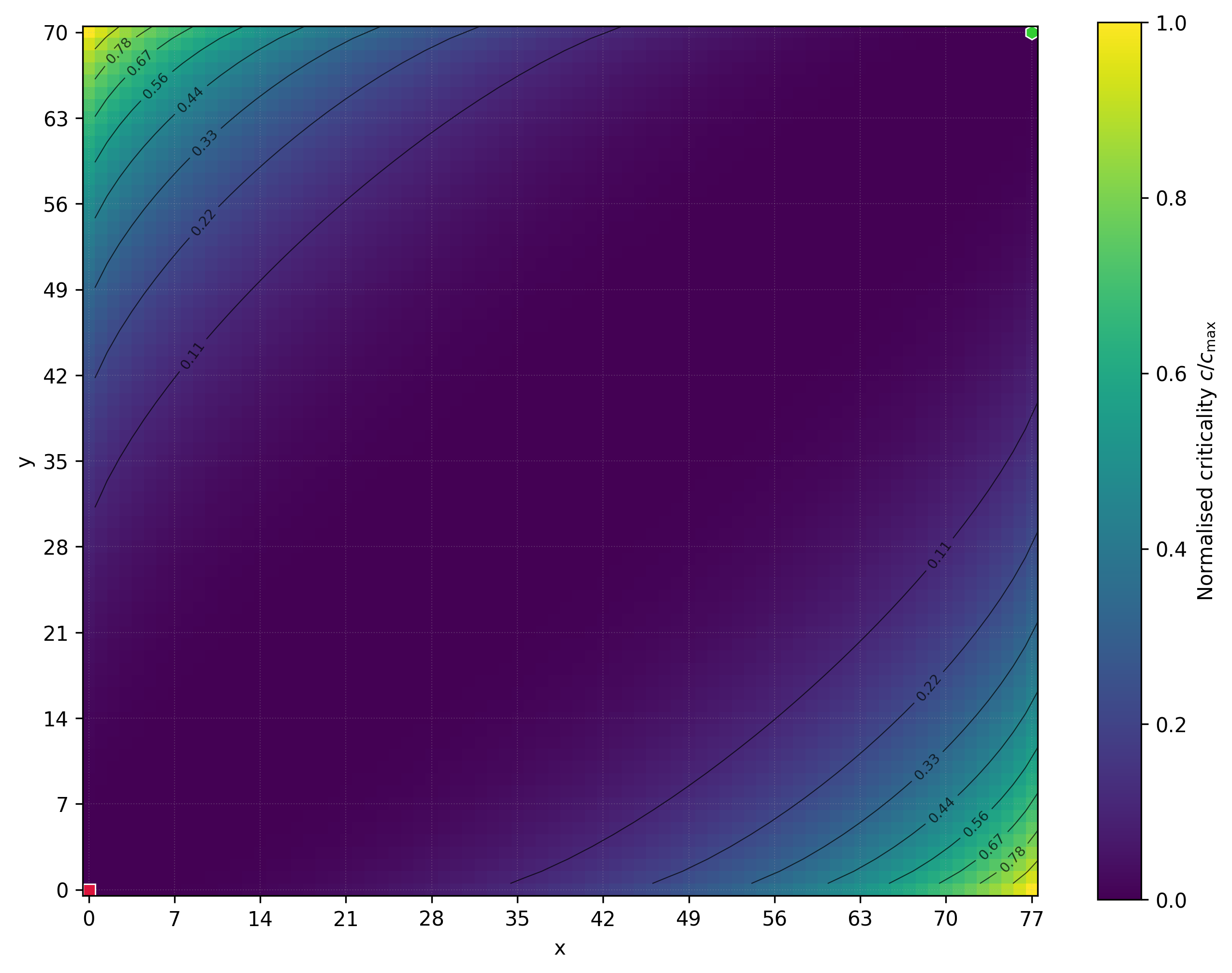}
		\caption*{(b) SPE}
	\end{minipage}
	\caption{The relative criticality $\cG/c_{max}$ of \eqref{e:cGdef} for every point on a $78 \times 71$ section of a square lattice with source node $s=(0,0)$ and sink node $t=(78,71)$. Panel (a) shows the normalised SPC criticality $\cspc/c_{max}$ and panel (b) shows the normalised SPE criticality $\cspe/c_{max}$. Both measures highlight regions close to the geodesic, the diagonal from bottom left to top right, as most critical.
	}
	\label{f:latcrit-78-71}
\end{figure}

The results in \figref{f:latcrit-78-71} show of a  $78 \times 71$ lattice ($5,538$ nodes, $10,991$ edges) as a heat map for and these are typical of what we find in general. While individual optimal paths (SPC and SPE main paths) pick out specific node sequences close to the geodesic, the criticality measures highlight entire regions of near-optimal nodes close to the diagonal, the geodesic, revealing the geometric structure of the solution space. 

\begin{figure}[htb]
	\centering
	\begin{minipage}[b]{0.45\textwidth}
		\centering
		\includegraphics[width=\textwidth]{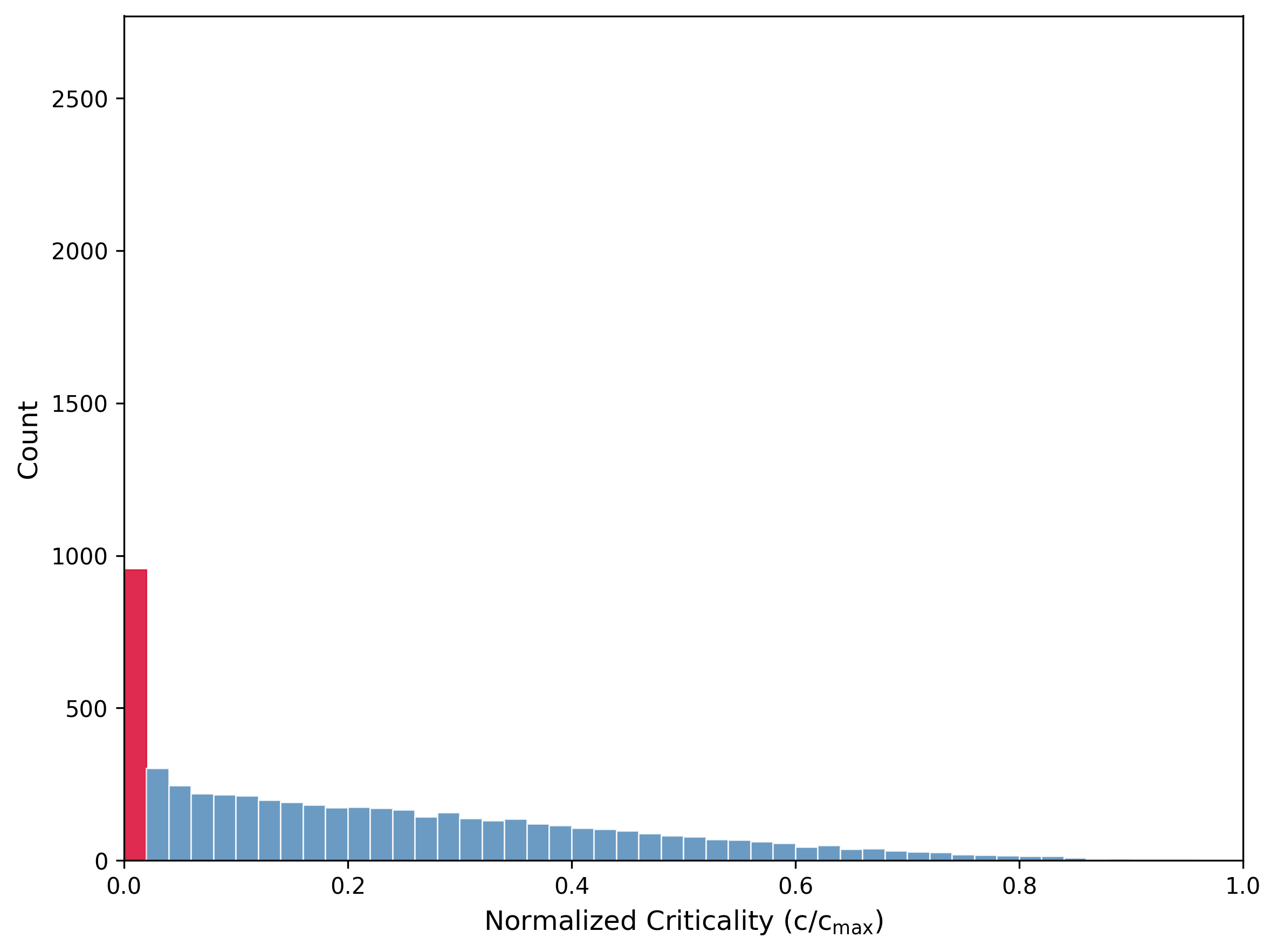}
		\caption*{(a) SPC Criticality Distribution}
	\end{minipage}
	\hspace*{0.05\textwidth}
	\begin{minipage}[b]{0.45\textwidth}
		\centering
		\includegraphics[width=\textwidth]{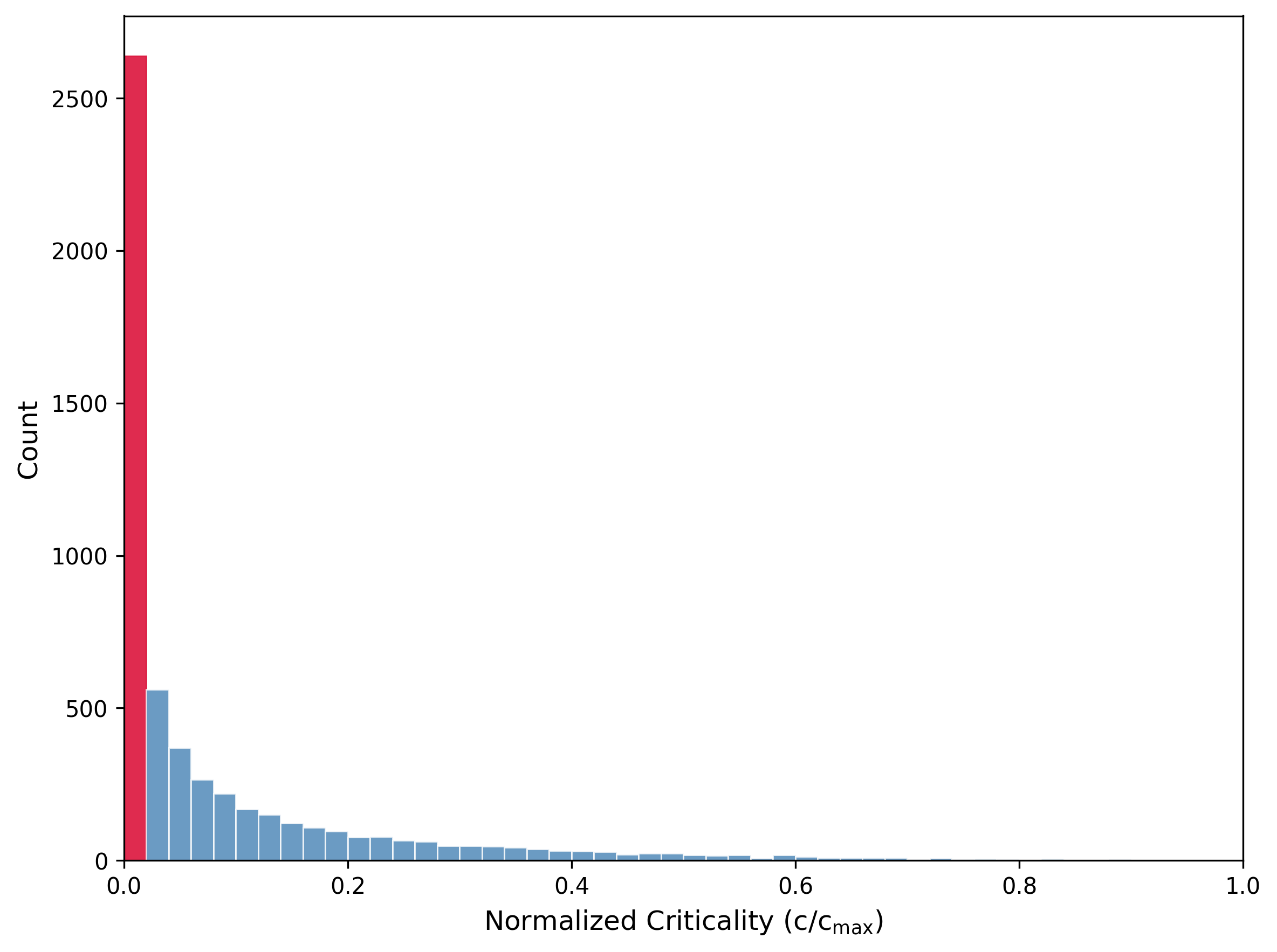}
		\caption*{(b) SPE Criticality Distribution}
	\end{minipage}
	\caption{Histograms of normalised criticality values for the $78 \times 71$ lattice. Panel (a) shows the SPC criticality distribution, which is highly concentrated with most nodes having zero criticality and only 2.7\% of nodes having any criticality. Panel (b) shows the SPE criticality distribution, which has a broader spread with 61.3\% of nodes having criticality $\le 0.05$ and 73.2\% having criticality $\le 0.10$.
	}
	\label{f:latcrit-hist-78-71}
\end{figure}

The criticality analysis such as in \figref{f:latcrit-78-71} shows that criticality behaves very differently under the SPC and SPE weighting schemes. For these lattice interval DAGs the nodes with low SPC criticality $\cspc$ are much more sharply peaked around the geodesic than the nodes with low SPE criticality. This is shown in more detail in the histograms in \figref{f:latcrit-hist-78-71}.


We can quantify this further in \tabref{tab:lattice-criticality-summary}.
This shows that SPC and SPE have similar numbers of critical nodes\footnote{See Section~\ref{s:numerics} for details on precision requirements, arbitrary-precision arithmetic, and SPE logarithmic transformation implementation.}, and that substantially more nodes fall below low relative criticality thresholds under SPE than SPC.
\begin{table}[htb]
  \centering
  \begin{tabular}{c||c|c|c|c}
    Path             & $c/c_{\max}\le 0.01$ & $c/c_{\max}\le 0.05$ & $c/c_{\max}\le 0.10$ & $c/c_{\max}\le 0.20$ \\ \hline
    Longest (SPC)    & 13.3\% & 24.8\% & 35.0\% & 52.3\% \\
    Longest (SPE)    & 38.5\% & 61.3\% & 73.2\% & 84.8\% \\
  \end{tabular}
  \caption{Fraction of nodes with criticality fraction $c/c_{\max}$ at or below 1\%, 5\%, 10\%, and 20\% for the $78\times71$ lattice ($N=5538$). Values are computed exactly over the entire lattice.}
  \label{tab:lattice-criticality-summary}
\end{table}



\clearpage

\section{Random geometric DAG}\label{s:rgdag}

While the hypercubic lattice model of \secref{s:lattice} provides valuable analytical insights, its highly regular structure is far removed from the irregular connectivity patterns observed in real citation networks. 
To address this issue, our second model moves towards greater realism by introducing randomness into the node placement while maintaining the essential geometric structure needed for our analysis.

Our approach builds on the well-established framework of random geometric graphs, originally due to \citet{G61} but see \citet{P03} for a more recent review, in which nodes of a network are embedded in a continuous space and pairs of nodes are connected based on their geometric proximity. 
The key difference in our model is the addition of an ordering that transforms these undirected geometric networks into directed acyclic graphs suitable for main path analysis. 
This order allows us to define a clear sense of causality, the arrow of time that is fundamental to understanding innovation flows in citation networks.

This model retains the essential feature of a defined geodesic in the continuous embedding space, providing a clear benchmark against which to test path-finding algorithms. 
Again, we are interested in the hypothesis that main path analysis identifies paths close to the geodesic even when the underlying network structure is irregular and without symmetries, features found in a real citation network.

\subsection{Definition of model}\label{s:pppdefn}

The model creates a DAG embedded in a $D$-dimensional space so each node $v$ will have $D$-real coordinates. 
For simplicity, we will work within a hypercube where points $x$ have $D$ coordinates $x_i$ between zero and one, $0\leq x_i< 1$ where $i \in \{ 1,2,\ldots,D\}$. 
In our DAG we always include a source node $s$ at the origin ($s_i=0$) and a sink node $t$ at the opposite corner of the hypercube ($t_i=1$). This sets the units of distance in the model such that the sides of our hypercube are always length $1.0$.  \tsenote{THIS COULD BE SAID SOMEWHERE, MAYBE NOT HERE. In general multi-source and multi-sink graphs, equivalent behaviour can be achieved by adding a global source and a global sink.}
We then add a further $\Nbox = (N-2)$ nodes $v$, called \tsedef{box points}, to our DAG where each is given $D$ coordinates $v_i$ chosen\footnote{This is not quite a Poisson point process as we fix the number of nodes. In a Poisson process, the average number of nodes is fixed, but the actual number can vary. The difference for large numbers of nodes is likely to be irrelevant here.} uniformly at random to lie between zero and one.

The edges in our DAG are then defined using two criteria. 
First, there is a directed edge from node $u$ to $v$ only if $u_i < v_i$ for all $D$ coordinates $i \in \{ 1,2,\ldots,D\}$, the cube-space rule\footnote{For two dimensions, this cube-space model is equivalent to $1+1$ dimensional Minkowski space-time, the cone-space model of \citet{BB91} but this is not true for higher dimensions. There is some numerical work on these Minkowski space-time based random geometric DAGs, for example see \citet{RW09,CE16,CE17,J22b}.} of \citet{BB91}. 
This defines the order inherent in a DAG. 
The cube-space condition means that the coordinate of each node in any one direction defines a total order for the DAG and so acts like a possible `time' coordinate present in real citation networks.

We then go beyond the standard cube-space model of \citet{BB91} by enforcing a second condition. 
Namely, we connect two nodes $u$ and $v$ only if they are less than a distance $R$ apart, $d^\mathrm{(Eucl)} (v,u)<R$, where we use the standard (Euclidean) distance for  $d^\mathrm{(Eucl)} (v,u)$.

So together, the criteria for adding an edge from node $u$ to node $v$ in our random geometric DAG model are
\beq
  \Big( \sum_{i=1}^D (v_i-u_i)^2 \Big)^{1/2} < R \, , \quad
  u_j < v_j \;\; \forall \; j \in \{ 1,2,\ldots, D \}
  \, .
  \label{e:connection}
\eeq

Work using the cube-space condition alone, such as \citet{BB91} and \citet{RW09}, is equivalent to taking $R \to \infty$, which leads to transitively complete DAGs. 
The reason for introducing this Euclidean separation condition is that DAGs found in the real world, such as citation networks, are rarely transitively complete so a finite $R$ is more realistic. 
However, we always choose a large enough $R$ so that the long paths we study are not very sensitive to this parameter\footnote{We do this by ensuring that $\texpect{\kin} \gg (1/D)\ln(N)$. This means that nodes that lie on the long paths of interest for $R \to \infty$ should have at least one edge for this finite $R$ value.}.
\tsenote{This is a criterion derived in \appref{as:rgdagnum}.} 
Also, by working with sparse DAGs (not transitively complete), we reduce the number of edges and so the computational difficulty. 
Shorter paths, such as the random path and shortest path, will depend on $R$.

We need to be clear about the number of points we are considering, as there are several such variables in our model. 
The number of points in our hypercube sections of Euclidean space $[0,1]^D$ is always fixed to be a given number $N$ in any one simulation, consisting of the source node $s$, the sink node $t$ and the $\Nbox = (N-2)$ box points placed at random. 
For any connection rule, here for any  $R$, we always have a DAG, which we will call the hypercube DAG $\Dcalhc(N,R)$ which has $N$ nodes. 
For cube-space with $R \to \infty$ this is always an interval as any point has a directed edge from source $s$ and a directed edge to $t$. 

For finite $R$, the hypercube DAG $\Dcalhc(N,R)$ may have many weakly-connected components. Multiple sources and sinks are also possible. 
However, we focus on all possible paths from $s$ to $t$, so we are in effect working with an interval from $s$ to $t$, the interval DAG $\Dcalint(N,R; s,t)$  with $\Nint$ nodes. This interval DAG is in general a subgraph of the full hypercube DAG $\Dcalhc(N,R)$ so the number of points $N$ in the hypercube DAG $\Dcalhc(N,R)$ is usually larger than the number of points $\Nint$ in the interval DAG $\Dcalint(N,R; s,t)$. 
We will always quote $N$, the parameter used in the construction of the full DAG $\Dcalhc(N,R)$ even when we operate within the interval DAG $\Dcalint(N,R; s,t)$ with $\Nint$ nodes. 

It is also important to note that this interval DAG $\Dcalint$ may not exist for finite $R$ as there may be no path from one corner of the hypercube to the opposite corner, from $s$ to $t$. In our numerical experiments, we choose $N$ so that we are far away from the percolation transition and so most have a path from $s$ to $t$. In any case, we condition on instances where at least one $s$ to $t$ path exists; realisations without such a path are resampled.

In practice, it makes more sense to compare DAGs with similar edge densities. 
To do this, rather than specify $R$ we will define a \tsedef{target out-degree} $\ktarget$ and the aim is to work with models where the average out-degree of a node in an infinite version of the model ($N \to \infty$) is exactly equal to this target degree $\ktarget$. 
This means that the original model parameter $R$ is now given by $R(\ktarget,N)$ where 
\beq
  R(\ktarget,N) = a \left[ \frac{\ktarget \cdot 2^{D}}{ V_D} \right]^{1/D} \, ,
  \quad
  \text{where} 
  \;\;
  a = (N-2)^{-1/D} \, ,
  \quad
  V_D = \frac{\pi^{D/2}}{\Gamma(1+(D/2))}
  \, .
  \label{e:radiusdef}
\eeq
Here $a$ is a distance scale comparable to the typical distance between nearest nodes so it plays the role of the lattice spacing in the previous model of \secref{s:lattice}. Both $R$ and $a$ are length scales measured in units where the sides of the hypercube are length $1.0$.
The volume of the unit sphere in $D$ dimensions is $V_D$. The factor $2^{D}$ accounts for the restriction by the cube-space rule of connections from any one node $v$ to be those in the future set 
of $v$, here the forward \href{https://en.wikipedia.org/wiki/Orthant}{orthant} of a sphere radius $R$ centred on $v$. 

By construction, $\ktarget$ is the expected out-degree for interior points (those points more than distance $R$ from any boundary). 
A few nodes $v$ close to the boundary, whenever $(1-v_i) <R$ for any coordinate $i$, will be expected to have fewer outgoing edges so the actual global mean out-degree of the DAG is smaller than $\ktarget$ where this boundary effect will depend on $R$, $N$ and $D$. 
This means there are still some differences in the average out-degree of our finite size random geometric DAGs and we will report the actual global averages in the results as well as the input parameter $\ktarget$.

We will denote the DAGs constructed by specifying $\ktarget$ as $\Dcalhc(N,\ktarget)$ and the corresponding interval DAGs as $\Dcalint(N,\ktarget;s,t)$.


\tsecomment{Not sure we have any Analytical Results worth talking about in main text? Well we do have the estimates in my notes but no time to follow those up. AFTER THESIS SUBMITTED???}

\subsection{Numerical results}\label{s:pppnumerics}

\begin{figure}[htb!]
  \centering
  \includegraphics[width=0.75\linewidth]{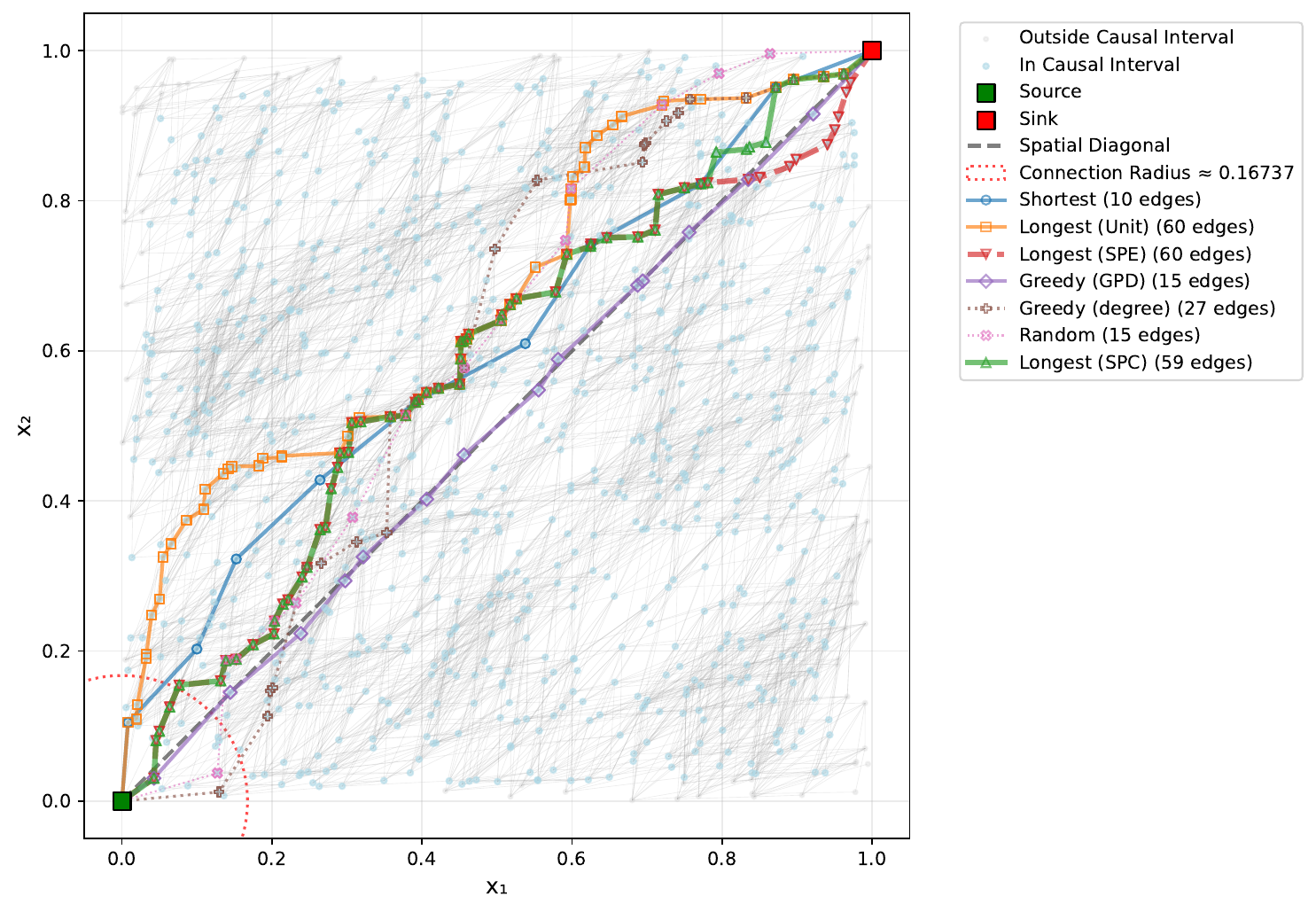}
  \caption{Random geometric DAG in two dimensions with $\Nbox=1000$ box points ($N=1002$ including source and sink) and target degree $\ktarget=22.0$ giving $R \approx 0.167$ showing various path types from source (bottom left) to sink (top right). The network demonstrates high connectivity with 916 nodes in the causal interval (91.6\% can participate in source--sink paths). Different path algorithms create diverse routing strategies with path lengths ranging from 10 to 60 edges. By way of comparison, the length scale for the inter-point spacing is $a = \Nbox^{-1/D} \approx 0.0316$ so $R \approx 5.30\, a$ and the diagonal is $\sqrt{2} \approx 44.8\, a \approx 8.47\, R$. The unit length of the shortest path, $10$, is therefore close to the lower bound given by the geodesic in units of $R$. The unit length of the longest paths (by unit, SPC or SPE lengths) at $60$ or $59$ are noticeably longer than the geodesic length $44.8\, a$ in units of the average inter-point separation $a$, consistent with the asymptotic values given in \citet{BB91}.}
  \label{f:ppp-dag-deg22-n1000}
\end{figure}

An example of one of these random geometric DAGs is shown in \figref{f:ppp-dag-deg22-n1000}. This shows that most of the paths are reasonably close to the diagonal, the geodesic, and to each other. 
Qualitatively the same pattern holds in three and four dimensions: normalised geodesic deviations, path-weight rankings, and criticality coverage follow the trends seen in two dimensions. We report results for higher dimensions in \appref{app:rgg_3d4d}.

\newpage
\subsubsection{Path weights}

In \tabref{t:ppp_weights_trials_deg22_n100000} we show the statistics for the path weights  $\LG(\Pcal)$ of \eqref{e:LGdef} for various paths and different weights $G$ across $100$ independent network realisations in two-dimensions with $\Ntilde=10^5$ box points.\tsenote{I think we need $D=3$ and $D=4$ results somewhere. Perhaps appendix once we have the rest in place.} 
Each trial uses a random geometric DAG with target degree $\ktarget=22.0$ to match that shown in \figref{f:ppp-dag-deg22-n1000}, but with different random node placements. 
Using length, measured in terms of number of edges (unit weight), SPC or SPE weights, the longest paths by all three measures perform very similarly. 
However, the difference between the mean unit length of the SPC longest path and the mean unit length of the other two longest paths (by SPE or unit length) is consistent across trials and this difference is larger than the standard error of the mean, with the SPC longest path typically about one edge shorter than the other two longest paths. 
In terms of their SPC and SPE weights, there is no significant difference.

\begin{table}[htb!]
	\centering
	\begin{tabular}{l|| c | c | c}
		Path $\Pcal$ & \multicolumn{3}{c}{Weight $\LG(\Pcal)$ (mean $\pm$ s.e.m.)} \\ 
				     & Unit & SPC & SPE \\ \hline
		Shortest           & 89.760 $\pm$ 0.043 & $(7.7803 \pm 5.3500) \times 10^{267}$ & 53945.403 $\pm$ 40.368 \\ 
		Longest (Unit)     & 622.470 $\pm$ 0.650 & $(5636.5 \pm 2966.6) \times 10^{267}$ & 378909.649 $\pm$ 640.840 \\
		Longest (SPC)      & 608.800 $\pm$ 0.908 & $(7577.9 \pm 4001.1) \times 10^{267}$ & 371351.465 $\pm$ 756.166 \\
		Longest (SPE)      & 622.290 $\pm$ 0.658 & $(6425.0 \pm 3299.6) \times 10^{267}$ & 379127.585 $\pm$ 633.022 \\
		Greedy (GPD)       & 144.830 $\pm$ 0.453 & $(60.911 \pm 32.191) \times 10^{267}$ & 87410.491 $\pm$ 279.150 \\
		Greedy (Degree)    & 185.493 $\pm$ 0.774 & $(316.547 \pm 151.738) \times 10^{267}$ & 112104.646 $\pm$ 497.131 \\
		Random             & 142.586 $\pm$ 0.572 & $(11.9285 \pm 10.4206) \times 10^{267}$ & 85923.567 $\pm$ 373.899 \\ 
		\end{tabular}
	\caption{Path weights $\LG(\Pcal)$ of \eqref{e:LGdef} averaged over $100$ independent network realisations in two-dimensions with $(N-2)=10^5$ box nodes and $\ktarget=22.0$. 
		Shown for various paths and for three different weights $G$. 
		Entries show mean $\pm$ standard error of the mean for $\LG(\Pcal)$  in each case. 
	}
	\label{t:ppp_weights_trials_deg22_n100000}
\end{table}

The unit length $\Lunit(\Pcallp)$ of the longest paths by unit length $\Pcallp$ can be compared against the asymptotic values given by \citet{BB91} for this model in the limits of $N,R \to \infty$. We find our results to be consistent with the asymptotic bounds of \citet{BB91}, see \appref{as:pathweights} for a longer discussion.

One last feature of note can be seen in these numbers for the path weights.
The numbers quoted in \tabref{t:ppp_weights_trials_deg22_n100000} for the SPC weights have up to 271 digits.
As these are sums of many many numbers, arbitrary precision integer arithmetic is required for reliable computation (see \secref{s:numerics} for implementation details).
The problem spills over into the SPE calculations as intermediate values used for SPC calculations are used for the SPE weights, though logarithmic transformation provides numerical stability.
One illustration of these extreme scales is the SPC weight of the SPC path for one network with $(N-2)= 2.5 \times 10^5$ box nodes and target out-degree of $\ktarget= 22.0$, which has 422 digits as shown in \figref{af:spc250k} of \appref{as:pathweights}.

To look at the large fluctuations of the SPC weights in more detail, we ran 500 independent trials, generating a new random geometric DAG in each case using the same parameters. For each trial, we computed the longest path according to three different edge-weight schemes: Unit \eqref{e:Lunitdef}, SPC \eqref{e:Gspcdef} and SPE \eqref{e:Gspedef}. 
We then examined the distribution of each path's unit/SPE/SPC native weights across the 500 realisations.
To get a sense of the size of the fluctuations, \tabref{tab:ranges} shows the largest and smallest value of each weight of the SPC main path found in the 500 realisations. 
\begin{table}[htbp]
	\centering
	\begin{tabular}{rl|cccc}
		\multicolumn{2}{c|}{Weight type} & Minimum              & Maximum               & Difference      & Frac.Diff.\ \\
		\hline
		Unit&$\Lunit(\Pcalspc)$     & 128                  & 144                   & 16              & 11.8\% \\
		SPE &$\Lspe(\Pcalspc)$     & 16{,}191             & 19{,}501              & 3{,}310         & 18.5\%\\
		SPC &$\Lspc(\Pcalspc)$     & $1.5 \times 10^{55}$ & $2.7 \times 10^{63}$  & --              & --    \\
		\multicolumn{2}{c|}{$\log_{10}(\Lspc(\Pcalspc))$} & 55.2                 & 63.4                  & 8.2             & 13.9\% \\
	\end{tabular}
	\caption{Range and variation of different weight values for the SPC main path $\Pcalspc$ across 500 realisations for random geometric DAG with $N=5,002$ nodes and target out-degree $\ktarget=22.0$. 
		Each row lists the largest and smallest values of the given weight for the SPC main path found across the 500 realisations.
		The ``Frac.Diff.'' column gives the difference between maximum and minimum values divided by the average of the maximum and minimum values. Since the SPC values span eight orders of magnitude, we also took the logarithm of these numbers in the last row and this compresses the numbers to give a more meaningful comparison.}
	\label{tab:ranges}
\end{table}

For the Unit and SPE weights of the SPC main path we find a reasonable variation. However, the problem we found when taking the means of SPC values is clear as we find eight orders of magnitude variation in the SPC values for very similar networks since they are created using Poisson statistics. Of course, one solution is to work with the logarithms of the SPC values (we use base 10 in our table) and then we do bring the fluctuations in the SPC values under control. This is similar to how the SPE works but for the SPE we take the logarithms of values used in the intermediate parts of the SPC calculation and apply logs there, whereas in the table, we only apply logs at the end of the SPC calculation. 

Intuitively, SPC multiplies the number of source-to-$u$ paths by the number of $v$-to-sink paths for every edge $(u,v)$, so Poisson fluctuations in local in- and out-degrees are compounded along the DAG. Even modest variance in the underlying point process therefore amplifies to factorial growth in the SPC totals, producing variations of up to eight orders of magnitude across otherwise similar networks, as shown in \tabref{tab:ranges}. 
Taking the logarithm of the SPC weight of a path converts these products into sums, compressing extreme values and stabilising the distribution.

\begin{figure}[htbp]
	\centering
	\includegraphics[width=\textwidth]{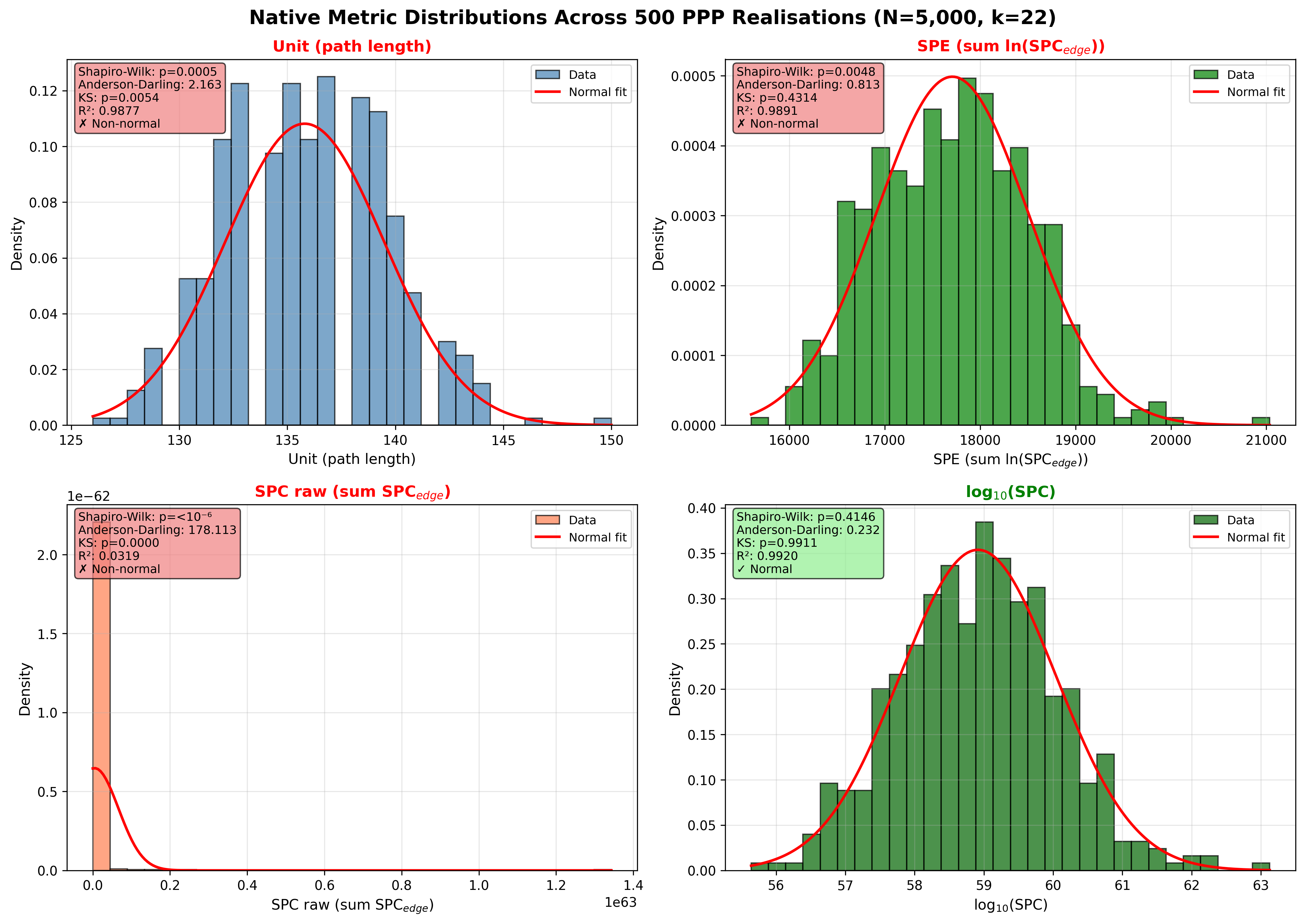}
	\caption{The distributions of the optimised weight value for longest paths by unit, SPC and SPE weight  across 500 random geometric DAG realisations, each with $N=5,002$ nodes and target out-degree $\ktarget=22.0$. Top left: The unit path weight for longest unit path ($\Lunit(\Pcalunit)$) exhibits high Q--Q linearity (R$^2$=0.99) despite borderline Shapiro-Wilk $p$=0.0005. Top right: SPE weight for SPE main path ($\Lspe(\Pcalspe)$) shows similar behaviour (R$^2$=0.99, $p$=0.005). 
		Bottom left: The SPC path weight for the SPC main path ($\Lspc(\Pcalspc)$) is highly non-Gaussian, spanning eight orders of magnitude (R$^2$=0.03). Bottom right: We show the distribution of the decimal log of the SPC path weights for SPC main path which does exhibit excellent normality (R$^2$=0.99, $p$=0.41). Each panel shows fit quality metrics: Shapiro--Wilk $p$-value, Anderson--Darling statistic, Kolmogorov--Smirnov $p$-value, and R$^2$ (Q--Q plot coefficient of determination).
	}
	\label{fig:native_metrics}
\end{figure}

We can be a little more precise about the distribution of these fluctuations and look at tests of normality over these 500 realisations with results shown in \figref{fig:native_metrics}. Essentially the distribution of unit lengths for the longest unit path ($\Lunit(\Pcalunit)$), and the SPE weights of SPE main path ($\Lspe(\Pcalspe)$), both show reasonable Gaussianity suggesting the absolute values of these path measures are well behaved and we can use standard methods of analysis on these measures.  
The test results are summarised in \tabref{tab:native_metrics_normality}. These tests also confirm that the absolute SPC path weights of the SPC main paths ($\Lspc(\Pcalspc)$) are non-Gaussian and difficult to handle. On the other hand, the logarithm of these SPC path weights is nicely distributed, roughly Gaussian as shown in \figref{fig:native_metrics}. So the suggestion is that provided you can calculate these SPC path weights and take their logarithms accurately, then it would also be straightforward to work with these logarithms of SPC path weights\footnote{Put another way, it means that we should be looking at relative SPC path weights values, not differences. However, probably simpler to work in log space with SPC values.}. Note that the SPE weight is the sum of logarithms of path counts $\Omega_{vu}$ in \eqref{e:Gspedef} while our suggestion to work with the logarithm of SPC values is the logarithm of the sum of path counts $\Omega_{vu}$ in \eqref{e:Gspcdef}. Since the operations of taking sums and logarithms do not commute, these two approaches are not the same.  However, in terms of smoothing out the fluctuations to produce a simpler measure, using the logarithm is playing a similar role in both cases.

\begin{table}[htbp]
	\centering
	\begin{tabular}{rl|cccc}
		\multicolumn{2}{c|}{Weight type}& Mean $\pm$ SEM              & Shapiro--Wilk $p$ & $R^2$ & Assessment \\
		\hline
		Unit & $\Lunit(\Pcalunit)$       & $135.8 \pm 0.17$             & 0.0005            & 0.988 & High linearity \\
		SPE  & $\Lspe(\Pcalspe)$       & $17{,}708 \pm 36$            & 0.0048            & 0.989 & High linearity \\
		SPC  & $\Lspc(\Pcalspc)$       & Median: $8.8 \times 10^{58}$ & $<10^{-6}$        & 0.032 & Non-normal \\
		\multicolumn{2}{c|}{$\log_{10}(\Lspc(\Pcalspc))$} & $58.9 \pm 0.05$ & 0.41        & 0.992 & Normal \\
	\end{tabular}
	\caption{Tests of normality for the distribution of a given path weight of the SPC main path found 500 realisations of the random Geometric model ($N=5,002$, $\ktarget=22.0$). 
		Summary statistics and measures of the quality of fit to a normal distribution quality are given.
		Unit weight of the longest unit path ($\Lunit(\Pcalunit)$) and the SPE weight of the SPE main path ($\Lspe(\Pcalspe)$) show high $R^2$ despite borderline Shapiro--Wilk $p$-values. The raw SPC weight of the SPC path ($\Lspc(\Pcalspc)$) categorically fails but the logarithm of this ($\log_{10}(\Lspc(\Pcalspc))$) passes all tests.
		\\
		\tsecomment{``Summary statistics and fit quality'' OF WHAT? Fitting what to what? I guessed here}	
		}
	\label{tab:native_metrics_normality}
\end{table}

\tsenote{Perhaps repeat this in the discussion? As a result we recommend: 
(i) report log$_{10}$(SPC) rather than raw SPC when summarising distributions; (ii) restrict SPC-based algorithm comparisons to paired, within-realisation evaluations; (iii) avoid interpreting absolute SPC values across independent trials.}

\subsubsection{Perpendicular Distance to Geodesic $\Delta$}\label{ss:ppp-perpendicular-distance}

\begin{figure}[htb!]
	\centering
	\includegraphics[width=0.7\linewidth]{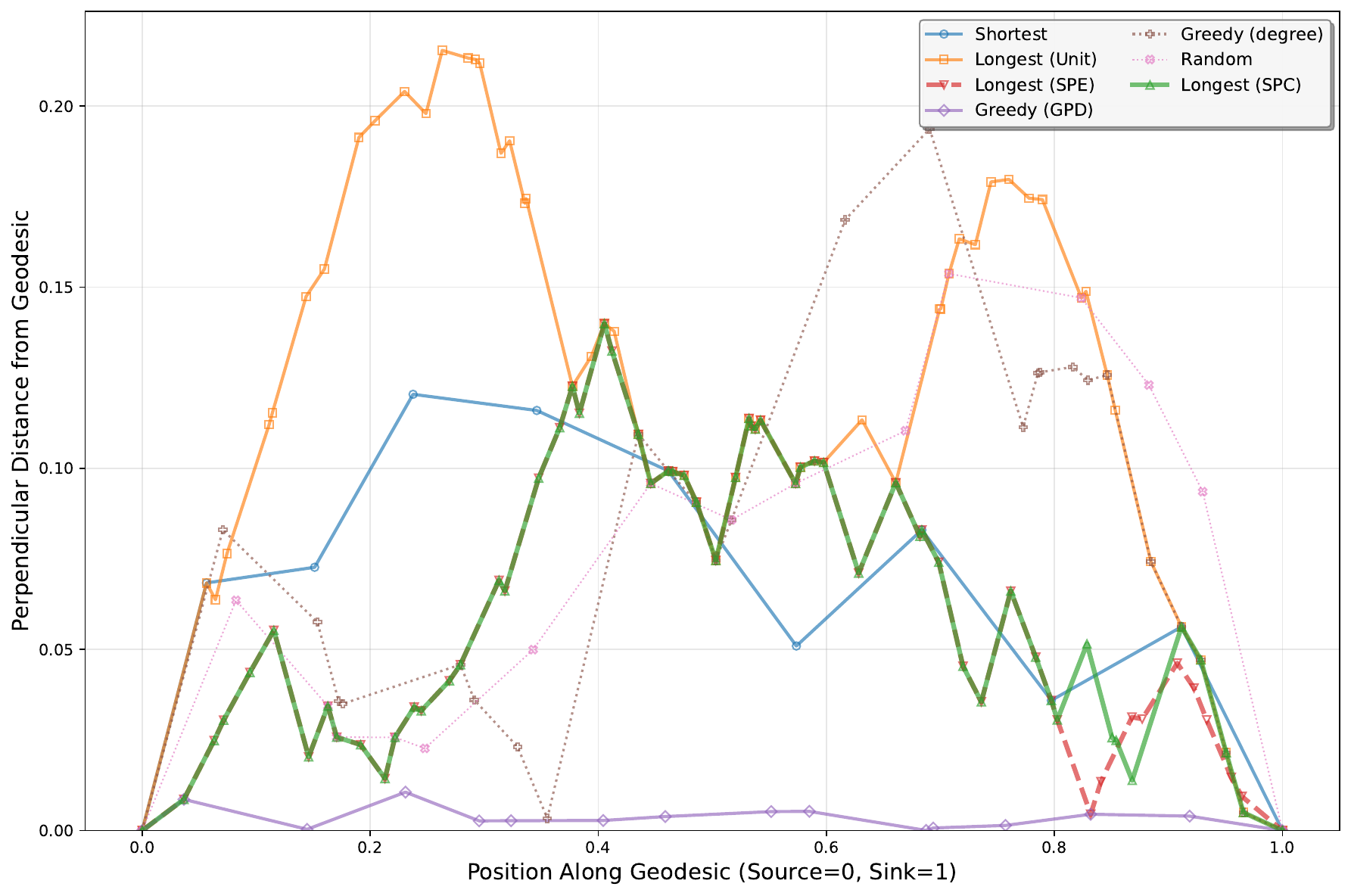} 
	\caption{Geodesic deviation analysis for a representative two-dimensional random geometric DAG with $N=1002$ nodes and target out-degree $\ktarget=22.0$. For each path we plot the perpendicular distance $\Delta(v)$ measured in units of the side length of the box ($1.0$) to the geodesic (diagonal) for each node $v$ as shown by the point symbols.  The node values from any one path are placed in path order from source node (far left) to the sink node (far right). Values for each path are shown with different point symbols and colours, straight lines between points are to aid visualisation. The greedy perpendicular distance (GPD) algorithm achieves minimal deviation while other algorithms demonstrate varying amounts of deviation from the geodesic.}
	\label{f:ppp_geodesic_deviation}
\end{figure}

One way to compare paths on the random geometric DAG to the geodesic in the embedding Euclidean space is to examine the perpendicular distance $\Delta(v)$ \eqref{e:perdistdef} of nodes on the discrete path to the geodesic. In \figref{f:ppp_geodesic_deviation} we visualise the results from a single instance of a relatively small network. Again, we show one path of each type even though there can be many that satisfy the criteria for each path. 

As expected, the greedy perpendicular distance (GPD) algorithm minimises perpendicular deviation by construction. 
This shows that paths which follow the geodesic extremely well can be found numerically when the geometry of the continuous space is known. With citation networks and other DAGs derived from data, the embedding space and its geodesic are unknown (though see \citep{CE16,CE17}) so we cannot define the GPD path.

The random path is surprisingly close to the geodesic, as for example seen in the visualisation in \figref{f:ppp-dag-deg22-n1000}. When looking at fixed target out-degree $\ktarget$ but varying $N$, we see the random path appears to get closer to the geodesic, not in accord with common-sense. The problem with this is that we are looking at this measurement in terms of units of distance where the length of the side of the hypercube is $1.0$. In these units, the scale for the perpendicular distance of the random path to the geodesic, $\Delta$, scales with $N^{-1/(2D)}$, i.e.\ in these units the random path \emph{does} get closer to the geodesic. However, a more useful way to look at this issue is in terms of units of $a$, the short distance scale set by the typical distance between neighbouring points. In these units, with fixed $\ktarget$, the density of points is fixed, $R$ is fixed, but as $N$ increases as the sides of the hypercube get longer (by $N^{1/D}$). In these units the perpendicular distance of the random path to the geodesic does indeed increase as $\Delta/a \sim N^{+1/(2D)}$. More details on these estimates are given in \appref{as:pppdeltarandom}.

Given the emphasis put on the longest unit path, it may appear surprising that the longest unit path in \figref{f:ppp_geodesic_deviation} shows some of the largest deviations from the geodesic.
However, a feature of longest unit paths is that there can be many such paths, forming a broad distribution. What we are seeing in \figref{f:ppp_geodesic_deviation} is an artefact of the deterministic tie-breaking algorithm  built into our code (to ensure reproducibility) which selects a particular path depending on lexicographic ordering of node indices. The chosen longest unit path may be an outlier with respect to geodesic proximity.

Next we compare the performance of our three type of longest path in terms of the normalised mean perpendicular distance $\Delta/a$ across different network sizes $N$ for fixed density with $\ktarget=22.0$. We also include the greedy geodesic algorithm (GPD) for comparison. Results are shown in \tabref{t:ppp_geodesic_multi_trial_a_norm}. 
\begin{table}[htb!]
	\centering
	\begin{tabular}{rc|| c | c | c | c}
		$N$ & $a$    & \multicolumn{4}{c}{Normalised mean perpendicular distance $\Delta/a \, \pm$ sem}  \\ 
		&        & Greedy (GPD)       & Longest (SPC)      & Longest (SPE) & Longest (Unit) \\ 
		\hline 
		502 & 0.0447 & 0.106 $\pm$ 0.0022 & 1.398 $\pm$ 0.0384 & 1.512 $\pm$ 0.0433 & 1.737 $\pm$ 0.0527 \\
		1002 & 0.0316 & 0.116 $\pm$ 0.0025 & 1.852 $\pm$ 0.0618 & 2.173 $\pm$ 0.0759 & 2.424 $\pm$ 0.0842 \\
		2002 & 0.0224 & 0.122 $\pm$ 0.0022 & 2.192 $\pm$ 0.0816 & 2.573 $\pm$ 0.0916 & 2.896 $\pm$ 0.1117 \\
		5002 & 0.0141 & 0.127 $\pm$ 0.0018 & 2.862 $\pm$ 0.0917 & 3.490 $\pm$ 0.1188 & 3.994 $\pm$ 0.1382 \\
		7502 & 0.0115 & 0.130 $\pm$ 0.0017 & 3.109 $\pm$ 0.0952 & 3.682 $\pm$ 0.1384 & 4.062 $\pm$ 0.1497 \\
		10002 & 0.0100 & 0.130 $\pm$ 0.0016 & 3.411 $\pm$ 0.1021 & 4.086 $\pm$ 0.1361 & 4.457 $\pm$ 0.1585 \\
		15002 & 0.0082 & 0.133 $\pm$ 0.0015 & 4.006 $\pm$ 0.1299 & 4.692 $\pm$ 0.1559 & 5.058 $\pm$ 0.1789 \\
	\end{tabular}
	\caption{The normalised mean perpendicular distance $\Delta/a$ for four paths in the random geometric DAG model in two dimensions with fixed $\ktarget=22.0$ but varying number of nodes. Here $a=(N-2)^{-1/2}$ is the characteristic internode separation scale.  Values are shown as mean $\pm$ standard error of the mean (sem). The connection radius $R$ is chosen to achieve target out-degree $\ktarget$, so $R/a \approx 5.293$ is constant. The number of trials for $N=502$ is $300$ while $200$ trials are used for larger $N$.}
	\label{t:ppp_geodesic_multi_trial_a_norm}
\end{table}

The greedy geodesic distance path (GPD) consistently achieves $\Delta/a \approx 0.1$, demonstrating excellent geodesic approximation across all network sizes, betyter than our three longest paths. This numerical value is consistent with the distance scale $\Delta_1$ where $R(\Delta_1)^{D-1} V_{D-1}=1/N$ set by the size of a hyper-cylinder of length $R$ and radius $\Delta_1$ around the geodesic that contains on average one point. For the parameter values of \tabref {t:ppp_geodesic_multi_trial_a_norm} we have that $\Delta_1/a = 1/[2 (R/a)] \approx 0.094$. The greedy geodesic distance path is designed to minimise $\Delta$ locally so it is not surprise it is an order of magnitude better than our other paths. It serves to show what is possible if we know the inherent geometry of our network.

The normalised geodesic deviation $\Delta/a$ for the three longest paths increases with $N$ but is always close to the geodesic in terms of the short distance scale $a$ as \tabref{t:ppp_geodesic_multi_trial_a_norm} shows. So while these are not as good as could be achieved, all of these paths are within a few points (a few $a$) of the closest paths. Given the algorithms do not exploit the geometry in the model, unlike the greedy geodesic distance path, these paths all give paths that are remarkably good approximations to the geodesic. There are clear differences, the SPC main path is closer to the geodesic, in a statistically significant sense, than the other two paths. However, the difference is only around one factor of $a$, one point, for our largest networks. Given the longest unit path is one of many possible such paths, it is quite possible there are unit longest paths that perform as well or even better than the SPC. So our main message is that all these longest paths, by unit, SPC or SPE weight, are very good at finding the geodesic without using the geometry of the embedding space of this model. We give more details these results in \appref{as:ppp_perpendicular_distance}.

\subsubsection{Degree of nodes on paths}

We have seen how the longest unit path, SPC and SPE main paths share many similar properties. Equally, we have seen that these paths are not identical. It would be interesting to know if these different paths favour different types of node in our networks. So we next ask if any of these paths favour high density regions in our random geometric DAG model. The simplest way to study this question is to look at the degree of the nodes on each path since a high degree means the path is passing through a region of high density. The model is based on Poisson spatial statistics so these spatial fluctuations in node, and hence edge, density are not large but they are noticeable as can be seen in any visualisation such as  \figref{f:ppp-dag-deg22-n1000}. Results for the typical node out-degree for 100 two-dimensional networks are shown in \tabref{t:path_degrees_trials_n50000}. 

In a practical context, degree is also one of the most common centrality measures so these statistics may be hinting that the importance of a node, as measured by degree, may be correlated with the choice of path used to study a network.

Poisson statistics in this model means that we expect most points (those not close to a boundary) to have an out-degree in a Poisson distribution with mean equal to $\ktarget$, here $22.0$, so with standard deviation $\sqrt{22} \approx 4.7$. Results for various paths are summarised in \tabref{t:path_degrees_trials_n50000}. In fact all the paths have degree statistics that are statistically different from the underlying Poisson statistics but some are much closer than others. Some difference is to be expected as only interior points, well away from the boundaries, have the underlying Poisson statistics statistics.

\begin{table}[htb!]
\centering
\begin{tabular}{l|| c | c | c | c}
Path 			   & Unit Length       & \multicolumn{3}{c}{Out Degree} \\ 
     			   &                   & Mean             & Median           & Std.Dev. \\ \hline
Shortest (unit)    &  $63.72 \pm 0.05$ & $22.14 \pm 0.06$ & $21.99 \pm 0.08$ & $4.79 \pm 0.04$ \\ 
Longest (unit)     & $438.02 \pm 0.50$ & $23.60 \pm 0.04$ & $23.61 \pm 0.06$ & $5.17 \pm 0.03$ \\ 
Longest (SPC)      & $429.15 \pm 0.64$ & $24.16 \pm 0.03$ & $24.16 \pm 0.05$ & $5.16 \pm 0.03$ \\ 
Longest (SPE)      & $437.85 \pm 0.51$ & $23.82 \pm 0.04$ & $23.81 \pm 0.06$ & $5.18 \pm 0.03$ \\ 
Greedy (GPD)       & $102.80 \pm 0.41$ & $21.26 \pm 0.06$ & $21.27 \pm 0.07$ & $4.84 \pm 0.04$ \\ 
Greedy (degree)    & $131.48 \pm 0.67$ & $26.50 \pm 0.07$ & $27.40 \pm 0.07$ & $6.55 \pm 0.07$ \\ 
Random             &  $97.22 \pm 0.46$ & $21.08 \pm 0.06$ & $21.41 \pm 0.07$ & $5.68 \pm 0.05$ \\ 
\end{tabular}
\caption{Node out-degree statistics along each path type in 100 realisations of the two-dimensional random geometric DAG model
	with $N=50002$ nodes and target out-degree $\ktarget=22.0$. 
	The mean values are quoted with the standard error of the mean and are based on the values for every node on the specified path exept for the initial and final node, the source and sink of the DAG.}
\label{t:path_degrees_trials_n50000}
\end{table}

The greedy degree path provides an extreme case for comparison as by design it has a strong bias towards high degree nodes so not surprisingly it has the highest degree statistics and is least like the Poisson statistics of random nodes in the interior of the model. 

Interestingly, the shortest path by unit length is closest to match the Poisson statistics of the points. This is probably because with edges of much longer Euclidean length ($\approx R$), the shortest path quickly moves away from the boundary so only the two points next to the end points are likely to be close to the boundary. These results suggests that the shortest path is essentially picking nodes uniformly at random and so does not seem to select nodes of much interest in this model.

We might have thought that the degree of nodes on a random path should be largely sampled from Poisson statistics. However, we see that the mean is below the Poisson distribution mean while the standard deviation is only the second lowest in \tabref{at:path_degrees_n10000}. In this case there are many more points close to the boundary at the end of the path, close to the source. This is because the random path drifts away from the geodesic until it hits the boundary which it does well before reaching the sink node. So a much larger proportion on nodes are expected to be close to the boundary with a lower degree. This is what we see reflected in the statistics of the degree of the random path.  

The longest paths are the most interesting. Pairwise $t$-tests with Bonferroni correction ($\alpha' \approx 0.00238$) on the results of \tabref{t:path_degrees_trials_n50000} confirm small but clear differences among the longest-path constructions. The SPC main path has a significantly higher mean out-degree than both the unit longest path (a difference of $\Delta\bar{k}^\mathrm{(out)} = 0.56$, $p < 10^{-24}$) and SPE main path ($\Delta \bar{k}^\mathrm{(out)} = 0.34$, $p \approx 1.3 \times 10^{-10}$). 
The unit and SPE longest paths also differ in their mean degree but it is a much smaller difference ($\Delta \bar{k}^\mathrm{(out)} = 0.22$, $p = 9.0 \times 10^{-5}$). In terms of path length, the SPC main path produces markedly shorter paths ($429.2 \pm 0.6$ edges) than both longest unit ($438.0 \pm 0.5$) and SPE ($437.9 \pm 0.5$), reductions of about $2\%$ ($p < 10^{-25}$), while Unit and SPE are statistically indistinguishable ($p = 0.81$). These results reinforce the idea that the SPC main paths prefer regions of slightly higher-degree nodes while also yielding slightly shorter source-sink connections when comparing them to the unit and SPE longest paths.

However, while we can detect statistically significant differences between the three longest paths (Unit, SPC, SPE), we have to emphasise that these differences in degree are again small. This again reinforces the view that these paths are passing through very similar regions and there is little to distinguish these paths.


\subsubsection{Baskets Not Paths}\label{s:basketnotpath}

The results so far have shown that relying on a single optimal path to identify key nodes makes the analysis simple but unreliable. 
For longest unit paths, unit-weight criticality is integer-valued, leading to many ties and multiple optimal paths; routines return only one example via deterministic tie-breaking. In contrast, Search Path Count main path (SPC) and Search Path Entropy main path (SPE) use very large integers and real-valued weights respectively and so typically yielding unique longest paths. However even these two main path variants rarely agree on which path is optimal, something we expect to see across all main path variants. The lack of a consistent robust answer from all these methods suggests all these paths are missing part of the story. For instance, it is quite possible for several key documents to be produced at similar times but they cannot all lie on one common path due to causality constraints. So in this section we seek a larger ``basket'' of nodes in a region of the citation network where papers are close to the cutting edge of innovation development. In our geometric models, these baskets of nodes would be close to the geodesic. 

\begin{figure}[htb]
  \centering
  \begin{tabular}{ccc}
    \includegraphics[width=0.3\textwidth]{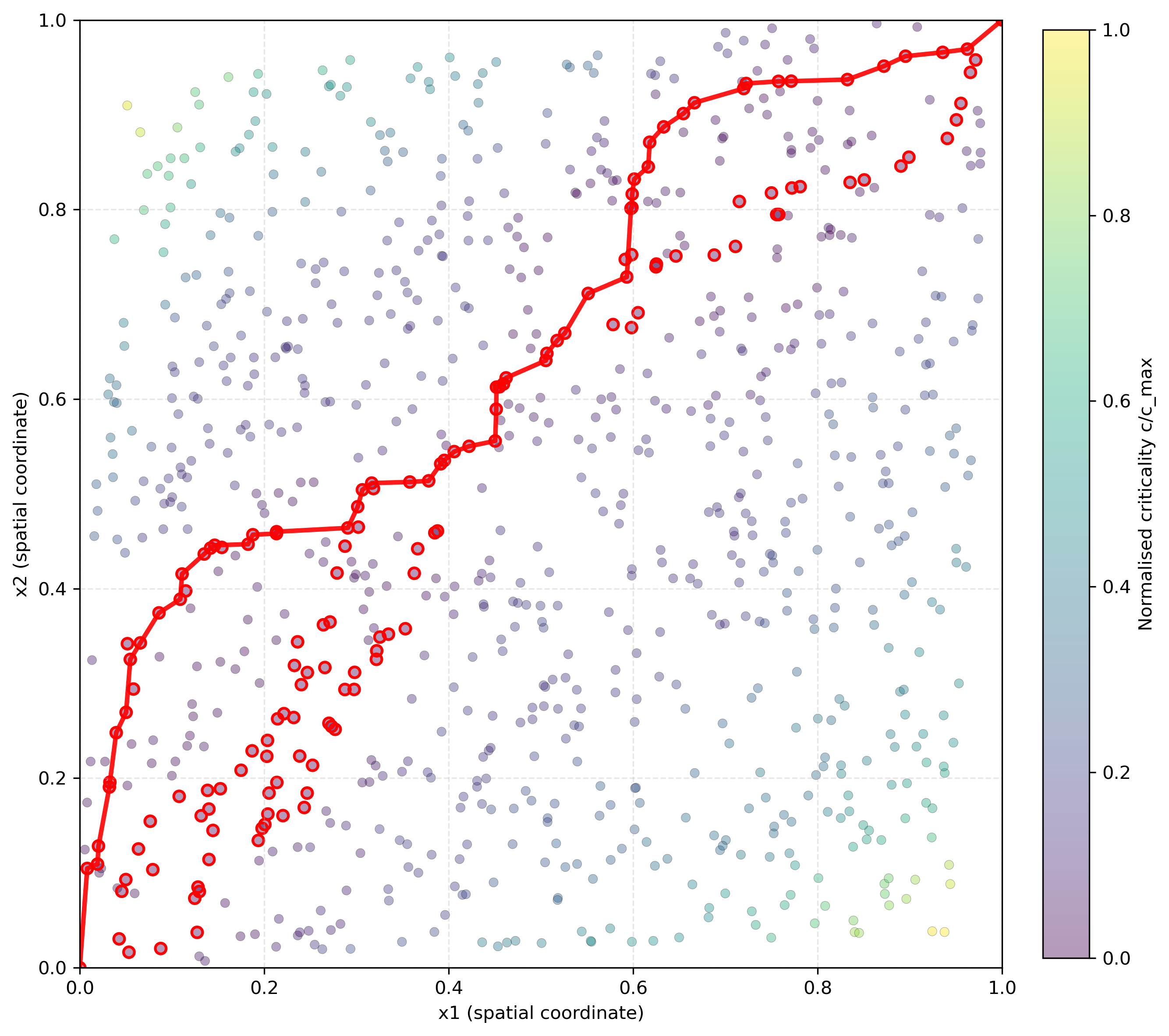}
    &
    \includegraphics[width=0.3\textwidth]{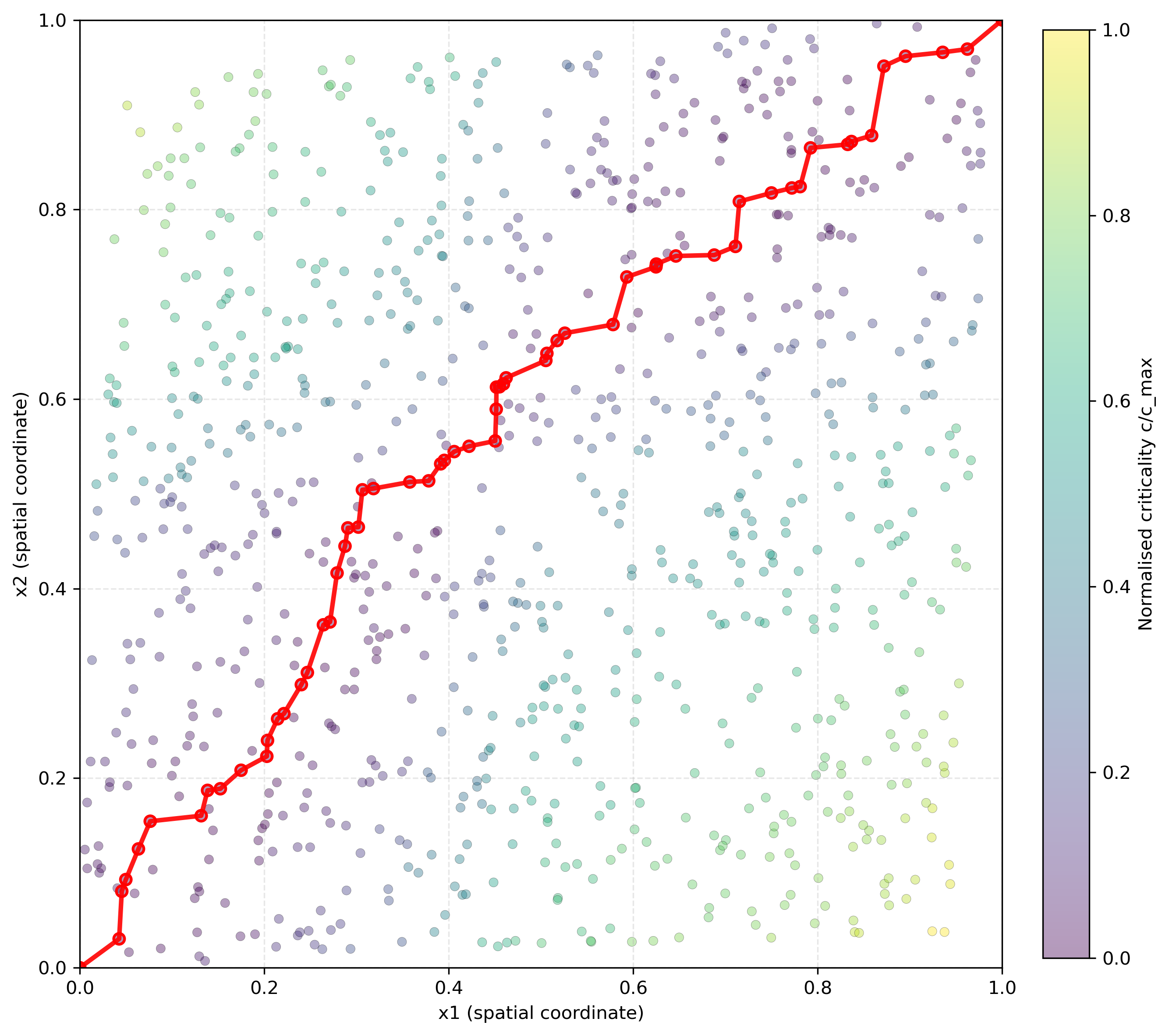}
    &
    \includegraphics[width=0.3\textwidth]{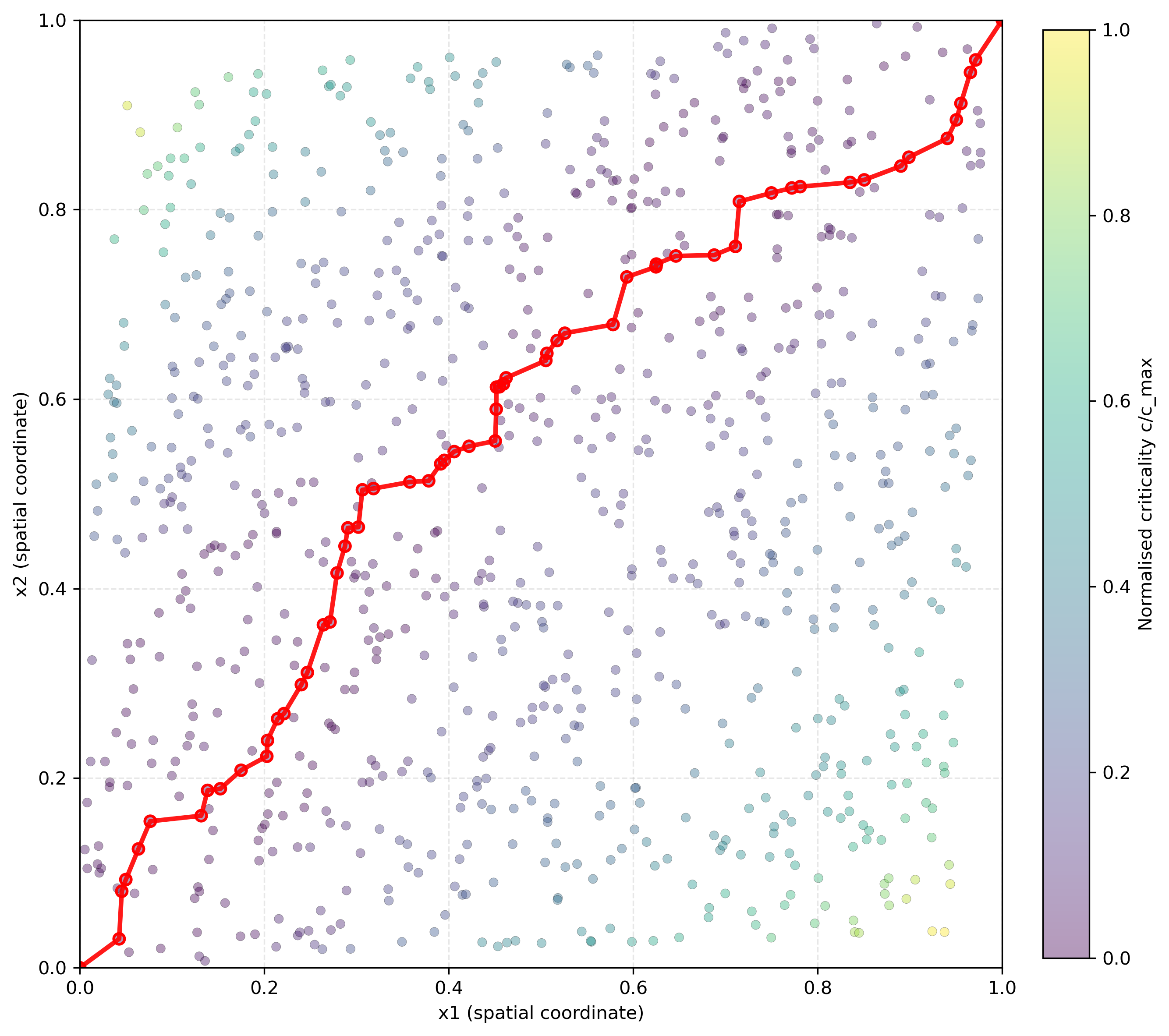}
    \\
    (a) Unit
    &
    (b) SPC
    &
    (c) SPE
  \end{tabular}
  \caption{Relative criticality heatmaps for the same network as in \figref{f:ppp-dag-deg22-n1000} with $1002$ nodes and target out-degree $\ktarget=22.0$. 
  	The relative criticality $\cG/\max(\cG)$ uses a different path weight $G$ in each case: 
  	in (a) we have unit-weight criticality $\Gunit = \Amat$ of eq{e:Lunitdef},
  	in (b) SPC-weight criticality $\Gspc$ of \eqref{e:Gspcdef},
  	and in (c) SPE-weight criticality $\Gspc$ of \eqref{e:Gspedef}.
  	Nodes with critical values are shown in red. In addition, we also show one example of a path that has the largest weight for that plot by adding in the edges for that path, also in red.
  	Note that in this small example, the SPE longest path coincides with a longest path by unit length, highlighting why SPE also appears among the most critical routes in this case.}
  \label{f:ppp-criticality-heatmaps-n1000}
\end{figure}

We start by replicating the study of analysis of criticality on the lattice DAG in \secref{s:lattice_criticality} on representative random geometric DAGs in two-dimensions. First we emulate the plots of relative criticality in \figref{f:latcrit-78-71} but here for one example of a random geometric DAG as shown in \figref{f:ppp-criticality-heatmaps-n1000}

When using the unit weight, we find a large number of nodes with critical values for the unit path weight, $\cunit=0$, in \figref{f:ppp-criticality-heatmaps-n1000}a. This is far more than lie on any of the paths with maximal unit weight such as the example shown in that figure. This demonstrates that even for small DAGs there are multiple distinct longest paths by unit weight, in part due to the integer nature of the measure which makes ties much more likely. Consequently, a single longest unit path, such as shown in \figref{f:ppp-criticality-heatmaps-n1000}a, does not capture all the critical nodes. 

On the other hand, the SPC main path of \figref{f:ppp-criticality-heatmaps-n1000}b  and the SPE main path of \figref{f:ppp-criticality-heatmaps-n1000}c are unique and so capture all critical nodes ($\cspc=0$ or $\cspe=0$ respectively) for their respective weights in this example. 

The SPC and SPE main paths here are very similar, but they are not identical, both are very different from the longest unit path we show. However, all paths pass through regions close to the diagonal (the geodesic), which contain many nodes with very low criticality, as indicated by the colour of the nodes which are not red. So in all cases, a single path fails to capture the many important nodes that have low criticality values if not exactly zero criticality.

Given our intuition that paths are often a poor way to pick our the most important nodes, we turn to examine \tsedef{baskets of nodes} $\Bcal^\mathrm{(G)}(f)$, defined in \eqref{e:Bdef} as the fraction $f$ of the nodes in the DAG that have the lowest generalised criticality values $\cG$ when using weight $\Gmat$. Thus $\Bcal^\mathrm{(G)}(f)$ is a set of nodes of size\footnote{Up to rounding issues as we are comparing integers here.} $|\Bcal^\mathrm{(G)}(f)| = f |\Vcal|$.

In terms of our visualisations in \figref{f:ppp-criticality-heatmaps-n1000},  these baskets would be all the points within a pair of contours starting and ending at the source and target nodes, running either side of the diagonal but bulging away from the diagonal. This region would get larger as $f$ increases.\tsecomment{A visualisation of these contours would be nice but it is not needed now and might just end up in the appendix.}

To understand how effective these baskets of nodes might be, compare the nodes in a basket $\Bcal^\mathrm{(G)}(f)$ to the nodes on a longest path for another weight $G'$.
We will look at unit, SPC and SPE weights to give us nine comparisons for each value of $f$.
We will do this over 100 realisations of random geometric DAGS created with the same parameter values.
In particular we look at the basket coverage of a path, $\phi(\Pcal,\Bcal)$ as defined in \eqref{e:phidef}.


In \tabref{t:ppp_criticality_coverage_trials} we report the coverage (\%) of the longest paths by unit, SPC and SPE weight and we compare that with the baskets defined using the same three weights. We show this for the basket of critical nodes, nodes where $\cG=0$ as well as for the 1\%, top 5\%and top 10\% of nodes, with statistics gathered from 100 trials of random geometric DAGs created with $N=1002$ nodes and a target out-degree of $\ktarget=22.0$. What \tabref{t:ppp_criticality_coverage_trials} shows is that baskets defined by unit weight capture all the nodes found by using any of the longest paths, be they longest by unit, SPC or SPE weight. So the unit weight basket is a more efective tool that using eitrher of the SPC or SPE weight baskets. \tsenote{ANYTHING ELSE WE CAN SAY?}
\begin{table}[htb!]
	\centering
	\small 
	\begin{tabular}{lc||ccc}
		\multicolumn{2}{c||}{Basket $\BG(f)$ } & \multicolumn{3}{c}{Path $\Pcal$ } \\
		$G$  & $f$    & Longest (Unit)             & Longest (SPC)              & Longest (SPE)               \\ \hline\hline
		Unit & $0$    & 100.0\% (100.0\%--100.0\%) & 100.0\% (78.6\%--100.0\%)  & 100.0\% (100.0\%--100.0\%)  \\
		Unit & $1\%$  & 100.0\% (100.0\%--100.0\%) & 100.0\% (78.6\%--100.0\%)  & 100.0\% (100.0\%--100.0\%)  \\ 
		Unit & $5\%$  & 100.0\% (100.0\%--100.0\%) & 100.0\% (78.6\%--100.0\%)  & 100.0\% (100.0\%--100.0\%)  \\ 
		Unit & $10\%$ & 100.0\% (100.0\%--100.0\%) & 100.0\% (100.0\%--100.0\%) & 100.0\% (100.0\%--100.0\%)  \\ 
		\hline
		SPC & $0$     & 62.4\% (47.5\%--77.0\%)    & 100.0\% (100.0\%--100.0\%) & 92.3\% (78.9\%--98.5\%)     \\
		SPC & $1\%$   & 62.4\% (47.5\%--77.0\%)    & 100.0\% (100.0\%--100.0\%) & 92.3\% (78.9\%--98.5\%)     \\ 
		SPC & $5\%$   & 62.4\% (47.5\%--77.0\%)    & 100.0\% (100.0\%--100.0\%) & 92.3\% (78.9\%--98.5\%)     \\ 
		SPC & $10\%$  & 83.5\% (56.5\%--98.3\%)    & 100.0\% (100.0\%--100.0\%) & 100.0\% (84.7\%--100.0\%)   \\ 
		\hline
		SPE & $0$     & 76.2\% (59.7\%--86.0\%)    & 93.2\% (81.0\%--98.5\%)    & 100.0\% (100.0\%--100.0\%)  \\
		SPE & $1\%$   & 76.2\% (59.7\%--86.0\%)    & 93.2\% (81.0\%--98.5\%)    & 100.0\% (100.0\%--100.0\%)  \\ 
		SPE & $5\%$   & 76.2\% (59.7\%--86.0\%)    & 93.2\% (81.0\%--98.5\%)    & 100.0\% (100.0\%--100.0\%)  \\ 
		SPE & $10\%$  & 100.0\% (92.1\%--100.0\%)  & 100.0\% (98.3\%--100.0\%)  & 100.0\% (100.0\%--100.0\%)  \\ 
	\end{tabular}
	\caption{Basket coverage $\phi(\Pcal,\Bcal)$ giving the median with quartile values over 100 trials of a random geometric DAGs with $N=1002$ nodes and target out-degree $\ktarget=22.0$. 
		On each row specifies one basket $\BG(f)$ for a given weight $G\in\{\text{Unit, SPC, SPE}\}$ and a given fraction $f$. The value $f=0$ indicates that the basket contains all critical nodes, i.e.\ with $\cG=0$. The last three columns on the right specify the path $\Pcal$ to be the longest path by unit, SPC or SPE weight. The entries in the table are then the coverage $\phi(\Pcal,\Bcal)$, the percentage of path nodes lying in the given basket.}
	\label{t:ppp_criticality_coverage_trials}
\end{table}


%

%


\section{Real-World Networks}
\label{s:realworldnetworks}

In this section we move away from the synthetic directed acyclic graphs of the last two sections to look at networks found in real-world examples.
Previously, our directed acyclic graphs were constrained to lie in highly symmetric spaces and as a result we do not have strong clustering of nodes, the degrees of nodes show relatively little variation and there are no hierarchical structures.
Looking at real data sets also means that there is no geodesic from an explicit embedding space to use as a reference path.
So here we rely on the criticality analysis of \secref{s:lattice_criticality} and \secref{s:basketnotpath} to identify important nodes in terms of the flow of information through the directed acyclic graphs.
The real world networks we use are a range of sizes, mostly citation networks but from different areas.
We will summarise their main characteristics here but a more detailed discussion is given in \appref{as:realworldnetworks}.

We have nine citation networks built from data sets of academic papers and their bibliographies.
Eight of these are based on preprints in different sections of the \texttt{arXiv} preprint database up to 2017 \citep{C17}, denoted ``arXiv \emph{section}'' where \emph{section} is the abbreviation used by arXiv for one section.
The last citation network of academic papers comes from the articles published in the journals of the American Physical Society up to 2019, denoted ``APS Physics''.\tsenote{The ``Physics'' is redundant, should just be `APS'.}
The citation network denoted ``US Patent'' comes from US patents from 1975--1999 \citep{HJT01}  with 3.8M nodes and 16.5M edges.
The ``SCOTUS'' citation network comes from the data of \citet{FJ08} covering all decisions and citations made by the US Supreme Court up to 2002.
This has 25K nodes and 216K edges.
We also have eight vaccine citation networks.
Each is derived from various documents (academic papers, medical trials, patents) found by tracking back from one document giving formal government approval of one vaccine \citep{HPEO24a,HPEO24,HPEO25}.
Here we have ``AstraZeneca'' (Vaxzevria), ``BioNTech'' (BioNTech/Pfizer BNT162b2), ``Dengue'' (DENGVAXIA), ``Imvanex'' (monkeypox), ``Moderna'' (mRNA-1273), ``Novavax'', ``Shingrix'' (zoster), and ``Zabdeno'' (Ebola).\tsenote{Is this our standard notation? There seems to be no consistency in the naming.  Sometimes it is ``brand name (disease)'', or it is ``disease (brand name)'' or something else again.}

A rather different type of DAG is the one representing the dependencies of code in a Linux distribution in 2014, denoted here as ``Linux Deps'' which has around 31K nodes and 230K edges.
This is one of only two networks here which are not document citation networks.
Finally we have two small DAGs.
One denoted ``Python Deps'' represents dependencies in a Python installation on one machine from \citet{VE19a} and this is the only other network which is not a document citation network.
It has 215 nodes and 3,813 edges.
Our last DAG comes from data provided by \citet{E25a} which is derived from the DNA historiograph shown in Fig.1 of \citet{HD89} which in turn is derived from the information in \citet{A62}.
This provided the first illustration for main path analysis and, with just 39 nodes and 59 edges, is provided for historical context.
It is denoted as ``DNA (HD89)''.

The raw data often produces directed networks with a few cycles for various reasons, such as the multiple ``publication'' dates assigned to documents \citep{HBC15} or simply errors in the data.
Such cycles are removed to produce a DAG as detailed in the various sources cited for each network.

Unconstrained by any explicit embedding space, all our networks have clusters of nodes, heterogeneous degree distributions, and emergent hierarchical structures that arise from real-world dynamics.
So these provide a diverse test environment for path analysis algorithms and critical node identification across a range of scales, from thirty-nine to over one hundred thousand nodes.

One new issue that arises with DAGs derived from real data is that they often have multiple source and sink nodes.
Standard main path analysis such as SPC, as well as the other methods considered here, longest unit path and SPE main path, are all well defined for multiple sources and sinks as allowed for in our definitions in \secref{s:methods}.
However, to make a fair comparison with the interval DAGs produced in our artificial models, we produce an \textbf{induced interval} DAG (see Section~\ref{s:numerics} for implementation details).
This does not change the path counts used in constructing the SPC and SPE edge weights.
It does add one to both the height and depth of any node when working with unit edge weights but this is of little consequence to our analysis.

\subsection{Criticality Analysis and Basket Coverage}
\label{sec:vaccine-criticality}

We look first at the criticality analysis of \secref{s:lattice_criticality} and \secref{s:basketnotpath} for representative vaccine citation networks.

\begin{figure}[htb!]
	\centering
	\includegraphics{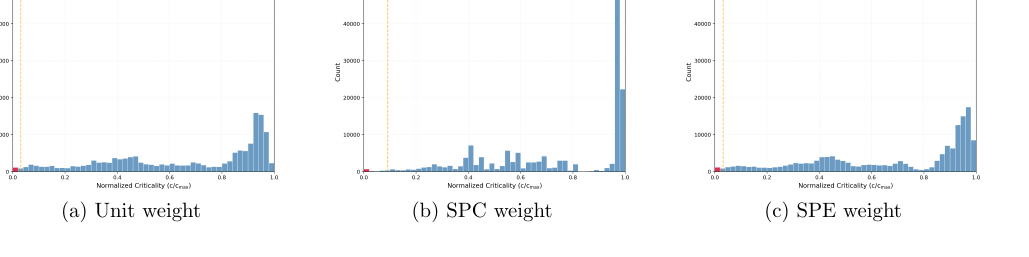}
	\caption{The distribution of criticality values $\cG$ for the Zabdeno vaccine network ($N=153,448$ nodes, $E=1,034,423$ edges) under three different edge weights: (a) unit criticality $\cunit$ (201 critical nodes, 0.13\%),
		(b) SPC criticality $\cspc$ (167 critical nodes, 0.11\%), and 
		(c) SPE criticality $\cspe$ (167 critical nodes, 0.11\%).
		All three measures show concentrated distributions with most nodes having low criticality, whilst critical nodes ($\cG(v)=0$, the red bars) form a small essential subset.
        The yellow dashed line marks the threshold separating the top 5\% most critical nodes (7,682 nodes) from the remainder of the network.
		Similar patterns are seen across all citation networks.
		\\
		\tsecomment{HP CHECK THESE NUMBERS}
        }
	\label{fig:zabdeno-criticality-histograms}
\end{figure}

In \figref{fig:zabdeno-criticality-histograms} we show histograms of the normalised criticality values for the Zabdeno vaccine, representative of large citation networks, under multiple weighting schemes (unit, SPC, SPE).
All three distributions exhibit the characteristic feature that critical nodes (those with $c(v)=0$) form a small but essential subset, whilst the vast majority of nodes have positive criticality values indicating they do not lie on the longest path under any of these edge weight schemes.
These criticality distribution patterns are consistent across the citation networks analysed, regardless of network size or domain, suggesting they are characteristic features of hierarchical citation topology.


We have argued that selecting a single path in a DAG to capture the most important nodes in a DAG is liable to be unreliable. So in \tabref{tab:simplified-interval-coverage}\tsenote{(and in an extended version \tabref{atab:simplified-interval-coverage} in the Appendix) }
we show how the basket $\Bcalunit$ based on unit weight criticality $\cunit$ performs in terms of the longest unit path and the SPC and SPE main paths.
We do this by looking at the coverage $\phi(\Pcal,\Bcalunit)$ of \eqref{e:phidef}, which is the percentage of the nodes each path $\Pcal$ is contained within the set of nodes defined by the unit basket $\Bcalunit(f)$ containing a given fraction $f$ of nodes in the network.
What we see is that in almost all cases, the set of critical nodes by unit weight $\Bcalunit(c=0)$ contains almost all of the nodes in the SPE path and most of those in the SPC path.
Generally, the set of unit critical nodes, all nodes lying on a longest unit weight path, captures all the nodes on the SPE path (i.e.\ the SPE main path is almost always one of the possible longest unit paths) and a large fraction of nodes on the SPC main paths.
If the basket with 1\% of unit critical nodes is larger, then pushing up to this larger set often gives much better results for the SPC.
All the networks with poor coverage of the SPE main path for the $f=1\%$ budget are examples where this basket $\Bcalunit(f=1\%)$ is smaller than the set of critical nodes.

It is noticeable that the SPC main path is less well covered than the SPE path.
Our experience from the geometric models is that the SPC path is more extreme, a larger proportion of nodes have high SPC criticality $\cspc$ as compared to unit and SPE criticality.
This is backed up looking at the distributions of criticality values in these networks, illustrated here by the example in \figref{fig:zabdeno-criticality-histograms}.
Although the SPC and SPE main paths are based on exactly the same counts of the number of paths, taking the logarithm in the construction of the SPE weights seems to smooth out the extreme fluctuations in the SPC counts, which typically vary factorially when we look at our simple models.

\begin{table}[htb!]
	\begin{tabular}{llrr|rr|rr|rr}
		& Network & Nodes & Edges 	& \multicolumn{2}{c}{$\cunit=0$} & \multicolumn{2}{c}{$f=1\%$} & \multicolumn{2}{c}{$f=5\%$}  \\
		& & & & SPE & SPC &  SPE & SPC & SPE & SPC \\
		\hline \hline
		\multicolumn{10}{l}{\textbf{Citation (Large)}} \\
		& US Patent            &  3,774,753 & 16,518,209 & 100\% & 82\% & 100\% & 100\% & 100\% & 100\% \\
		& APS Physics          &    527,271 &  5,987,551 & 100\% & 61\% & 100\% & 100\% & 100\% & 100\% \\
		& arXiv Cond-Mat       &    118,496 &    765,970 & 100\% & 67\% & 100\% & 100\% & 100\% & 100\% \\
		\midrule
		\multicolumn{10}{l}{\textbf{Citation (Medium)}} \\
		& SCOTUS               &     25,378 &    216,198 & 100\% &  74\% &  61\% &  42\% & 100\% &  99\% \\
		& arXiv Quant-Ph       &     36,955 &    263,235 & 100\% &  85\% & 100\% & 100\% & 100\% & 100\%  \\
		& arXiv hep-ph         &     34,530 &    419,528 & 100\% &  86\% & 100\% &  87\% & 100\% & 100\% \\
		& arXiv Physics        &     19,237 &     33,675 & 100\% & 100\% & 100\% & 100\% & 100\% & 100\%  \\
		& arXiv CS             &     10,352 &     11,114 &  10\% &  10\% &  90\% &  90\% & 100\% & 100\%  \\
		& arXiv Math-Phys      &      8,847 &     17,127 & 100\% & 100\% & 100\% & 100\% & 100\% & 100\%  \\
		\midrule
		\multicolumn{10}{l}{\textbf{Citation (Small)}} \\
		& arXiv Stats          &        984 &        825 & 100\% & 100\% &  67\% & 67\% & 100\% & 100\%  \\
		& arXiv Chaos          &        630 &        828 & 100\% & 100\% &  46\% & 46\% & 100\% & 100\%  \\
		\midrule
		\multicolumn{10}{l}{\textbf{Vaccine Networks}} \\
		& Zabdeno              &    153,448 &  1,034,423 & 100\% & 30\% & 100\% & 100\% & 100\% & 100\%  \\
		& Moderna              &    112,861 &    800,567 & 100\% & 97\% & 100\% & 100\% & 100\% & 100\%  \\
		& AstraZeneca          &    113,387 &    672,241 & 100\% & 55\% & 100\% & 100\% & 100\% & 100\%  \\
		& BioNTech             &     65,626 &    348,945 & 100\% & 94\% & 100\% & 100\% & 100\% & 100\%  \\
		& Imvanex              &     58,383 &    365,823 & 100\% & 84\% & 100\% & 100\% & 100\% & 100\% \\
		& Shingrix             &     27,046 &    179,952 & 100\% & 100\% & 100\% & 100\% & 100\% & 100\%  \\
		& Novavax              &     23,703 &    107,379 & 100\% & 100\% & 100\% & 100\% & 100\% & 100\%  \\
		& Dengue               &     15,284 &     85,828 &  65\% &   5\% &  65\% &   5\% & 100\% & 41\%  \\
		\midrule
		\multicolumn{10}{l}{\textbf{Non-Citation}} \\
		& Linux Deps           &     30,839 &    226,318 & 100\% &  98\% & 100\% & 99\% & 100\% & 100\%  \\
		& Python Deps          &        215 &      3,813 & 100\% & 100\% & 25\%  & 25\% &  50\% &  50\%  \\
		& DNA (HD89)           &         39 &         59 & 100\% & 100\% & 0\%   &  0\% &  15\% &  15\%  \\
	\end{tabular}
	\caption{The coverage $\phi(\Pcal,\Bcalunit)$ of \eqref{e:phidef} of the unit criticality basket which is the fraction of nodes from a path $\Pcal$ lie in the set of nodes in the basket. Results are shown for the SPC and SPE main paths, and for three different budgets: $\cunit=0$ (critical nodes only), $f=1\%$, and $f=5\%$. Note the results for $f=1\%$ can be worse than for the basket of critical nodes whenever the number of critical nodes is more than 1\% of the total number of nodes in the network which happens for some of our small networks.}
	\label{tab:simplified-interval-coverage}
\end{table}

There are still some notable deviations from our general rule as we would expect since no single prescription can ever work for all networks.
In this case we see the Dengue vaccine network and \texttt{arXiv CS} stand out.
What this suggests is that these two networks may have a different macroscopic topology of the other networks.
The main point here is that we have not ``cherry picked'' our data sets so that we can illustrate how our approach generally works well without additional work.
Where there are issues, such as we can see with our Dengue and \texttt{arXiv CS}  networks, further investigation is needed as there are likely to be special features in the data which have to be identified and taken into account.

\clearpage
\section{Discussion}\label{s:discuss}

Main path analysis has been widely applied to citation networks for over three decades \citep{HD89}, yet it has always lacked a proper theoretical foundation.  
The choices behind traversal counts and optimal paths have been justified mainly by post hoc validation rather than reasoning from first principles.  
Here we have filled that gap through a systematic investigation using two artificial models, a lattice and a random geometric DAG, before testing the methods on a range of real networks.

\subsection{Theoretical Analysis}\label{ss:theoretical}

On the theoretical side, we established an information-theoretic basis for traditional main path analysis.  
By recasting path counts in terms of entropy, the SPE main path of \eqref{e:Gspedef} links path-finding directly to information measures.  
The lattice model allowed explicit expressions for the SPC weights and accurate approximations for the SPE weights, illustrating the same behaviour we later saw in the other models and data sets.

We also generalised the definitions of height $\hG(v)$ and depth $\dG(v)$ for any weighting scheme $G$, and used them to define a simple measure of criticality, $\cG(v) = \HG - \hG(v) - \dG(v)$ in \eqref{e:cGdef}.  
This provides a way to distinguish important nodes from peripheral ones without relying on a single path.  
We used baskets of nodes $\Bcal(f)$ \eqref{e:Bdef}, selecting a fraction $f$ of the lowest-criticality nodes to capture the region of high structural importance around the core paths rather than just one trajectory.

In both artificial models, the longest unit path and the two main paths, SPC and SPE, approximate the geodesics of the underlying Euclidean geometry well. All three paths therefore identify the key geometric structure in a DAG.  
This generalises what had previously only been observed for the longest unit path in cube space and Minkowski spacetime \citep{BB91}.

The numerical work also made a simple point clear: if we focus only on nodes belonging to a single optimal path we miss many that lie just as close to the geodesic.  
Most analysis is based on a single main path which gives one narrative even though the citation network supports many near-optimal routes.  
When the weights are small integers (unit paths) multiple optima occur exactly; when the weights are large integers or floating-point values (SPC or SPE weights respectively) the ties vanish, but many paths have weights indistinguishably close to the maximum.  
Our basket formulation $\Bcal(f)$ \eqref{e:Bdef} captures this multiplicity cleanly, providing robustness against both algorithmic tie-breaking choices and uncertainties in the network data itself.

\subsection{Empirical Evaluation}\label{ss:empirical}

Real-world networks have none of the symmetry of our lattices in \secref{s:lattice} or the random geometric DAGs of \secref{s:rgdag} and so real-worl networks have no well-defined geodesic to use as a benchmark in our work.  
So in \secref{s:realworldnetworks}, we looked at over twenty different DAGs derived from data and compared the three longest paths (by unit, SPC and SPE weight) against each other and against our baskets of nodes.  
What we find is working with the basket defined to be the just the nodes of zero unit criticality, $\cunit=0$ in \eqref{e:cGdef}, those lying on one of the many longest unit paths, is a good compromise between having a larger number of nodes to consider while picking out the most important nodes. This unit critical basket $\Bcalunit(c=0)$ almost always contains the nodes lying a main path, SPC or SPE, and these unit critical nodes are all close to the geodesic where that is defined. This approach was exploited by \citet{HPEO24,HPEO24a,HPEO25}.

One practical issue is that SPC edge weights grow factorially with network size. This means that even networks with only a few tens of thousands of nodes produce integers well beyond $64$-bit precision. Our Moderna vaccine network ($112K$ nodes) reaches $68$ digits for the SPC weight of the SPC main path, and eighteen of our twenty-two networks require arbitrary-precision arithmetic for the SPC and SPE main path calculations. 
Approximation schemes are unlikely to help, since SPC weights are built from sums of many small similar numbers, not a few dominant terms.  
The SPE formulation softens the extremes through logarithmic compression but requires the SPC weights so this calculation has similar problems.
The current version of \texttt{python} handles these issues cleanly as integers are arbitrary-precision, but most languages do not. 

In terms of computation, SPC and SPE need only one extra sweep compared with the longest unit path, so they scale about twice as slowly as the calculation of the unit path weights, so SPC and SPE main path calculations are still entirely feasible.  
Nevertheless, the unit longest path and unit criticality baskets are faster, simpler, and easier to implement, and give similar or better results than the main path methods. So in our opinion, there is little reason to persist with SPC or SPE main path approaches unless one specifically needs their theoretical framing such as the information theory basis of SPE main path.

Our results also highlight that concentrating on a single path is too restrictive and often misleading.  
Real networks contain multiple routes of near-equal importance, and the notion of one privileged trajectory is more artefact than reality.  
Working with the set of unit critical nodes $\Bcalunit(c=0)$, or with small-budget baskets $\Bcalunit(f)$, captures the same core information, avoids spurious precision, and provides a stable basis for interpretation.  
In practice this approach is simple, fast, and empirically reliable.

So, while SPC and SPE have their uses for theoretical comparison, we see no advantage in using them for routine work.  
The unit critical node set gives the same insight at a fraction of the cost and with far less numerical baggage.

\subsection{Concluding Remarks}\label{ss:conclusion}

This study ties together formal theory, artificial models of DAGs, and empirical analysis to place main path methods on a solid foundation.  
Across all settings the same pattern emerges: unit weight criticality $\cunit$ \eqref{e:cGdef} identifies a compact, stable region of important nodes, while single-path approaches overstate precision and understate uncertainty.  
The basket framework acknowledges the multiplicity of plausible paths and provides a practical, interpretable way to describe the backbone of directed acyclic graphs.  
In short, it turns main path analysis from a heuristic convenience into a coherent, defensible method.


\clearpage
\begin{center}
	\Large\textbf{Appendices}
\end{center}
\appendix
\renewcommand{\thesection}{\Alph{section}}
\setcounter{equation}{0}
\setcounter{section}{0}
\counterwithin{equation}{section}
\counterwithin{figure}{section}
\counterwithin{table}{section}
\renewcommand{\theequation}{\thesection.\arabic{equation}}
\renewcommand{\thefigure}{\thesection.\arabic{figure}}
\renewcommand{\thetable}{\thesection.\arabic{table}}
\renewcommand{\theHsection}{\Alph{section}}
\renewcommand{\theHequation}{\thesection.\arabic{equation}}
\renewcommand{\theHfigure}{\thesection.\arabic{figure}}
\renewcommand{\theHtable}{\thesection.\arabic{table}}



\tsecomment{Start to construct the actual appendices supplementary material we will need for the paper.}


\section{Additional Background Notes}

\subsection{Use of Main Path Analysis in the Literature}\label{as:mpalit}

Main path analysis was first proposed by \citet{HD89} is a widely used method used in bibliometric studies to find the most important trajectory in a given citation network. Studies include
\citet{HD89,CHH93,B03d,V07a,LL08,FNV09,LL12,YKLK13,CB14a,XLLZ14,BDFK14,BGGMN15,LK15,LPC17,F21,KJKS22} but a longer list and a review\footnote{Some online resources at the time of writing include
    \href{http://vlado.fmf.uni-lj.si/pub/networks/Data/cite/default.htm}{pajek}, 	
    \href{http://www.garfield.library.upenn.edu/histcomp/index.html}{garfield},
	\href{https://www.sussex.ac.uk/webteam/gateway/file.php?name=sewp166&site=25}{sussex},
	\href{https://davincierlab.weebly.com/list-of-main-path-articles.html}{da Vincier Lab Main Path} web page.} 
can be found in \citet{LLH19}.

\begin{figure}[htbp]
    \centering
    \includegraphics[width=0.9\textwidth]{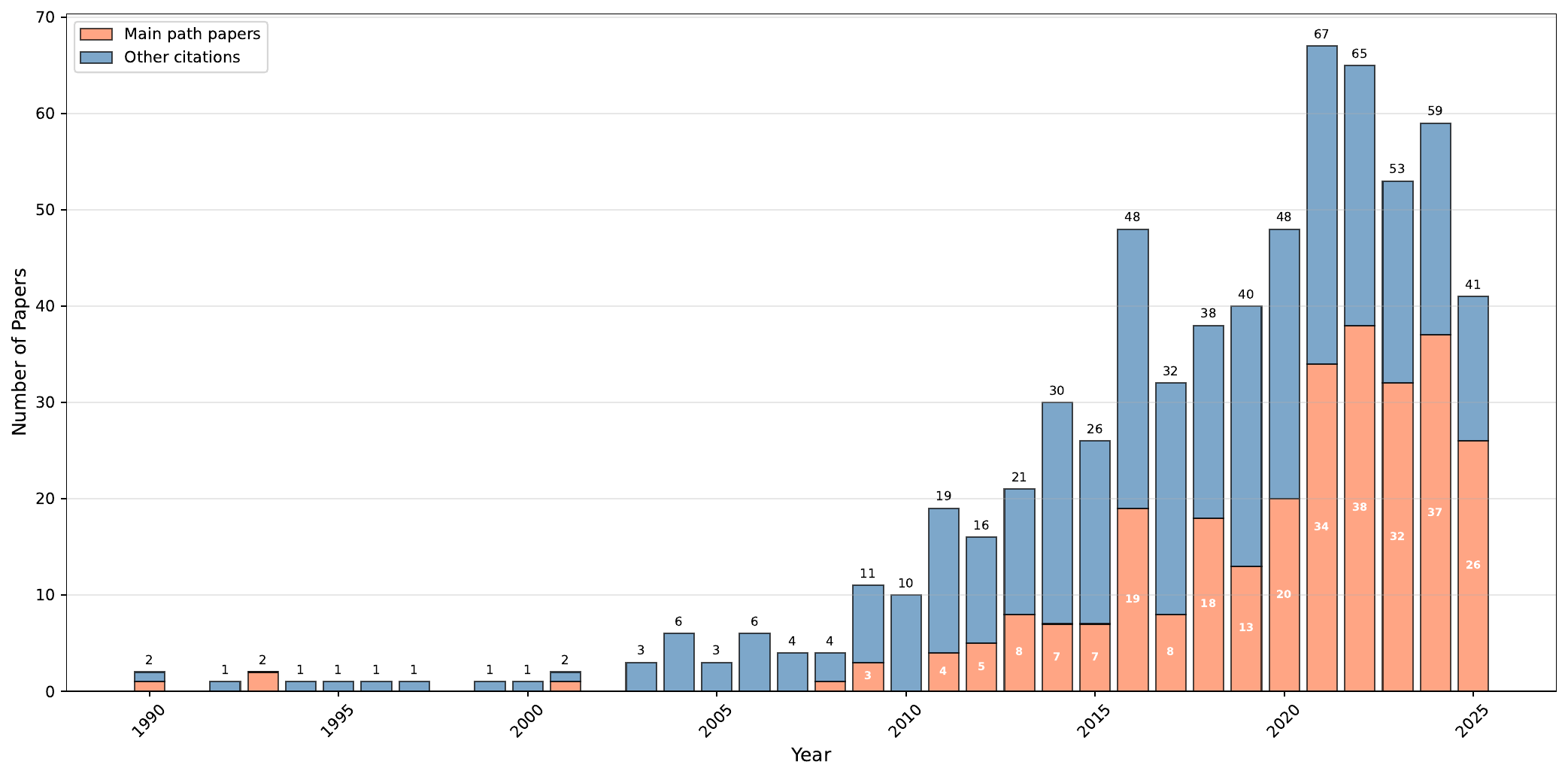}
    \caption[Temporal distribution of main path citations]{Temporal distribution of papers citing \citep{HD89}. The total height, given in black above each bar, represents the number of papers citing \citep{HD89} in each year according to \href{http://dimensions.ai}{\texttt{dimensions.ai}} on 13\textsuperscript{th} December 2025.  The orange segments indicate the subset of papers that explicitly mention ``main path'', ``main-path'', or ``mainpath'' in their 
    title or abstract with that number given in white in the centre of the orange bars. 
    This visualisation shows how the main path methodology has been adopted over time, with a notable concentration in the last fifteen years.}
    \label{f:mpapapercount}  
\end{figure}

To find the wider impact of main path analysis we looked at all citations of \citep{HD89} on \href{http://dimensions.ai}{\texttt{dimensions.ai}} and found 663 papers of which 284 papers also contained the phase ``main path'', ``main-path'', or ``mainpath'' in their 
title or abstract. The distribution by year is shown in \figref{f:mpapapercount}.

As shown in \figref{f:mpapapercount}, the popularity of main path analysis has increased substantially over time, with a particularly notable acceleration in recent decades. The growing proportion of papers explicitly referencing main path methodology among all citations to \citep{HD89} reflects both the expanding adoption of these techniques and their increasing recognition as a standard tool in citation network analysis.

Further evidence of the utility of main path analysis is that it is built into one of the most popular numerical  packages for social network analysis \href{https://mrvar.fdv.uni-lj.si/pajek/}{\texttt{pajek}} \citep{BM98,BM03}. Several other implementations are discussed in \secref{as:software}.

\subsection{Main Path Analysis Software Ecosystem}\label{as:software}


We give a brief review of some of the packages and libraries that can calculate the main path for a DAG. The range of numerical packages available along with the extensive literature shows that main path analysis is a well established technique in bibliometrics.

One of the oldest packages for social network analysis which is still under active development is 
\href{https://mrvar.fdv.uni-lj.si/pajek/}{\texttt{Pajek}} \citep{BM98,BM03} so this is widely used, well tested, well documented \citep{NMB18}, fast and efficient. It can also handle extremely large networks \citep{mrvar2016,pajekxxl2018}.
The integration of efficient main path algorithms \citep{B03d} into \texttt{Pajek} means this is probably the main tool used for main path analysis in the literature. The main drawback here is that it is closed software so we are unsure how this package handles the numerical issues highlighted in our work.


\href{https://jdvelasq.github.io/techminer/}{\texttt{TechMiner}}\footnote{\url{https://jdvelasq.github.io/techminer/main-path-analysis.html}} is an open-source \textsc{python} package with a \href{https://jdvelasq.github.io/techminer/main path-analysis.html}{Main Path} algorithm which you can find in the open code at \href{https://github.com/jdvelasq/techminer/blob/main/src/core/main_path.py}{\texttt{main\_path.py}}.

There are various open-source implementations of main path analysis across multiple programming languages to be found in open repositories such as on \href{https://github.com/}{\texttt{GitHub}} but they are not widely used and they are not under active development\footnote{For example,
	a \texttt{Julia} implementation \url{https://github.com/jfb-h/MainPaths.jl},
	a \texttt{python}/\texttt{NetworkX} version \url{https://github.com/felipefrr/mpa_splc},
	and a \texttt{C++} web application \url{https://github.com/bencabrera/wikiMainPath}.
}.
Such implementations offer code snippets and specific algorithmic implementations  which might be valuable resources for researchers requiring customisation or integration into larger analytical workflows.

Several other established bibliometric tools can import citation data and produce citation networks but they lack any builtin main path analysis software. Typically they recommend exporting it to \texttt{pajek} if main path analysis is required.

\subsection{Methodological Variants and Applications}\label{as:variants}

The framework distinguishes between sources (foundational documents cited but not citing), sinks (recent documents citing others but not yet cited), and intermediates, capturing the hierarchical and temporal nature of knowledge evolution.
Methodological variants include backward local search (revealing convergent streams), multiple global paths, and key-document paths \citep{LL12}.
Recent advances include semantic MPA that integrates text embeddings to ensure topical coherence and discover multiple developmental trajectories simultaneously \citep{KJKS22}.
\emph{SimSPC} handles cycles by counting simple search paths and extracting main paths directly on cyclic networks \citep{jiang2020}.
Hybrid schemes combine traversal counts with prestige/centrality (e.g., SPC-PR, SPLC-PR, SPNP-PR) to weight citations by citing-paper influence \citep{yu2021}.

MPA has traced backbones in scientific literatures across disciplines and has seen extensive use in patent citation networks to map technological trajectories.
Applications include studies of semiconductor manufacturing \citep{F21}, fuel cell research \citep{V07a}, data communication standards \citep{FNV09}, thin-film solar cells \citep{lai2021}, and environmental innovation \citep{BGGMN15}.
Such applications reveal paradigm shifts, methodological innovations, and cross-field knowledge diffusion.
Methodological critiques and diagnostics examine path-length biases and coverage \citep{kuan2023,kuan2024}.

\newpage
\section{Additional Material on Methods}\label{as:methods}

\subsection{Greedy paths}\label{as:greedypaths}

The \tsedef{Greedy Perpendicular Distance path} (GPD path) $\Pcalgpd(s,t) = \big( v_0=s,v_1,\dots,v_L=t \big)$  running from node $s$ to node $t$ in a DAG is defined using a greedy algorithm in which we grow our path $\Pcal$ of length $L$ by adding the successor node $w$ of the last node $v_L$ which has the smallest perpendicular distance $\Delta(w)$ (illustrated in \figref{af:pd}).

In more detail we can define the Greedy Perpendicular Distance path as follows. It is useful to define the set of successors $\Ncal^+(v)$ of a node $v$ to be the set of nodes at the head of edges leaving node $v$, so $\Ncal^+(v) = \{ w | (v_n,w) \in \Ecal\}$ of a node $v$. 
\begin{enumerate}
	\item Set the index $j$ of the last node in our path to zero, $j=0$.
	\item Define the initial node $v_0$ in our path to be the source node $s$, so that $v_0=s$.
	\item \label{l:loop} Look at the successors $\Ncal^+(v_j)$ (within the interval) of the last node $v_j$ in our current path.
	\item \label{l:greedyselect} Select a successor node $v_{j+1} \in \Ncal^+(v_j)$ to be the one closest to the geodesic.
		\beq
		   v_{j+1} \;=\; \arg\min_{w\in \Ncal^+(v_j) } \Delta(w) \, .
		\eeq
	\item Increment the counter $j$ by one.
	\item If the last node in our path (now $v_j$) is not the target node, i.e.\ if $v_{j} \neq t$, return to step  \ref{l:loop}.
	\item \label{l:stop} The greedy perpendicular distance path (GPD path) $\Pcalgpd(s,t)$ is the sequence of nodes $\Pcalgpd(s,t) = (v_0,v_1,\ldots,v_j)$.
\end{enumerate}

In the same way the greedy degree path just replaces step \ref{l:greedyselect} in the GPD algorithm by selecting a successor node $v_{j+1}$ to be the one with smallest degree
\beq
	v_{j+1} \;=\; \arg\min_{w\in \Ncal^+(v_j) } k_w \, .
\eeq

Our random paths are also defined using the same greedy algorithm but now step \ref{l:greedyselect} in the GPD algorithm is replaced by selecting the a successor node $v_{j+1}$ uniformly at random from the set of possible successor nodes $\Ncal^+(v)$.

Note that this random path algorithm is not sampling from the set of all possible paths uniformly at random. The random paths chosen do not even respect the direction reversal symmetry discussed at the end of \secref{s:DAGformal}.
That is if we used the same algorithm in the reversed DAG, $\Dcalrev$, and if we then reversed the order of the nodes the random path found, we would now have a random path $\Pcalrev$ in the original DAG $\Dcal$. However such paths $\Pcal$ found in the original are being sampled differently from the paths $\Pcal$ found from random searches in the reverse random graph, i.e.\ for the same path $\Pcal=\Pcalrev$ the probability of choosing the path from the reversed DAG is not guaranteed to be the same as finding this path in the original DAG. Here we are not interested in exactly how we sample the set of paths. Random paths are used in this paper to provide a more generic path for comparison and our iterative definition is sufficient for our purposes.


\subsection{Numerical and Computational Implementation}\label{as:numerics}

\subsubsection{Precision-Sensitive Motifs in General DAGs}\label{as:motifs_precision}

\begin{figure}[htb]
	\centering
	\includegraphics[width=0.8\textwidth]{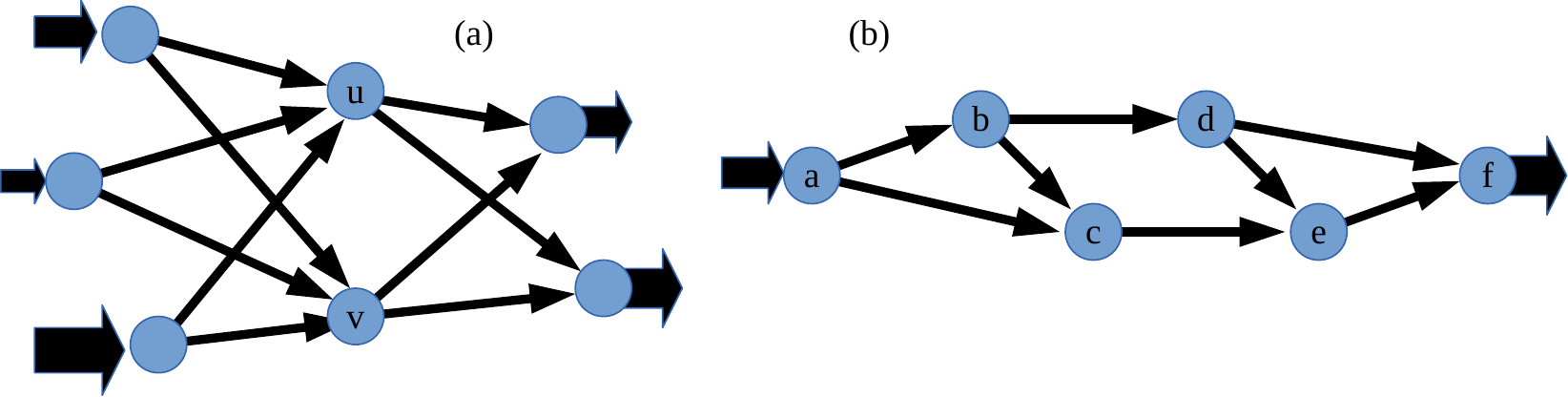}
	\caption{Precision-sensitive motifs that induce degenerate longest paths by unit, SPC or SPE weight. In both cases a diamond motif can be seen which causes the problems. The large arrows on the left (on the right) of each network represent the number of paths arriving (leaving) these boundary nodes provided these networks are sub-DAGs induced by the nodes shown.
		In case (a), nodes $u$ and $v$ are the centre of six different diamond motifs. Since nodes $u$ and $v$ have exactly the same neighbours, if we swap nodes $u$ and $v$ we will get the exactly same DAG (a graph automorphism). Any path $\Pcal$ through $u$ has an equivalent path $\Pcal^\prime$  where node $u$ is switched for node $v$ yet the path weights for any weight will be identical, $\LG(\Pcal) = \LG(\Pcal')$. In case (b) we show a ladder structure which also has a diamond motif defined by nodes $\{b,c,d,e\}$. This gives a pair of paths $\Pcal=(\ldots,a,b,c,e,f,\ldots)$ and $\Pcal'=(\ldots,a,b,d,e,f,\ldots)$. There are no graph automorphisms, for instance swapping $c$ and $d$ do not give the same network as $f$ is a successor of $d$ but not $c$. However, we still have $\LG(\Pcal) = \LG(\Pcal')$ for unit, SPC and SPE weights and further, these are the paths with largest path weights of these types, regardless of the number of paths arriving and leaving this sub-DAG.
	}
	\label{fig:precision_motifs}
\end{figure}

The precision problems seen in all our models arise whenever a DAG contains simple motifs that create 
two or more routes with either exactly the same or approximately the same path weights. Common problems come from 
diamond structures (two paths from $u$ to $v$ of equal weight), long near-parallel chains that differ only on a few edges, and reconvergence hubs where many incoming edges have almost identical weights. For example, see \figref{fig:precision_motifs} for some examples. In each case, the path found to have the longest numerical value for very large integer valued SPC weights or the real valued SPE weights can flip with rounding errors or the ties can be broken in a way that depends on the details of the numerical implementation\footnote{For instance, the order in which nodes are stored in data structures representing the DAG are not always guaranteed to be the same for every instance of the same code run on the same machine with the same parameter values. This then lead to strange changes in results.}.

\tsecomment{??? ``, so insisting on a unique main path overstates certainty.''}

More generally, any symmetry or near-symmetry in a network produces path degeneracy: 
balanced forks, automorphism-related subgraphs, or bottlenecks with almost-equal edges all create situations where different paths are indistinguishable at the resolution of the model. 
Hypercubic lattices provide the cleanest examples, but the same issues can arise in real citation DAGs.

However, these numerical issues in trying to find a unique path in a reproducible manner highlight the fact that there are actually many paths with similar if not exactly the same weight. So while careful coding can deal with these issues, a better response is to move away from finding one path which is inherently unstable. A better response is to move to our basket-based approach of \secref{s:basket}: rather than report one arbitrary ``winner,'' we collect and analyse the set of paths that are indistinguishable within a chosen tolerance.

\subsubsection{Arbitrary-Precision Arithmetic Requirements}

The computational implementation of main path analysis presents numerical challenges, particularly in the rapid growth of search-path counts with network size.
Search-path computations require path multiplicities $\Omega_{ui}$ from initial nodes $i \in \mathcal{I}$ to each node $u$ and $\Omega_{fv}$ from each node $v$ to final nodes $f \in \mathcal{F}$, as defined in \equationref{e:Omegadef}. 

For large networks, these multiplicities reach extreme values: vaccine citation networks routinely exceed $10^{40}$, random geometric DAGs with $N=250,002$ nodes produce SPC weights with 422 digits (see \figref{af:spc250k}), and networks with $N=100,000$ nodes yield SPC weights with up to 271 digits.
IEEE 754 double precision represents integers exactly only up to $2^{53} \approx 9\times 10^{15}$; hence we use arbitrary-precision integer arithmetic throughout the numerical passes described in \eqref{e:Wspcdef} and \eqref{e:Xspcdef}, maintaining exact computation until the final edge weights $G^{\text{SPC}}_{vu} = W^{\text{SPC}}_u X^{\text{SPC}}_v$ from \equationref{e:Gspcdef} are computed for the longest-path routine.

The search-path entropy (SPE) variant operates in logarithmic space for numerical stability.
SPE weights are derived from SPC path counts using logarithmic transformation: for each edge $(u,v)$, we compute $w_{\text{SPE}}(u,v) = \ln(w_{\text{SPC}}(u,v))$ where $w_{\text{SPC}}(u,v)$ represents the exact arbitrary-precision path count.
This logarithmic dampening prevents numerical overflow whilst preserving the multiplicative path-count structure essential for identifying high-connectivity routes.
The SPE edge weights are then used directly in the longest-path computation following \equationref{e:Gspedef}.

\subsubsection{Precision Issues and Floating-Point Artifacts}

Even double-precision (f64) arithmetic fails to distinguish paths at high-symmetry points, sometimes inflating the SPE-critical count beyond the true value.
Only exact arithmetic isolates the true degenerate structure.
For example, in 3D lattice analysis (see \secref{s:latticenumerics}), floating-point computations can report up to 58 SPE-critical nodes, but exact arithmetic reveals exactly 54 critical nodes and 2 degenerate optimal paths.
These artefacts vanish under higher or arbitrary precision.

In precision-sensitive motifs (see \secref{as:motifs_precision}), simple structures such as symmetric diamonds or near-parallel ladders can cause path selection to flip with rounding errors.
This motivates our basket-based approach (Section~\ref{s:basket}): rather than report one arbitrary ``winner,'' we collect and analyse the set of paths that are indistinguishable within a chosen tolerance.

For SPE computations, we use scaled high-precision logarithms to handle near-ties where multiple paths are indistinguishable at standard precision.
Near-ties mean multiple paths may be equally optimal at the chosen precision level, requiring deterministic tie-breaking.
For certified uniqueness, the implementation can increase the log scale or compare path weights multiplicatively with BigInt arithmetic, though our reported statistics remain unchanged under such stricter checks.

\subsubsection{Algorithmic Complexity and Implementation Details}

Edge weight computation proceeds via topological ordering with $O(N + E)$ complexity, where $N$ represents the number of nodes and $E$ represents the number of edges.
The implementation maintains separate computational pipelines for different weight types: exact arbitrary-precision arithmetic for SPC weights, scaled integer representation for SPE weights to avoid floating-point precision issues in logarithm computation, and standard precision for baseline methods.

Path finding proceeds woprks through nodes in topological order (see \secref{s:DAGformal}).
The longest-path routine maintains distance arrays for maximal weights and predecessor arrays for reconstruction.
For a path $\mathcal{P}$, we seek to maximize either $L_{\text{SPC}}(\mathcal{P})$ from \equationref{e:Mspcdef} or $L_{\text{SPE}}(\mathcal{P})$ from \equationref{e:Lspedef}, depending on the chosen metric.

Path construction algorithms vary in computational complexity according to their optimisation strategy.
Local search methods achieve $O(N)$ complexity through greedy construction, whilst global optimisation maintains $O(N + E)$ complexity.

Ties are broken in a deterministic manner to ensure reproducibility. Care is needed as access to some data structures is not guaranteed to produce elements in the same order even on the same machine. 
If two predecessors $u,u'$ yield identical scores for a node $v$,
\beq
\mathrm{dist}[u] + w(u,v) \;=\; \mathrm{dist}[u'] + w(u',v),
\eeq
where $w(u,v)$ represents either $G^{\text{SPC}}_{vu}$ or $G^{\text{SPE}}_{vu}$ depending on the method, we select the path whose sequence of node indices is lexicographically minimal.
Concretely, we iterate candidate predecessors of $v$ in increasing index order and relax with a strict inequality ($>$).
On ties, the current choice is retained, thereby preferring the earlier (smaller-indexed) predecessor and propagating lexicographic minimality through the DP.
For example, between optimal paths $0\!\to\!3\!\to\!7\!\to\!10$ and $0\!\to\!4\!\to\!7\!\to\!10$ of equal weight, the former is selected since $3<4$.

\subsubsection{Network-Specific Implementation Approaches}

All analyses in this paper employ an \tsedef{induced interval} approach, where artificial global source and sink nodes are added to create a single-source single-sink structure.
For lattice models, the source and sink are naturally defined as opposite corners of the lattice interval.
For random geometric DAGs and real-world citation networks with multiple sources and sinks, we add artificial global source and sink nodes connected to all original source and sink vertices respectively.
This modification enables comprehensive traversal algorithms to operate across the entire network, including disconnected components, whilst maintaining the essential DAG structure for path-finding analysis.

All path-finding algorithms used in this paper are source-to-sink algorithms operating on these induced intervals.
The \tsedef{Unit Path algorithm} identifies the longest path by unit length.
The \tsedef{SPC-weighted path algorithm} employs weighted longest path computation using exact arbitrary-precision SPC weights.
The \tsedef{SPE-weighted path algorithm} optimises paths using logarithmically transformed traversal counts.

\subsubsection{Validation and Reproducibility}

Validation includes comparisons with analytical results on small lattices where exhaustive enumeration is feasible.
On the lattice, numerical path counts match exact enumeration available because of the lattice's combinatorial structure.

Reproducibility is ensured via deterministic random number generation with seeds determined by network size, connection parameters, and spatial dimension, so identical parameters yield identical networks while differing parameters produce statistically independent realizations.

Alternative baselines include an unweighted shortest path (breadth-first search with unit edge weights), a random path generated by uniform local choices constrained to reach the sink as defined in Section~\ref{s:randompath}, and a greedy perpendicular distance (GPD) path that iteratively selects the successor minimizing perpendicular distance $\Delta(v)$ to the geodesic $\Gamma(s,t)$ between source and sink as defined in Section~\ref{s:gpdpath}.

\section{Additional Material on the Hypercubic Lattice Model}\label{as:lattice}

\subsection{Order-Reversal Symmetry on the Lattice}\label{as:orderrevsymm}

We noted in the main text that there are features in some of the numerical results which come from a symmetry of the DAGs defined on intervals of the square lattice model. In particular there is an \tsedef{order-reversal symmetry} $\That$ in which we reverse the direction of all the edges in our hypercubic DAGs to create a second DAG $\Dcalrev$. If the nodes are unlabelled, this transformation gives us exactly the same DAG. Measurements on our lattice models then lie in a representation of a $Z_2$ symmetry group. 

\begin{figure}[htb]
	\begin{center}
		\includegraphics[width=0.45\textwidth]{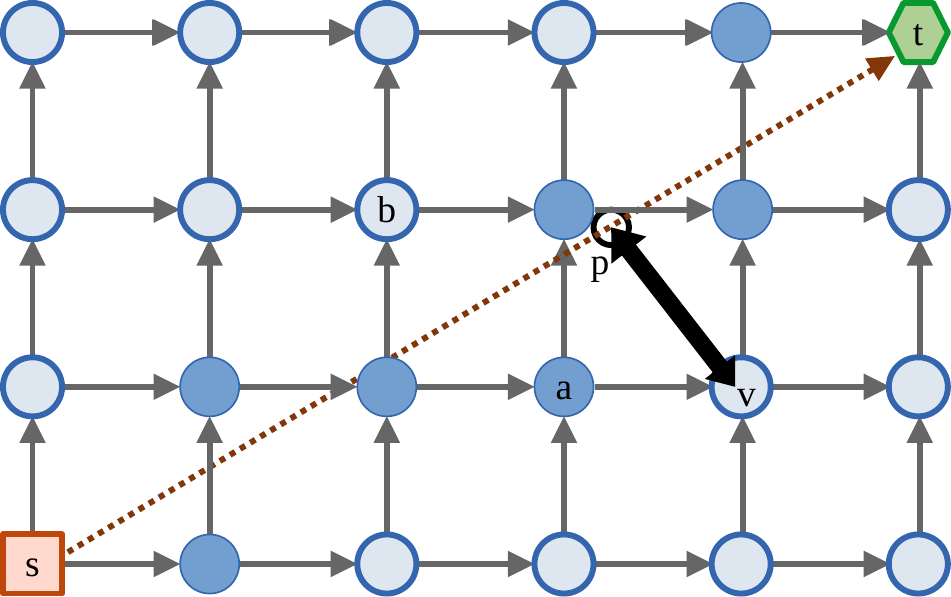}
	\end{center}
	\caption{Illustration of the perpendicular distance $\Delta(v)$ from a node $v$ to the geodesic as defined in \eqref{e:perdistdef}. The geodesic from the source node $s=(0,0)$ to the target node $t=(5,3)$ for this section of Euclidean space $\Ibb(s,t)$ is the diagonal shown as a red dashed line. The directed square lattice within this is the interval $\Dcalint(s,t)$ whose node set is the nodes shown as circles (filled or unfilled) plus the source node $s$ (red square) and the sink node $t$ (green hexagon). The perpendicular distance from a node $v$ to the geodesic, to a point $p$ (a black open circle), is the length of the thick black double-headed arrow. We also show the greedy perpendicular distance path (GPD, filled circles plus nodes $s$ and $t$) where the average distance of the nodes in the path to the diagonal is minimised. 
		\\
		\tsecomment{We probably do not need this figure. In any case, this figure is for the lattice model and not for a generic path on a generic DAG.}
	}
	\label{af:pd}
\end{figure}

Every node $v$ is paired up with a second node by this symmetry, $\vbar=\That(v)$, with coordinates $\vbar_i = t_i-v_i+s_i$ where in the original DAG $\Dcal$ the source node has coordinates $s_i$ and the sink node has coordinates $t_i$. These pairs of nodes $v$ and $\vbar=\That(v)$ linked by order reversal symmetry are always distinct except when the length of the sides of the hypercube are all even when there is one node at the centre mapped onto itself, a case we avoid in all our illustrations. 
An example of this is shown in \figref{af:pd}.

For example in \figref{af:pd} we can exchange node $a$ for node $b$ on this path (by symmetry of the rectangle, i.e.\ by order-reversal symmetry of these DAGs, $b=\That(a)$), but these are the only two examples of paths which minimise the perpendicular distance from path to diagonal. Since we have chosen a lattice of size $(L_1,L_2) = (5,3)$ in units of the lattice spacing, these $L_i$ have no common denominator and so the geodesic from $s$ to $t$ does not pass through any other lattice points.

\subsection{Additional Information on Criticality of Points}

\tsecomment{Concepts need to be defined properly}

Here we report a view  complementary to that shown in \tabref{tab:lattice-criticality-summary} by using an inverse CDF in \tabref{tab:lattice-criticality-invcdf} which shows the node with the $x$-th lowest criticality, what is its relative value? Using our analysis tool, we compute the $1$st, $5$th, $10$th, and $20$th percentiles of $c/c_{\max}$:
\begin{table}[htb]
	\centering
	\begin{tabular}{c||c|c|c|c}
		Path          & $x=0.01N$             & $x=0.05N$             & $x=0.10N$             & $x=0.20N$ \\ \hline
		Longest (SPC) & $24.38\times 10^{-6}$ & $541.4\times 10^{-6}$ & $4,308\times 10^{-6}$ & $20,880\times 10^{-6}$ \\
		Longest (SPE) & $1.090\times 10^{-6}$ & $17.12\times 10^{-6}$ & $142.2\times 10^{-6}$ & $784.7\times 10^{-6}$ \\
	\end{tabular}
	\caption{Inverse-CDF of $c/c_{\max}$ for the $78\times71$ lattice at the 1st, 5th, 10th, and 20th percentiles. Values are given to 4 significant figures.
	}
	\label{tab:lattice-criticality-invcdf}
\end{table}
These inverse-CDF values complement the fixed-threshold fractions in \tabref{tab:lattice-criticality-summary} and show that SPE concentrates much more mass near zero than SPC. In the next section we build on this with our ``basket'' coverage of \eqref{e:Bdef}.

\section{Additional Material on Random Geometric DAGs}\label{as:rgdag}

\tsecomment{I suggest we replicate the same subsections as in the main text in \secref{s:rgdag} (as far as we can) and then extra material for each of those sections can be placed here. We can cut and paste some of the explanations from the main text too as this may have to be a stand alone pdf for the journal.}


\tsecomment{Definition of model -Nothing here yet.}


\subsection{Additional Numerical Results $D=2$}\label{as:pppnumericsD2}

\subsubsection{Distance to Geodesic of Random Path}\label{as:pppdeltarandom}

In some of our visualisations such as \figref{f:ppp-dag-deg22-n1000} and in some of our measurements, the random path is appears to be surprisingly close to the geodesic. In particular, we look at fixed target out-degree $\ktarget$ and vary $N$, we the random path appears to get closer to the geodesic, not in accord with common-sense. The problem with this is that we are looking at this measurement from the point of view of a visualisation such as \figref{f:ppp-dag-deg22-n1000} and we are working in terms of distance units where the length of the side of the hypercube is $1.0$. In these units the scale for the perpendicular distance of the random path to the geodesic, $\Delta$, scales with $N^{-1/(2D)}$, i.e.\ in these units the random path \emph{does} get closer to the geodesic. However, the correct way to look at this is in terms of units of $a$, the short distance scale set by the typical distance between neighbouring points. In these units, with fixed $\ktarget$, the density of points is fixed, $R$ is fixed, but as $N$ increases the sides of the hypercube get longer (by $N^{1/D}$). In these units the perpendicular distance of the random path to the geodesic does indeed increase as $\Delta/a \sim N^{+1/(2D)}$. We can see this as follows.

The standard deviation $\sigma_1$ in the change in the perpendicular distance of the path from the geodesic in one step is $\sigma_1 = \surd \langle (\Delta(v_i)-\Delta(v_{i-1}))^2 \rangle = c_1 R$. Here $c_1$ is a constant which is determined by the geometry and so depends on the dimension but it will always have an upper bound of a half. The largest value for the perpendicular distance of the random path to the geodesic, $\Delta^\mathrm{(max)}$, can be estimated to be the standard deviation after $L$ steps which scales as $R \sqrt{L}$ if the random walk is $L$ steps long. We can estimate the number of steps $L$ to scale as $1.0/R$ in units where the sides of the hypercube interval are equal to $1.0$. So $\Delta^\mathrm{(max)} = c^\mathrm{(max)} \sqrt{R}$. In fact the geometry here means the change in perpendicular distance is usually very limited so the coefficient $c^\mathrm{(max)}$ is often quite small. We see in many examples that $\Delta^\mathrm{(max)} \approx \sqrt{R}$ after fifteen or so steps.

We usually work with models where there is a fixed $\ktarget$. So to see how the random path behaves as we vary $N$ we use \eqref{e:radiusdef} to find that $\Delta^\mathrm{(max)} \sim (\ktarget/N)^{1/(2D)}$ as measured in units of the side of the hypercube. So for fixed $\ktarget$, and for hypercubes of side $1.0$, the random path will appear to be getting closer to the geodesic when viewed, as we do in \figref{f:ppp-dag-deg22-n1000}, with a fixed side for the hypercube but increasing point density. However, we are using a bad length scale, the side of the hypercube, to view this problem.  It is better to use  $a \approx N^{-1/D}$ which is a measure of the spacing between nearest neighbour points in this model. In these units $R/a$ is independent of $N$ so $\Delta^\mathrm{(max)} \sim N^{-1/2D}$. That is the random path is actually diverging slowly\footnote{We have neglected several issues but we do not expect this will change the scaling. We expect the number of steps in the random walk will be on average $L = c_t 2\sqrt{D}/R$ where $c_t$ is some constant set by geometry with an upper bound of one half.  In fact, the random path will hit the boundary of the interval in a little less than $L$ steps most of the time and after that the random walk will be forced to move closer to the geodesic. So when estimating the largest deviation, this $L$ is a slight over estimate.} from the geodesic when working in units of distance given by the interparticle scale $a$.

This is a nice illustration of one of the drawbacks of this Euclidean RGG based model. That is, unlike generic networks with an exponential growth in the number of neighbours $L$ steps away from each node (for example see the discussion in \cite{EC22}), here we are constrained by a finite dimensional Euclidean space where the number of neighbours grows much slower as $L^D$.  The forward orthant that defines the possible future set of each node is tightly constrained in space. So even a random path cannot escape very quickly. Contrast that with the classic ``six degrees of separation'' of complex networks, where the number of neighbours $L$ steps from a node grows exponentially so the distance to any one node is likely scaling with the logarithmic of the number of nodes in a network. So on the surface, it might appear the Euclidean space of the random geometric DAG model is unrealistic. However, there is some evidence that citation networks may not be as unconstrained as random networks \citep{CE16} so perhaps the constraints of Euclidean space in this model may not be so unreasonable. Perhaps the random path \emph{within an interval} is not such a bad null model for paths?

\subsubsection{Path weights}\label{as:pathweights}

\tsecomment{This discussion is not so central to the paper. Need to decide if we keep it here.}


The numerical results for the unit length $\Lunit(\Pcallp)$ of the longest paths by unit length $\Pcallp$ in this random geometric DAG model with finite number of nodes $N$ and a finite cutoff $R$ can be compared against the exact asymptotic values given by \citet{BB91} for this model in the limits of $N,R \to \infty$. 
They state that\footnote{Since the sides of the hypercube between source $s$ and sink $t$ ($s_i=0$, $t_i=1$) are all length one in these models, and $a = \Nbox^{-1/D}$ is a measure of the internode distance scale, then $\Nbox^{1/D}$ is the length of a side in internode distance units $a$. So the results of \citet{BB91} are comparing the unit length of the longest path and the length of the \emph{sides}. They are not comparing against the length of the geodesic $\surd D$ bigger.} $\Lunit(\Pcallp)/\Nbox^{1/D}$ should lie between $1.596$ and $e \approx 2.718$. 
Using the mean values from Tables~\ref{t:ppp_weights_trials_deg22_n1000}, \ref{t:ppp_weights_trials_deg22_n5000}, and \ref{t:ppp_weights_trials_deg22_n100000} gives us
\begin{subequations}
	\begin{align}
		\Lunit(\Pcallp)/\Nbox^{1/D} = 59.35/\sqrt{1000}    \approx 1.897 \, , &  \quad (D=2, \Ntilde = 1000) \\
		\Lunit(\Pcallp)/\Nbox^{1/D} = 134.7/\sqrt{5000}    \approx 1.905 \, , &  \quad (D=2, \Ntilde = 5000) \\
		\Lunit(\Pcallp)/\Nbox^{1/D} = 622.47/\sqrt{100000} \approx 1.968 \, , &  \quad (D=2, \Ntilde = 10^5)		
	\end{align}
\end{subequations}
consistent with the asymptotic bounds of \citet{BB91}.

In \tabref{t:ppp_weights_trials_deg22_n1000} and \tabref{t:ppp_weights_trials_deg22_n5000} we show some additional results 
for the path weights  $\LG(\Pcal)$ of \eqref{e:LGdef} for various paths across $100$ independent network realisations in two-dimensions.
Each trial uses a random geometric DAG with the same parameters as shown in \figref{f:ppp-dag-deg22-n1000}, but with different random node placements. 
Using length, measured in terms of number of edges (unit weight), SPC or SPE weights, the longest paths by all three measures perform very similarly.

\begin{table}[htb!]
	\centering
	\begin{tabular}{l|| c | c | c}
		Path & \multicolumn{3}{c}{Weight (mean $\pm$ s.e.m.)} \\ 
		& Unit & SPC & SPE \\ \hline
		Shortest           &  9.370 $\pm$ 0.048 &  $(7.6599 \pm 3.3169) \times 10^{23}$ &  458.970 $\pm$ 2.552 \\ 
		Longest (Unit)     & 59.350 $\pm$ 0.275 &  $(3437.3 \pm 1249.4) \times 10^{23}$ & 3252.471 $\pm$ 25.330 \\
		Longest (SPC)      & 58.490 $\pm$ 0.302 &  $(3964.9 \pm 1472.9) \times 10^{23}$ & 3225.334 $\pm$ 26.094 \\
		Longest (SPE)      & 59.310 $\pm$ 0.277 &  $(3754.5 \pm 1334.7) \times 10^{23}$ & 3262.599 $\pm$ 25.034 \\
		Greedy (GPD)       & 14.420 $\pm$ 0.138 &  $(194.18 \pm 111.49) \times 10^{23}$ &  738.384 $\pm$ 8.252 \\
		Greedy (Degree)    & 24.515 $\pm$ 0.237 &  $(952.09 \pm 306.23) \times 10^{23}$ & 1299.132 $\pm$ 15.540 \\
		Random             & 15.113 $\pm$ 0.169 & $(220.178 \pm 60.016) \times 10^{23}$ &  769.615 $\pm$ 10.151 \\ 
		\end{tabular}
	\caption{Path weights across 100 independent trials in a 2D random geometric DAG with $\Nbox=1000$ nodes ($N=1002$ including source and sink) and $\ktarget=22$ (forward degree). Entries show mean $\pm$ standard error of the mean. Unit is edge count; SPC is sum of exact integer path counts along each path (shown in scientific notation); SPE is sum of ln(SPC) values.}
	\label{t:ppp_weights_trials_deg22_n1000}
\end{table}

\begin{table}[htb!]
	\centering
	\begin{tabular}{l|| c | c | c}
		Path & \multicolumn{3}{c}{Weight (mean $\pm$ s.e.m.)} \\ 
		& Unit & SPC & SPE \\ \hline
		Shortest           &  20.500 $\pm$ 0.112  &  $(18.7519 \pm 13.9157) \times 10^{56}$ &  2533.827 $\pm$ 17.622 \\ 
		Longest (Unit)     & 134.700 $\pm$ 0.765  &  $(4072.63 \pm 1779.50) \times 10^{56}$ & 17489.105 $\pm$ 165.329 \\
		Longest (SPC)      & 132.900 $\pm$ 0.883  &  $(5121.58 \pm 2343.06) \times 10^{56}$ & 17342.884 $\pm$ 176.227 \\
		Longest (SPE)      & 134.600 $\pm$ 0.782  &  $(4950.87 \pm 2312.98) \times 10^{56}$ & 17535.896 $\pm$ 166.656 \\
		Greedy (GPD)       &  32.250 $\pm$ 0.429  &  $(94.7763 \pm 39.2278) \times 10^{56}$ &  4056.626 $\pm$ 59.939 \\
		Greedy (Degree)    &  46.400 $\pm$ 0.782  &  $(671.422 \pm 293.460) \times 10^{56}$ &  5888.251 $\pm$ 112.952 \\
		Random             &  31.846 $\pm$ 0.472  & $(140.6395 \pm 75.0225) \times 10^{56}$ &  3962.665 $\pm$ 56.660 \\ 
		\end{tabular}
	\caption{Path weights across 100 independent trials in a 2D random geometric DAG with $\Nbox=5000$ nodes ($N=5002$ including source and sink) and $\ktarget=22$ (forward degree). Entries show mean $\pm$ standard error of the mean. Unit is edge count; SPC is sum of exact integer path counts along each path (shown in scientific notation); SPE is sum of ln(SPC) values.}
	\label{t:ppp_weights_trials_deg22_n5000}
\end{table}

Finally, working with exact integer arithmetic in our code, we give one illustration of the SPC weight of the SPC main path for one network with $N= 250,002$ nodes and target out-degree of $\ktarget= 22$ in \figref{af:spc250k}.
\begin{figure}[htb] 
	\centering
	\parbox{0.9\textwidth}{
		\texttt{\raggedright
			13532564859792006573036893538406665218616100620960906150911617214167002\\
			85450251094217907030018149473012717009362718875244583364741802788561075\\
			68574198300007838002870002902036912572605022988849130781610542648405907\\
			90393652809626666803192813040228500122465356385712758503172348424414781\\
			32832935378387105101770122447755832307958837020922601788879196945777221\\
			2102470267975038850675850237058202804965395006502086143039782052869}
	}
	\caption{The SPC weight of the SPC path for one network with $N= 250,002$ nodes and target out-degree of $\ktarget= 22.0$.
		This has 422 digits and illustrates the extreme scales encountered in SPC computation.
		See \secref{s:numerics} for implementation details on arbitrary-precision arithmetic requirements.}
	\label{af:spc250k}
\end{figure}

\clearpage

\subsubsection{Perpendicular Distance to Geodesic $\Delta$}
\label{as:ppp_perpendicular_distance}

\tsecomment{Nothing here yet.}

In \secref{ss:ppp-perpendicular-distance} we compared the performance of our three types of longest path, based on Unit, SPC and SPE weights, in terms of the normalised mean perpendicular distance $\Delta/a$ across different network sizes $N$ for fixed density with $\ktarget=22.0$.  We also used several other paths for comparison. Here we give additional information on similar simulations.

\begin{figure}[htb!]
	\centering
	\includegraphics[width=0.95\textwidth]{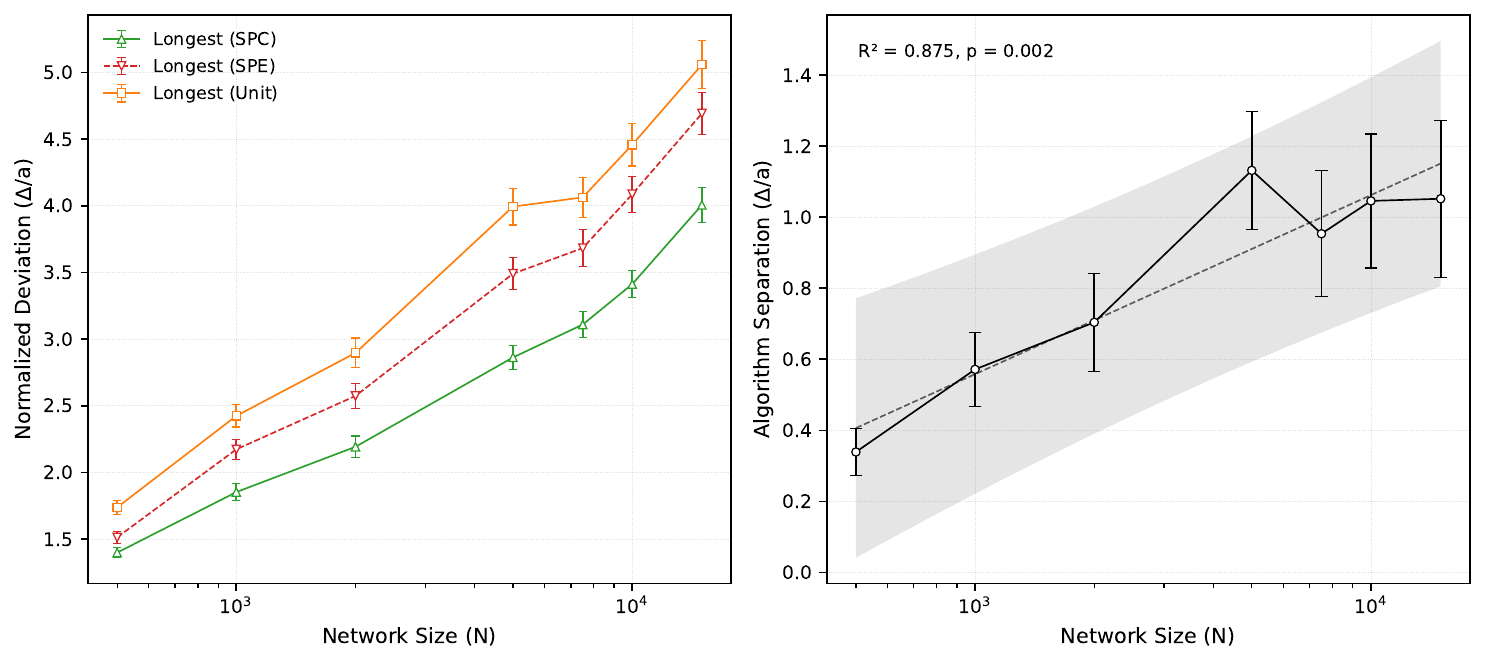}
	\caption{Geodesic proximity scaling for longest-path algorithms. (Left) Normalised deviation $\Delta/a$ vs network size $N$: algorithms converge at small $N$ (500) and diverge at large $N$ (15000). (Right) Algorithm separation scales logarithmically with $N$ (slope = $0.219 \pm 0.037$, $R^2 = 0.875$, $p = 0.002$, two-tailed $t$-test). Error bars show $\pm$1 SEM. Data from 300 trials ($N=500$) and 200 trials ($N \geq 1000$) with $\ktarget=22.0$.}
	\label{fig:geodesic_scaling}
\end{figure}

At a small number of points ($N =502$), finite-size effects constrain all longest paths similarly; differences between SPC and SPE are small and not significant after Bonferroni correction (Table~\ref{t:geodesic_statistical_tests}). As $N$ increases, the differences in the perpendicular distances to the geodesic of the different paths becomes statistically significant: the SPC  main path is closest to the geodesic (lowest $\Delta/a$), the longest unit path drifts furthest away (highest $\Delta/a$), and the SPE main path is intermediate. Effect sizes grow from negligible at $N=502$ to medium (Cohen's $d=0.68$ for SPC vs Unit at $N=5002$), and separation scales logarithmically with $N$ (\tabref{t:geodesic_separation_scaling}: $R^2=0.875$, $p=0.002$). This convergence-to-divergence pattern (\figref{fig:geodesic_scaling}) suggests algorithm choice becomes more consequential for larger networks, as increased routing options allow paths to optimise for other criteria at the expense of geodesic proximity.

\begin{table}[htb!]
	\centering
	\scriptsize
	\begin{tabular}{l||c|c|c|c|c|| c | c | c | c | c}
		Pair  & \multicolumn{5}{c||}{$p$-value ($\Ntilde$)} & \multicolumn{5}{c}{Cohen's $d$ ($\Ntilde$)} \\
		& 500 & 1K & 5K & 10K & 15K & 500 & 1K & 5K & 10K & 15K \\ \midrule
		SPC vs SPE & $>0.05$ & 0.003 & $<0.001$ & $<0.001$ & 0.002 & 0.16 & 0.33 & 0.42 & 0.40 & 0.34 \\
		& (n.s.) & $\star\star$ & $\star\star\star$ & $\star\star\star$ & $\star\star$ & (neg.) & (sm) & (sm) & (sm) & (sm) \\ \midrule
		SPC vs Unit & $<0.001$ & $<0.001$ & $<0.001$ & $<0.001$ & $<0.001$ & 0.42 & 0.55 & 0.68 & 0.55 & 0.48 \\
		& $\star\star\star$ & $\star\star\star$ & $\star\star\star$ & $\star\star\star$ & $\star\star\star$ & (sm) & (med) & (med) & (med) & (sm) \\ \midrule
		SPE vs Unit & 0.003 & $>0.05$ & 0.018 & $>0.05$ & $>0.05$ & 0.27 & 0.22 & 0.28 & 0.18 & 0.15 \\
		& $\star\star$ & (n.s.) & $\star$ & (n.s.) & (n.s.) & (sm) & (sm) & (sm) & (neg.) & (neg.) \\ 
	\end{tabular}
	\caption{Statistical analysis of geodesic deviation differences between longest-path algorithms across network sizes $\Ntilde=N-2$. Welch's two-sample $t$-tests (two-tailed, with Bonferroni correction for 3 comparisons per $N$) were performed; results were robust to non-normality in spot checks. At $N=502$, SPC vs SPE is not significant ($p>0.05$, Cohen's $d=0.16$), while SPC vs Unit and SPE vs Unit are significant. At larger N (1000+), SPC and SPE become significantly different; SPE vs Unit becomes non-significant at very large $N$ ($10002$ and $15002$). Significance: $\star$ $p<0.05$, $\star\star$ $p<0.01$, $\star\star\star$ $p<0.001$ (Bonferroni-corrected). Effect sizes: neg.=negligible ($|d|<0.2$), sm=small ($0.2\leq|d|<0.5$), med=medium ($0.5\leq|d|<0.8$). Trial counts: $300$ ($N=502$), $200$ ($N \geq 1000$).}
	\label{t:geodesic_statistical_tests}
\end{table}

\begin{table}[htb!]
	\centering
	\footnotesize
	\begin{tabular}{r|cc}
		N & Separation & $\Delta_{\max}/\Delta_{\min}$ \\
		\hline
		  502 & 0.339 & 1.24 \\
		 1002 & 0.572 & 1.31 \\
		 2002 & 0.704 & 1.32 \\
		 5002 & 1.132 & 1.40 \\
		 7502 & 0.953 & 1.31 \\
		10002 & 1.046 & 1.31 \\
		15002 & 1.052 & 1.26 \\
	\end{tabular}
	\caption{Scaling of algorithm separation with network size. The separation between longest-path algorithms increases logarithmically with $N$ ($R^2=0.875$, $p=0.002$), indicating that algorithm choice becomes increasingly consequential for larger networks. Separation is defined as maximum minus minimum normalised deviation $\Delta/a$ among SPC, SPE, and Unit algorithms. Linear regression: slope $0.219 \pm 0.037$, $R^2=0.875$, $p=0.002$ (two-tailed $t$-test, significant logarithmic scaling). The ratio $\Delta_{\max}/\Delta_{\min}$ increases from $1.24$ ($N=502$) to $1.40$ ($N=5002$), then stabilises around $1.26--1.31$ at very large $N$, suggesting saturation of divergence effects.}
	\label{t:geodesic_separation_scaling}
\end{table}

\subsubsection{Degree of nodes on paths}

We have seen how the longest unit path, SPC and SPE paths share many similar properties. Equally, we have seen that these paths are not identical. It would be interesting to know if these different paths favour different types of node in our networks. So we next ask if any of these paths favour high density regions in our random geometric DAG model. The simplest way to study this question is to look at the degree of the nodes on each path since a high degree means the path is passing through a region of high density. Since the model is based on Poisson spatial statistics, these spatial fluctuations in node, and hence edge, density are not large but they are noticeable as can be seen in any visualisation such as  \figref{f:ppp-dag-deg22-n1000} in \secref{s:pppnumerics}. Results for the typical node out-degree for 100 two-dimensional networks are shown in \tabref{at:path_degrees_n10000}.

\begin{table}[htb!]
	\centering
	\begin{tabular}{l|| c | c | c | c}
		Path		      & Unit Length       & \multicolumn{3}{c}{Out Degree} \\ 
		&                   & Mean             & Median           & Std.Dev. \\ \hline
		Shortest          & $28.96 \pm 0.20$  & $22.02 \pm 0.09$ & $21.99 \pm 0.12$ & $4.97 \pm 0.06$ \\
		Longest (Unit)    & $193.19 \pm 0.45$ & $23.11 \pm 0.07$ & $23.39 \pm 0.08$ & $5.63 \pm 0.04$ \\
		Longest (SPC)     & $189.71 \pm 0.55$ & $23.49 \pm 0.06$ & $23.83 \pm 0.07$ & $5.65 \pm 0.04$ \\
		Longest (SPE)     & $193.17 \pm 0.45$ & $23.31 \pm 0.06$ & $23.63 \pm 0.08$ & $5.64 \pm 0.04$ \\
		Greedy (GPD)      & $45.85 \pm 0.31$  & $21.08 \pm 0.08$ & $20.96 \pm 0.11$ & $5.05 \pm 0.06$ \\
		Greedy (Degree)   & $63.25 \pm 0.36$  & $25.05 \pm 0.11$ & $26.80 \pm 0.10$ & $7.84 \pm 0.07$ \\
		Random            & $43.67 \pm 0.31$  & $20.44 \pm 0.08$ & $21.08 \pm 0.09$ & $6.41 \pm 0.08$ \\
	\end{tabular}
	\caption{Node out-degree statistics along each path type in 100 realisations of the two-dimensional random geometric DAG model
		with $N=10002$ nodes and target out-degree $\ktarget=22.0$. 
		The means values are quoted with the standard error of the mean and are based on the values for every node on the specified path except for the initial and final node, the source and sink of the DAG.}
	\label{at:path_degrees_n10000}
\end{table}

Poisson statistics means that we expect most box points (those not close to a boundary) to have an out-degree in a Poisson distribution with mean equal to $\ktarget$, here $22.0$, so with standard deviation $\sqrt{22} \approx 4.7$.

Pairwise $t$-tests with Bonferroni correction ($\alpha' \approx 0.00238$) reveal significant differences in mean out-degrees between path types. The greedy degree path, with a mean out-degree of $25.05 \pm 0.11$, significantly exceeds all others ($p < 10^{-6}$), reflecting its systematic bias toward high-degree nodes; the effect size depends on the comparator (Cohen's $d \approx 0.23$–0.64 across path types). This is hardly surprising as it was designed to be biassed towards high degree nodes. The standard deviation also unrelated to the underlying Poisson statistics as one would expect.  

We might have thought that the degree of nodes on a random path should be largely sampled from Poisson statistics. However, we see that the mean is low while the standard deviation is only the second lowest in \tabref{at:path_degrees_n10000} and is not statistically equivalent to the expected value for a random interior point. What is happening is that the last few nodes at the end of a random path hit the boundary before the path is pushed towards the sink node. So these last few nodes will have a lower degree. Given that, and the number of nodes on a random path, these results seem fairly consistent and give a benchmark for the other paths.

The shortest path is the one whose degree statistics are closest to the Poisson statistics we expect. The mean degree is statistically indistinguishable from $22.0$ and it also has the standard deviation closest to the expect value, if still statistically a little high. So this suggests there is nothing `optimal' about this path, it is essentially passing through nodes that are almost indistinguishable from a set of nodes chosen at random. 
One explanation for this is as follows. The results suggest there is no strong bias to low or high degree nodes for these shortest unit paths. The low number of edges in the shortest path, the long Euclidean length of each edge, means the path has no reason to be close to the boundary except perhaps for the nodes in the shortest path which are next to the source and target. So boundary effects are much smaller than for the random path. 

The three longest paths (Unit, SPC and SPE) all have very similar degree statistics.
The degree statistics for these longest paths are significantly different from the shortest, greedy perpendicular (GPD) and random paths ($p < 10^{-6}$) and so higher than the degree of random interior nodes. This does then suggest that these longest paths are also finding the higher degree nodes, nodes with higher centrality, even though high node degree was not an explicit part of their definition. Presumably a high degree node represents a high density region with lower inter-node distances so longest paths favour such regions.    

Within the longest group, the degree of the SPC main path exceeds longest unit path ($p = 3.8 \times 10^{-5}$, $d \approx 0.07$) but does not differ significantly from SPE ($p = 0.034 > \alpha'$). The SPC also produces slightly shorter paths by unit length ($p \approx 10^{-6}$), a reduction of $\sim 1.8\%$ in unit length. 

Taken together, these results show that SPC consistently yields paths with slightly fewer edges and slightly higher degree than the other longest-path constructions, with the SPC–SPE difference emerging only at larger network sizes. Effect sizes remain small, but statistically robust, indicating that SPC's degree sensitivity is a subtle yet persistent feature of random geometric DAGs. 

However, the broader message is that the longest paths (Unit, SPC, SPE) all have nearly identical means and spreads in terms of degree, reflecting their tendency to traverse the same interior regions close to the diagonal geodesic.

\clearpage
\subsubsection{Baskets Not Paths}
\label{as:basketnotpath}

\begin{figure}[htb!]
	\centering
	\includegraphics[width=0.8\linewidth]{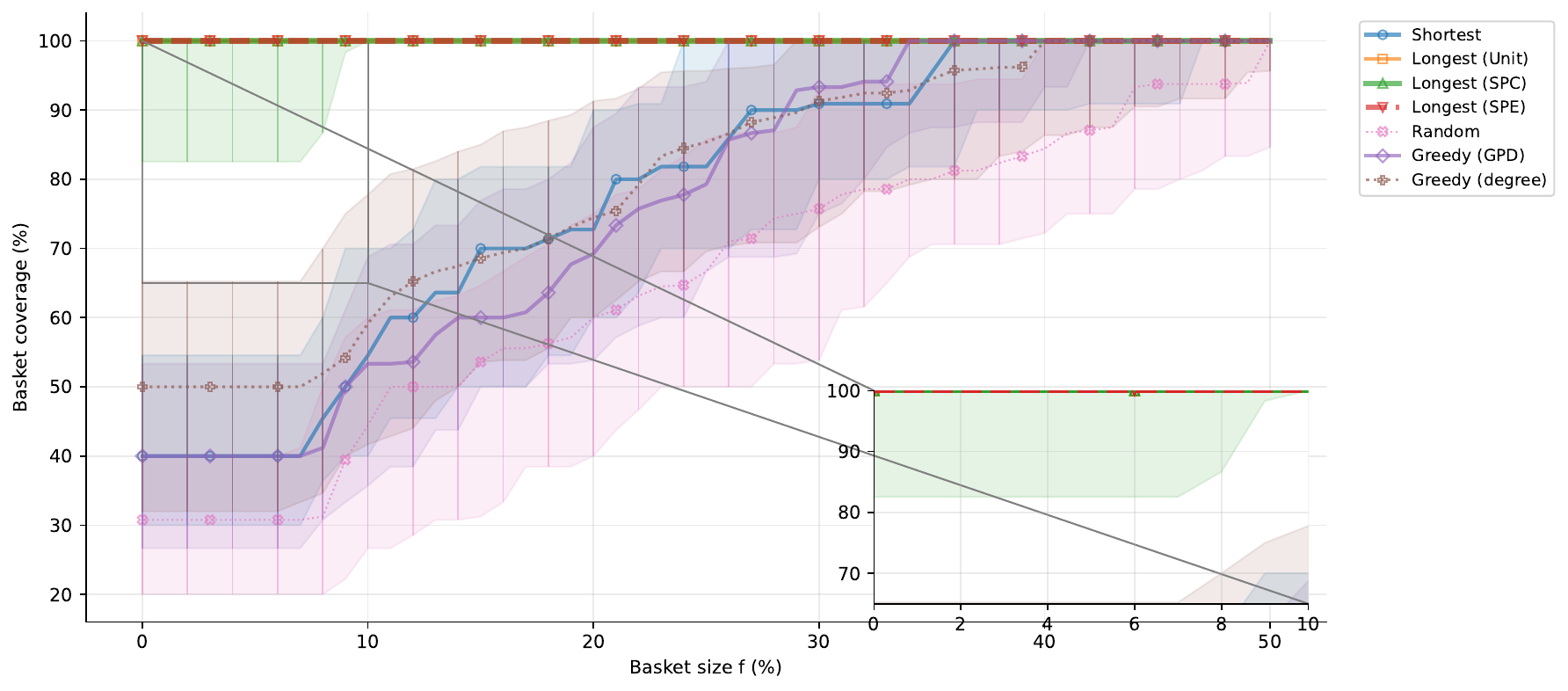}
	\caption{Plot to show how the coverage by the basket defined using unit weights changes as the fraction of nodes included in the basket, $f$ is changed. This is done for various paths; the longest by unit, SPC and SPE weight, the shortest path by unit weight, two greedy paths, one using degree the other using perpendicular distance, and finally a random path. This is for 100 instances of random geometric DAGs with $N=1002$ nodes and target out-degree $\ktarget=22.0$. For each path $\Pcal$ a line connects the median values of the coverage $\phi(\Pcal,\Bcal^\mathrm{(unit)})$ with the error bars and shaded regions representing first and their quartile ranges, calculated over 100 networks. For the basket defined using unit weight criticality, the plost shows that the coverage of the three types of longest paths is excellent.}
	\label{f:coverage-overlay-unit}
\end{figure}

In \figref{f:coverage-overlay-unit} we show how changing the fraction of nodes included in a basket changes the coverage of various paths, including the longest unit path, and the PSC and SPE main paths. The main result is to confirm that the coverage by the basket defined using unit weight criticality of the three types of longest paths is excellent.\tsecomment{WHY IS THE GPD COVERAGE SO POOR?}

\tsecomment{I am not sure \figref{f:coverage-overlay-unit} tells us anything new.  It looks good but actually most of the information is obscured by other features in the plot so I'm not sure it is a good scientific plot.}

\tsecomment{TO BE EDITED.}

\begin{figure}[htb]
	\centering
	\begin{tabular}{ccc}
		\includegraphics[width=0.3\textwidth]{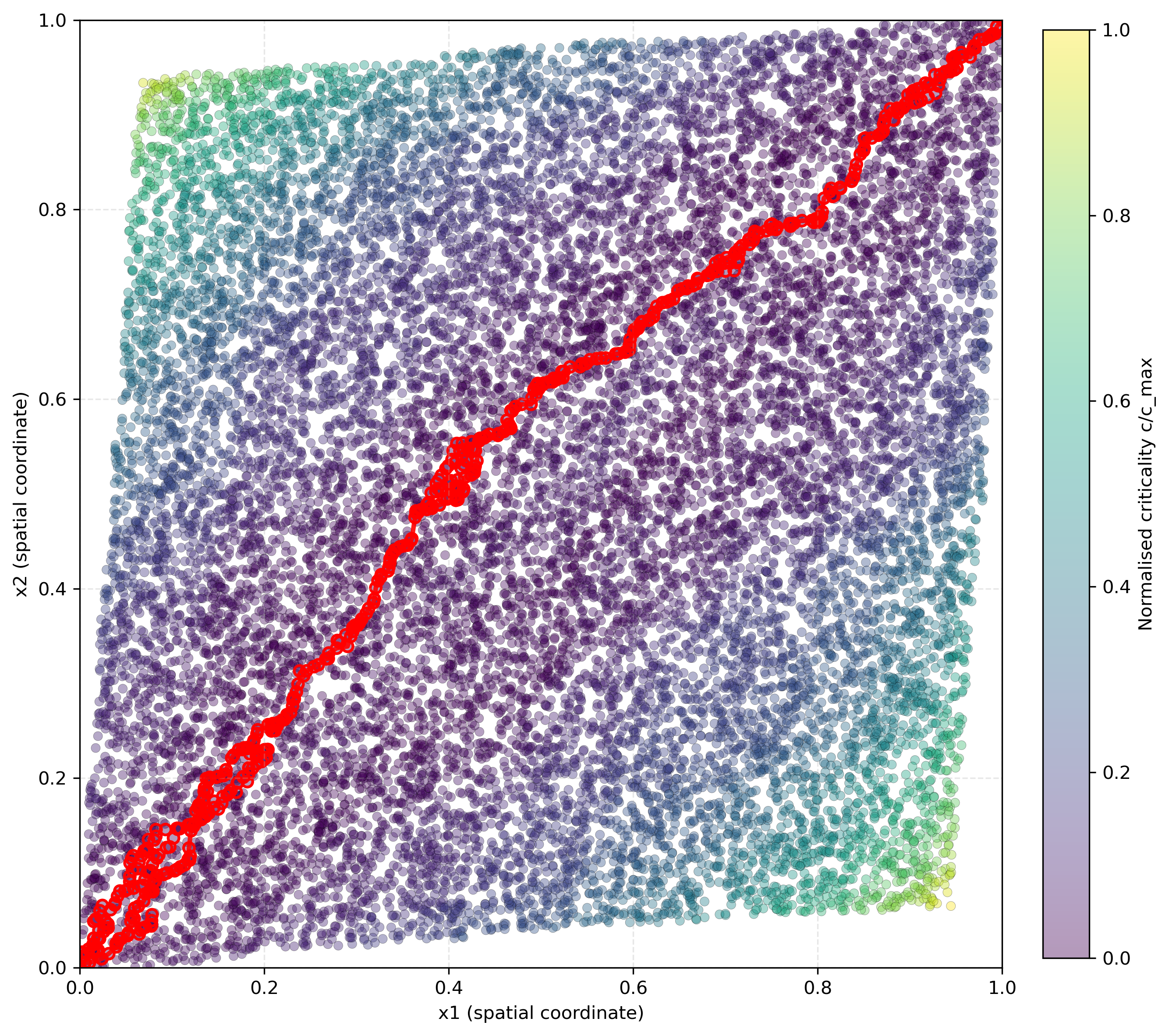}
		&    
		\includegraphics[width=0.3\textwidth]{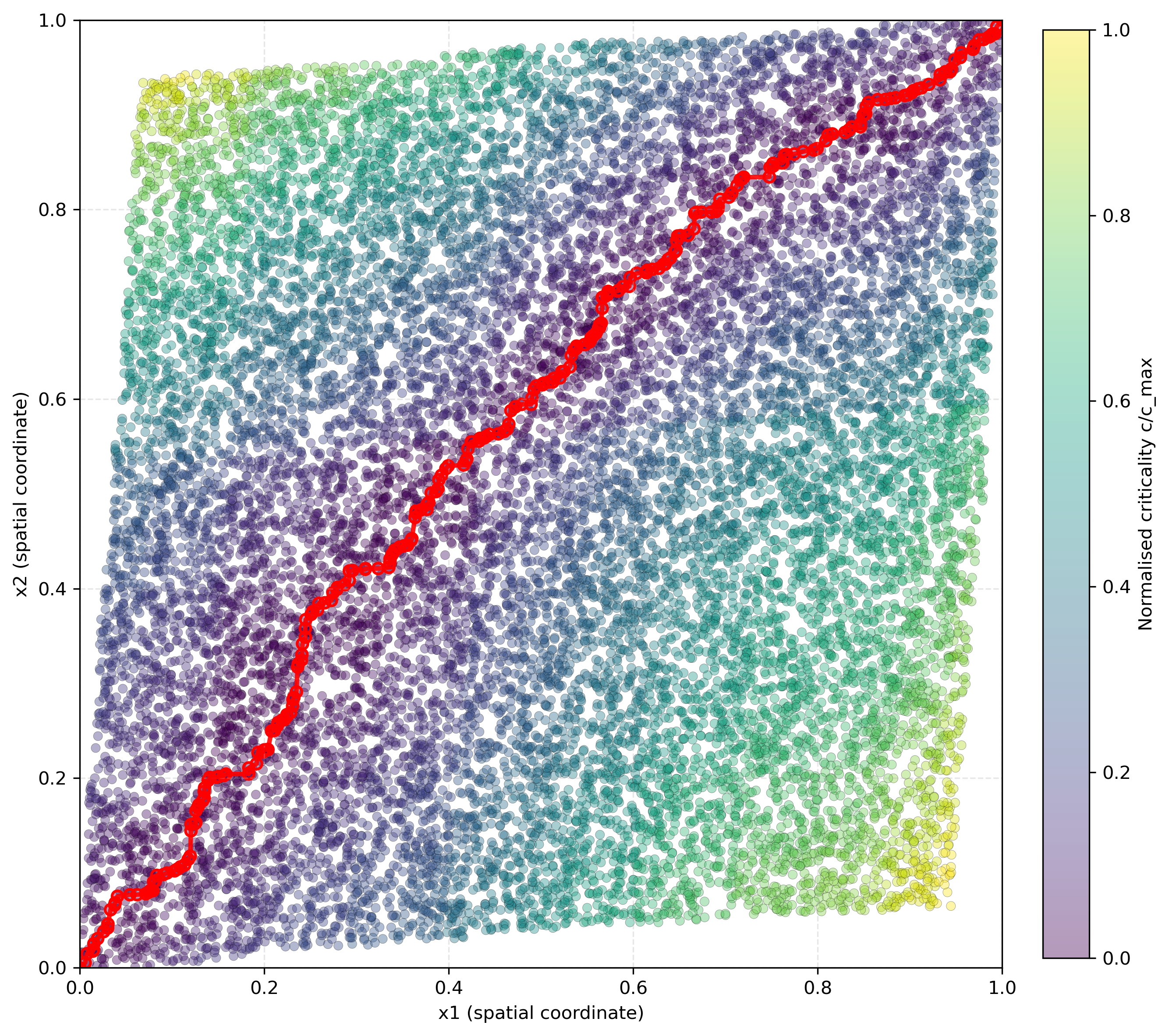}
		&
		\includegraphics[width=0.3\textwidth]{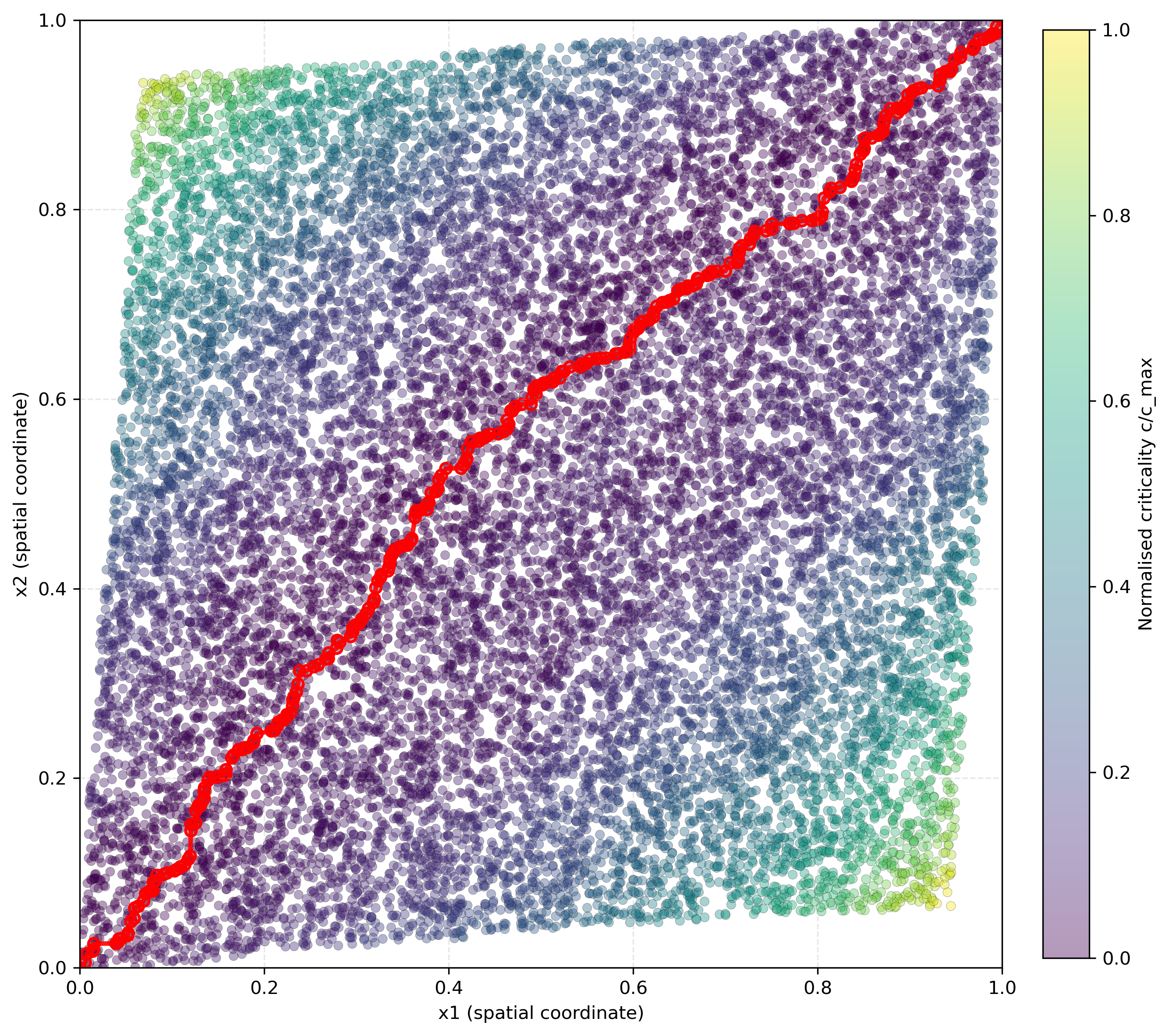}
		\\
		(a) Unit
		&
		(b) SPC
		&
		(c) SPE
	\end{tabular}
	\caption{Node criticality $\cG$ of \eqref{e:cGdef} on one representative random geometric DAG with $N=20002$ nodes and target out-degree $\ktarget=22.0$. Each node is coloured by its normalised criticality value, with critical nodes (c=0) highlighted by red outlines and longest paths traced in red. The network demonstrates good spatial coverage and clear criticality structures. We show results for the same network but for relative criticality $\cG/\max(\cG)$ using different path weights $G$: 
		in (a) we have unit-weight criticality, in (b) SPC-weight criticality, and in (c) SPE-weight criticality.
		  	Nodes with critical values are shown in red. In addition, we also show one example of a path that has the largest weight for that plot by adding in the edges for that path, also in red.
		 Note the SPC and SPE longest paths are not the same in the top right regions close to the sink node at $t=(1.0,1.0)$.
		\\
		\tsecomment{VERY NICE PLOTS! Shame to hide them in the appendix but they say less than the same plot for a smaller DAG.}
		}
	\label{af:ppp-criticality-heatmaps}
\end{figure}

We give a criticality heatmap in \figref{af:ppp-criticality-heatmaps} for a one random geometric DAG with $N=20002$ nodes and target out-degree $\ktarget=22.0$ which is a considerably larger network than the heatmap shown in \figref{f:ppp-criticality-heatmaps-n1000}. The increased density hides the similarities and differences between different paths but now produces a visualisation more like the relative criticality heatmap for the lattice DAG in \figref{f:latcrit-78-71}.  

\clearpage

\begin{figure}[htb]
	\centering
	\begin{tabular}{ccc}
		\includegraphics[width=0.3\textwidth]{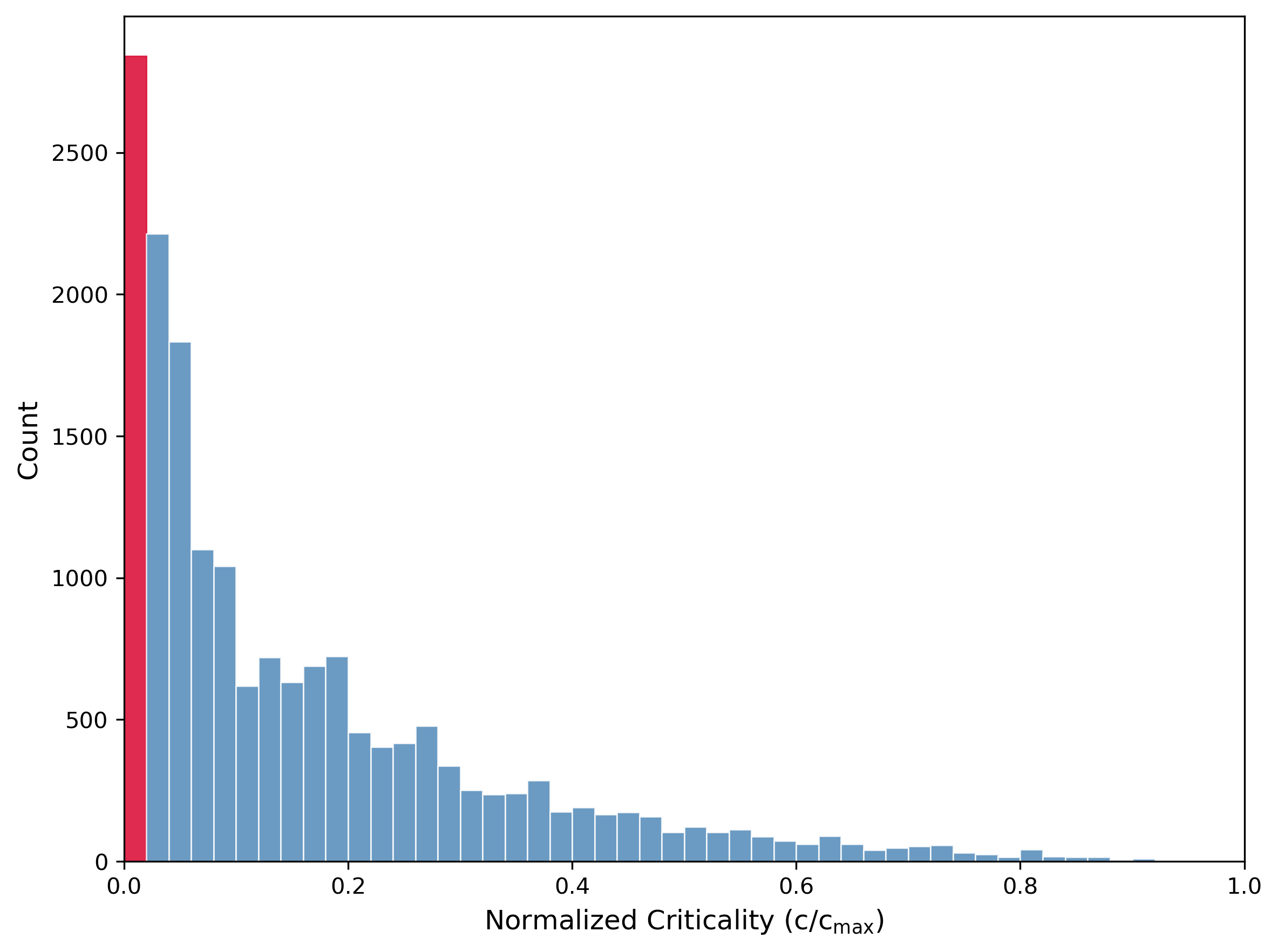}
		&
		\includegraphics[width=0.3\textwidth]{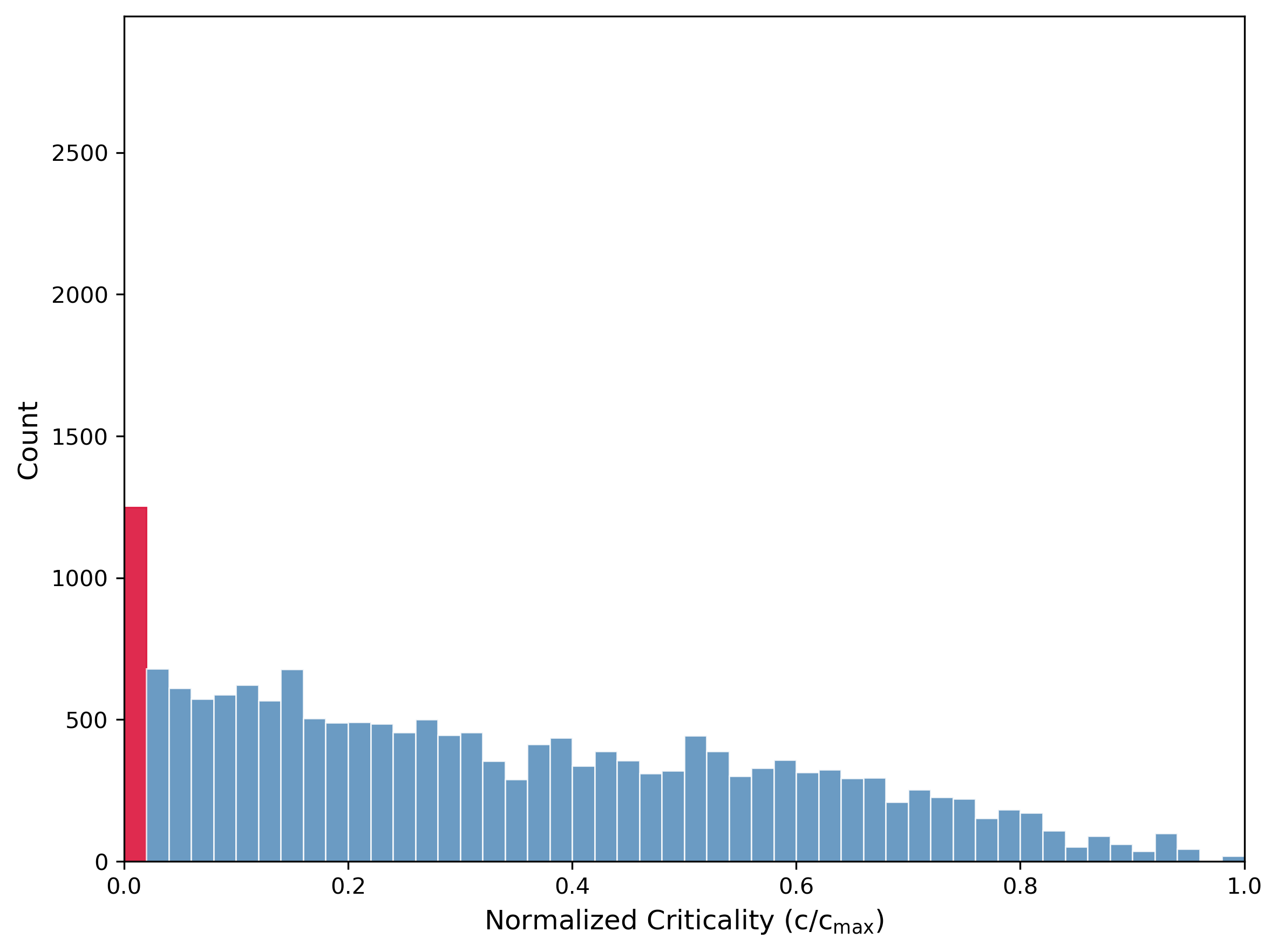}
		&
		\includegraphics[width=0.3\textwidth]{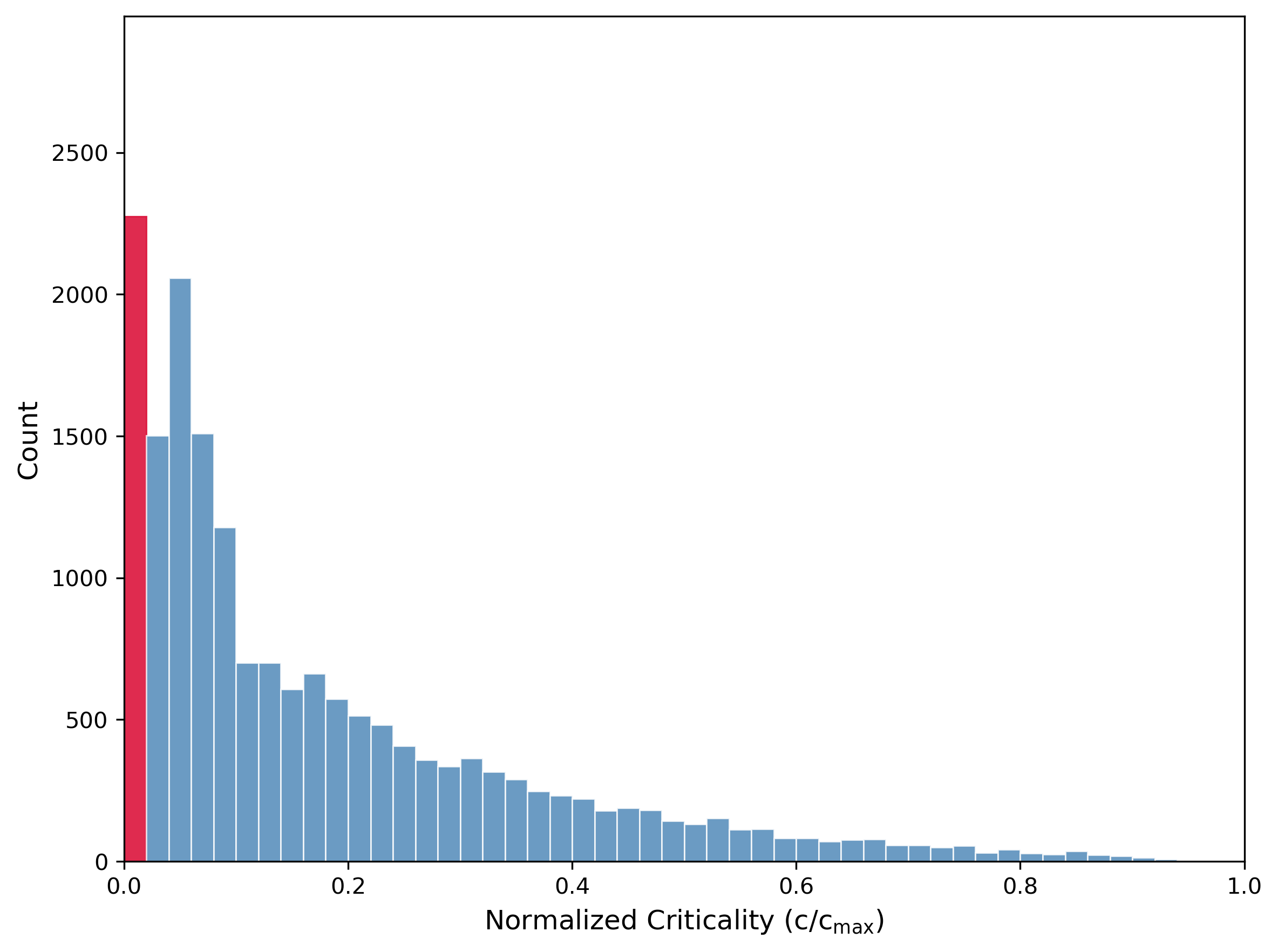}
		\\
		(a) Unit
		&
		(b) SPC
		&
		(c) SPE
	\end{tabular}
	\caption{The distribution of nodes by normalised criticality values for a random geometric DAG with $N=20002$ nodes and target out-degree $\ktarget=22.0$. For the unit path weight we have: median $\approx 0.093$, mean $\approx 0.162$, $2.92\%$ critical nodes ($\cunit=0$). For the SPC weight we have: median $\approx 0.293$ (mean $\approx 0.336$), $1.54\%$ critical nodes ($\cspc=0$). For the SPEweight we have: median $\approx 0.109$ (mean $\approx 0.179$), $1.63\%$ critical nodes ($\cspe=0$).}
	\label{f:ppp-criticality-histograms}
\end{figure}

In \figref{f:ppp-criticality-histograms}, we show histograms of the normalised criticality values under the three weighting schemes (unit, SPC, SPE) for one random geometric DAG.

In \tabref{at:ppp_criticality_coverage_trials_b} we report the coverage (\%) of four paths (shortest unit path, greedy degree, greedy perpendicular distance and a random path) and we compare that with the baskets defined using by unit, SPC and SPE weights. We show this for the basket of critical nodes, nodes where $\cG=0$ as well as for the 1\%, top 5\%and top 10\% of nodes, with statistics gathered from 100 trials of random geometric DAGs created with $N=1002$ nodes and a target out-degree of $\ktarget=22.0$.

\begin{table}[htb!]
	\centering
	\footnotesize
	\begin{tabular}{lc||cccc}
		\multicolumn{2}{c||}{Basket $\BG(f)$ } & \multicolumn{4}{c}{Path $\Pcal$ } \\
		$G$  & $f$    & Shortest & Greedy (Degree) &  Greedy (GPD) & Random \\ \hline
		Unit & $0$ & 40.0\% (30.0\%--50.0\%) & 50.9\% (32.0\%--66.7\%) & 39.4\% (25.0\%--53.3\%) & 31.2\% (21.1\%--42.9\%) \\
		Unit & $1\%$ & 40.0\% (30.0\%--50.0\%) & 50.9\% (32.0\%--66.7\%) & 39.4\% (25.0\%--53.3\%)& 31.2\% (21.1\%--42.9\%) \\ 
		Unit & $5\%$ & 40.0\% (30.0\%--50.0\%) & 50.9\% (32.0\%--66.7\%) & 39.4\% (25.0\%--53.3\%)& 31.2\% (21.1\%--42.9\%) \\ 
		Unit & $10\%$ & 50.0\% (36.4\%--63.6\%) & 61.1\% (41.7\%--78.3\%) & 51.3\% (37.5\%--66.7\%) & 42.1\% (28.6\%--57.1\%) \\ 
		\hline
		SPC & $0$ & 30.0\% (20.0\%--40.0\%) & 38.0\% (22.7\%--52.4\%) & 28.2\% (20.0\%--35.7\%) & 22.9\% (12.5\%--31.2\%) \\
		SPC & $1\%$ & 30.0\% (20.0\%--40.0\%) & 38.0\% (22.7\%--52.4\%) & 28.2\% (20.0\%--35.7\%) & 22.9\% (12.5\%--31.2\%) \\ 
		SPC & $5\%$ & 30.0\% (20.0\%--40.0\%) & 38.0\% (22.7\%--52.4\%) & 28.2\% (20.0\%--35.7\%) & 22.9\% (12.5\%--31.2\%) \\ 
		SPC & $10\%$ & 45.5\% (30.0\%--50.0\%) & 53.8\% (31.8\%--66.7\%) & 43.8\% (29.4\%--56.2\%)& 31.0\% (20.0\%--43.8\%) \\ 
		\hline
		SPE & $0$ & 30.0\% (27.3\%--40.0\%) & 35.3\% (23.8\%--52.0\%) & 26.7\% (18.8\%--35.7\%) & 23.5\% (12.5\%--31.2\%) \\
		SPE & $1\%$ & 30.0\% (27.3\%--40.0\%) & 35.3\% (23.8\%--52.0\%) & 26.7\% (18.8\%--35.7\%)& 23.5\% (12.5\%--31.2\%) \\ 
		SPE & $5\%$ & 30.0\% (27.3\%--40.0\%) & 35.3\% (23.8\%--52.0\%) & 26.7\% (18.8\%--35.7\%) & 23.5\% (12.5\%--31.2\%) \\ 
		SPE & $10\%$ & 40.0\% (30.0\%--54.5\%) & 54.2\% (32.1\%--68.0\%) & 40.0\% (28.6\%--53.3\%)& 31.2\% (21.4\%--43.8\%) \\ 
	\end{tabular}
	\caption{Basket coverage $\phi(\Pcal,\Bcal)$ giving the median with quartile values over 100 trials of a random geometric DAGs with $N=1002$ nodes and target out-degree $\ktarget=22.0$. 
		On each row specifies one basket $\BG(f)$ for a given weight $G\in\{\text{Unit, SPC, SPE}\}$ and a given fraction $f$. The value $f=0$ indicates that the basket contains all critical nodes, i.e.\ with $\cG=0$. The last four columns on the right specify the path $\Pcal$ to be the shortest unit path, greedy degree, greedy perpendicular distance and a random path. The entries in the table are then the coverage $\phi(\Pcal,\Bcal)$, the percentage of path nodes lying in the given basket.}
	\label{at:ppp_criticality_coverage_trials_b}
\end{table}

\begin{figure}[htb!]
	\centering
	\includegraphics[width=0.8\linewidth]{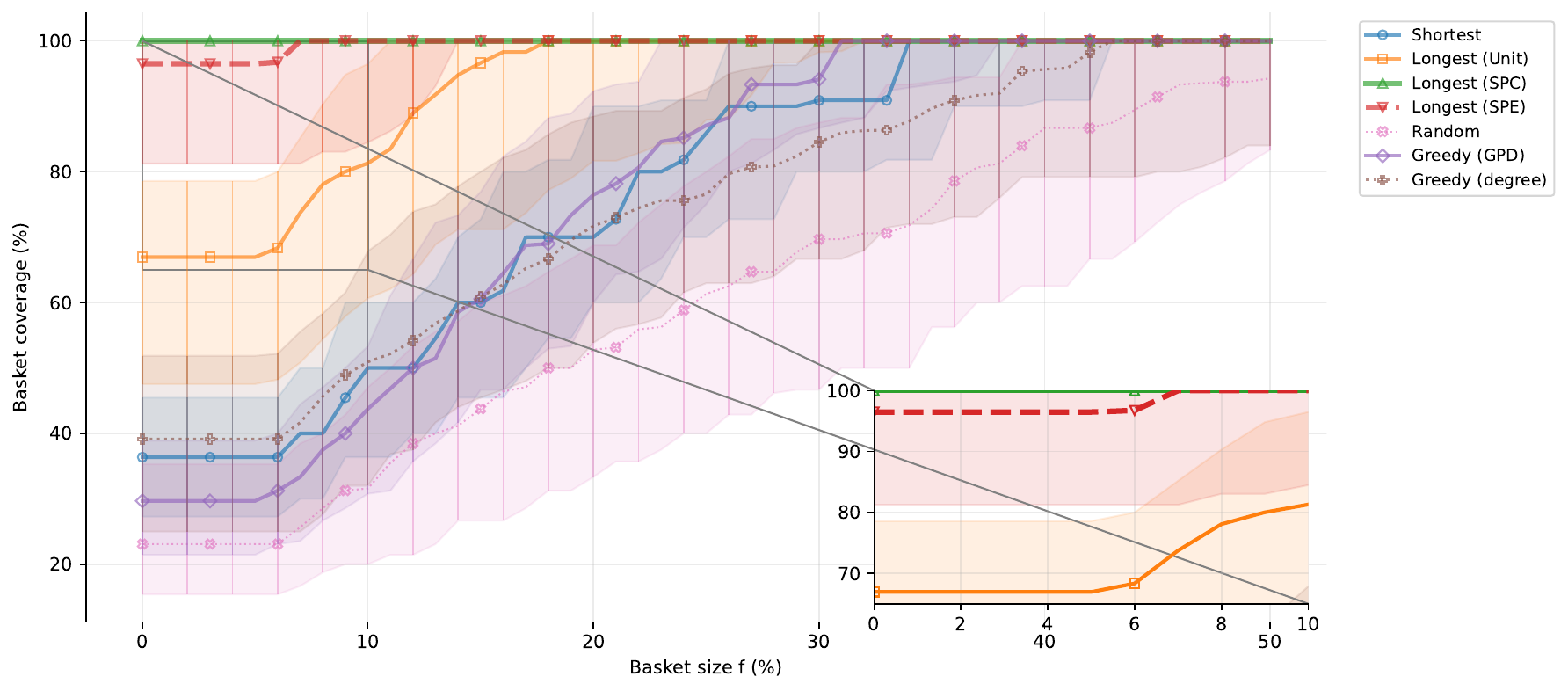}
	\caption{Plot to show how the coverage by the basket defined using SPC weights changes as the fraction of nodes included in the basket, $f$ is changed. This is done for various paths; the longest by unit, SPC and SPE weight, the shortest path by unit weight, two greedy paths, one using degree the other using perpendicular distance, and finally a random path. This is for 100 instances of random geometric DAGs with $N=1002$ nodes and target out-degree $\ktarget=22.0$. For each path $\Pcal$ a line connects the median values of the coverage $\phi(\Pcal,\Bcal^\mathrm{(unit)})$ with the error bars and shaded regions representing first and their quartile ranges, calculated over 100 networks. }
	\label{af:coverage-overlay-spc}
\end{figure}

\begin{figure}[htb!]
	\centering
	\includegraphics[width=0.8\linewidth]{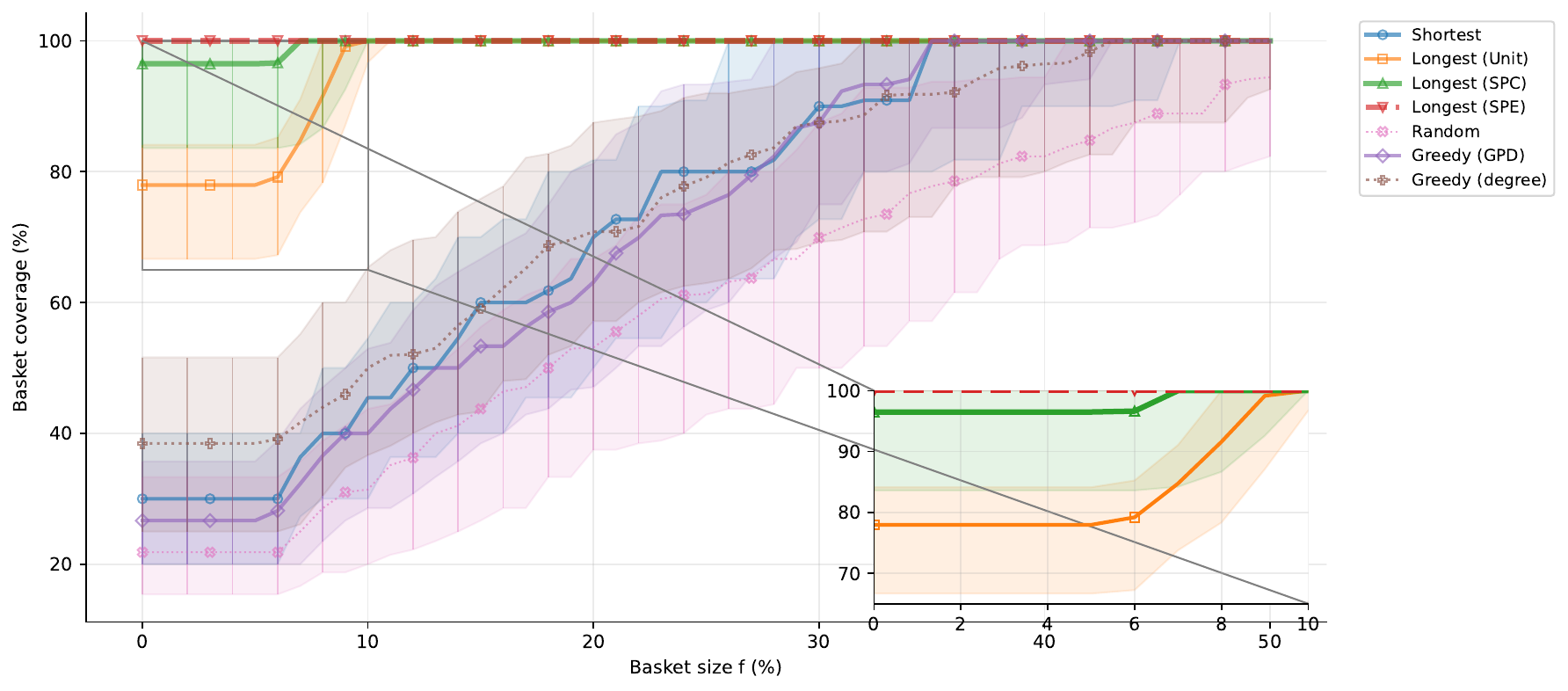}
	\caption{Plot to show how the coverage by the basket defined using SPE weights changes as the fraction of nodes included in the basket, $f$ is changed. This is done for various paths; the longest by unit, SPC and SPE weight, the shortest path by unit weight, two greedy paths, one using degree the other using perpendicular distance, and finally a random path. This is for 100 instances of random geometric DAGs with $N=1002$ nodes and target out-degree $\ktarget=22.0$. For each path $\Pcal$ a line connects the median values of the coverage $\phi(\Pcal,\Bcal^\mathrm{(unit)})$ with the error bars and shaded regions representing first and their quartile ranges, calculated over 100 networks.}
	\label{af:coverage-overlay-spe}
\end{figure}

\clearpage
Finally, we generalise \figref{f:coverage-overlay-unit} to show how the coverage of various paths changes the fraction of nodes included in the baskets defined using SPC node criticality in \figref{af:coverage-overlay-spc} and using the SPE node criticality values in \figref{af:coverage-overlay-spe}. The main result is to confirm that the coverage by the basket defined using unit weight criticality of the three types of longest paths is far better.\tsecomment{WHY IS THE GPD COVERAGE SO POOR?}

\subsection{Additional Numerical Results $D>2$}\label{as:pppnumericsDgt2}\label{app:rgg_3d4d}

\tsecomment{DO NOT DO THESE NOW. However, I expect a referee will want to see some testing of $D=3$ and $D=4$. I suggest we focus on getting a complete version of the paper done with $D=2$ and then focus on thesis. Only after that could we add $D=3$ and $D=4$. I would suggest we could simply repeat the same tables, figures, etc as used in the main text for $D=3$ and $D=4$ but we place them here.}

\section{Additional Material on Real-World Networks}\label{as:realworldnetworks}\label{a:data}

\subsection{Description of Network Data}\label{ss:network-inventory}

We used a number of datasets as the source of the real-world DAGs studied in \secref{s:realworldnetworks}.

To validate the generality of unit criticality baskets beyond synthetic models, we analyse twenty-two real-world DAG networks spanning multiple domains and network types. 
The networks divide into two fundamental categories based on their structural properties and coverage behaviour.

\tsecomment{Merge the next bit, just copied currently}

The nineteen citation-type networks represent information flow through hierarchical citation relationships across four distinct scale categories.
At the largest scale, we analyse three massive citation networks: APS citation network (from American Physical Society journals, has 527K nodes), US Patent citations (3.7M nodes), and arXiv Cond-Mat (118K nodes), which together demonstrate that our methods scale effectively to networks containing millions of nodes and edges.
The medium-scale category comprises six networks ranging from 8.8K to 37K nodes, including Student HepTh, SCOTUS legal cases, and four arXiv category networks (Quant-Ph, Math-Phys, Physics, and CS), representing diverse academic and legal citation domains.
Three small citation networks (arXiv Statistics, Chaos, and hep-ph, ranging from 630 to 984 nodes) provide test cases for understanding how the criticality basket approach performs on networks with limited path redundancy.
Finally, eight vaccine citation networks (AstraZeneca, BioNTech/Pfizer, Dengue, Imvanex, Moderna, Novavax, Shingrix, and Zabdeno, spanning 15K to 153K nodes) offer a focused case study within the medical research domain, allowing us to examine consistency of results within a single application area.

In contrast, three non-citation networks\tsenote{It is really not worth highlighting these as different as they may be too small to be important? Really the only use of DNA is it is small enough for us to visualise.} exhibit fundamentally different topological structures.
The Python Dependencies network (215 nodes, 3,813 edges) \citep{VE19a} represents a software dependency DAG where edges encode package requirements rather than information citations, creating a different pattern of connectivity with higher edge density relative to node count.
The DNA network (39 nodes, 59 edges) \citep{E25a}, despite being technically a citation network based on Fig.1 of the seminal main path analysis paper of \citet{HD89}, is included as a canonical small-scale test case to validate our methods against established benchmark networks in the literature.

The complete network inventory demonstrates scale invariance across five orders of magnitude (39 to 3.7M nodes) and domain diversity (physics, computer science, medicine, law, patents, dependencies).
All citation networks achieve 100\% Unit/SPE coverage at 5\% budget, whilst non-citation networks show 9--100\% coverage (DNA 9--31\%, Python Deps 50--71\%, Linux Deps 99--100\%), revealing that the ``5\% rule'' is topology-specific.

\begin{itemize}
	\item \textbf{APS Physics}. The APS dataset is the citation network from the journals of the American Physical Society. This data can be obtained on request from the American Physical Society, currently from \url{https://journals.aps.org/datasets}. Our data was downloaded on 9\textsuperscript{th} December 2014 and it has 527K nodes and 6.0M edges.
	
	\item \textbf{US Patent}. This is the ``NBER U.S.\ Patent Citations Data File'', data used in \citet{HJT01} which is publicly available such as from \url{https://www.nber.org/research/data/us-patents-1975-1999}.  It covers US patents from 1975--1999 and has 3.8M nodes and 16.5M edges.

	\item \textbf{SCOTUS}. This is the data set of \citet{FJ08} covering all decisions and citations made by the US Supreme Court from its inception in 1754 to 2002. This has 25K nodes and 216K edges. \tsecomment{I cannot find this online anymore.}	

	\item \textbf{arXiv}.  The \href{https://arXiv.org/}{\texttt{arXiv}} preprint server is a curated research-sharing platform which provides open access to its data. The data used here comes from the the paperscape project of \citet{GK17} which provides open access to the data at \url{https://github.com/paperscape/paperscape-data}. This covers the period from the start of the arXiv in 1991 till the end of 2017. This was processed by \citet{C17} to produce a citation network for each separate section as defined by arXiv (e.g.\ ``hep-th'', ``cond-mat'') in order to create smaller citation networks with a consistent naming convention for each paper.
	
	\item \textbf{Vaccine Networks}.  
	This is the data used in \citep{HPEO24}. It was derived from \tseurl{dimensions.ai}, \tseurl{lens.org} and  \tseurl{ClinicalTrials.gov}. However, restrictions apply to the availability of the raw data, which were
	used under licence for the current study, and so are not publicly available. Data are however available from the
	authors upon reasonable request and with permission of the mentioned sources. 
	
	\tsenote{A summary of some basic properties of the vaccine DAGs is shown in \tabref{tab:vaccine-network-properties}.}
	
	\item {Linux dependencies}. This is a DAG where the nodes are Linux source code files and the directed edges indicate the dependency of one file on another. This was derived from \url{https://www.kernel.org/} in 2014, and downloaded from the public repository \texttt{http://konect.uni-koblenz.de/networks/linux} in 2016. It has around 31K nodes and 230K edges.

	\item {Python dependencies}. This DAG comes from the package dependencies for an version of python installed on one computer in January 2019 and can be downloaded from \citet{VE19a}. This was extracted using the \href{https://pypi.org/project/pipdeptree/}{\texttt{pipdeptree}} python package \citep{pideptree}. It has 215 nodes and 3,813 edges. 
	
	
	
	\item \textbf{DNA}. This network comes from \citet{E25a} but is derived from the historiograph shown in \citet{HD89} which in turn is derived from the information in \citet{A62}. This provided the first illustration for main path analysis and with 39 nodes and 59 edges is provided for historical context.
	\\
	\tsecomment{TSE to make this available online.}
\end{itemize}

The basic properties of other DAGs built from real-world data are shown in \tabref{at:otherDAGproperties}.
\begin{table}[htb!]
	\centering
	\footnotesize
	\begin{tabular}{l|cccccccc} 
		Network        & Nodes   & Edges    & Avg.Deg.    & No.\   & No.\    & No.\ Weak  & Size of   & Fraction \\ 
                       & $N$     & $E$      &$\texpect{k}$& Sources& Sinks   & Comp.      & LWCC      & in LWCC  \\ \hline
		US Patent      & 3774753 & 16518209 &        8.75 & 515980 & 1685419 &       3630 &   3764094 & 0.997   \\ 
		APS Physics    &  527271 &  5987551 &       22.71 &  70050 &   15318 &        327 &    526470 & 0.998   \\ 
		SCOTUS         &   25378 &   216198 &       17.04 &   2157 &    6351 &         10 &     25350 & 0.999   \\ \hline
		arXiv Cond-Mat &  118496 &   765970 &       12.93 &  42788 &   12905 &       1124 &    115661 & 0.976   \\ 
		arXiv Quant-Ph &   36955 &   263235 &       14.25 &  11814 &    3636 &        258 &     36279 & 0.982   \\ 
		arXiv hep-ph   &   34530 &   419528 &       24.30 &  11885 &    4862 &         75 &     76663 & 0.998   \\ 
		arXiv Physics  &   19237 &    33675 &        3.50 &   8620 &    5327 &       1969 &     12742 & 0.662   \\ 
		arXiv CS       &   10352 &    11114 &        2.15 &   4572 &    3864 &       2182 &      3248 & 0.314   \\ 
		arXiv Math-Phys&    8847 &    17127 &        3.87 &   3321 &    2410 &        845 &      6115 & 0.691   \\ 
		arXiv Stats    &     984 &      825 &        1.68 &    453 &     387 &        252 &       182 & 0.185   \\ 
		arXiv Chaos    &     630 &      828 &        2.63 &    263 &     213 &         77 &       380 & 0.603   \\ \hline
		Linux Deps     &   30839 &   226318 &       14.68 &  18806 &    5149 &         12 &     30817 & 0.999   \\
        Python Deps    &     276 &      547 &        3.96 &    114 &      76 &          2 &       276 & 0.993   \\ 
		DNA (HD89)     &      39 &       60 &        3.08 &     11 &       9 &          4 &        35 & 0.854   \\ \hline 
		Zabdeno        &  153448 &  1034423 &       13.48 &      1 &   32880 &          1 &    153446 & 1.000   \\ 
		AstraZeneca    &  113387 &   672241 &       11.86 &      1 &   25191 &          1 &    113385 & 1.000   \\ 
		Moderna        &  112861 &   800567 &       14.19 &      1 &   18498 &          1 &    112859 & 1.000   \\ 
		BioNTech       &   65626 &   348945 &       10.63 &      1 &   10037 &          1 &     65624 & 1.000   \\ 
		Imvanex        &   58383 &   365823 &       12.53 &      1 &    8699 &          1 &     58381 & 1.000   \\ 
		Shingrix       &   27046 &   179952 &       13.31 &      1 &    5643 &          1 &     27044 & 1.000   \\ 
		Novavax        &   23703 &   107379 &        9.06 &      1 &    3795 &          1 &     23701 & 1.000   \\ 
		Dengue         &   15284 &    85828 &       11.23 &      1 &    4431 &          1 &     15282 & 1.000   \\ 
	\end{tabular}
	\caption{Basic properties of the real-world DAGS. The average degree is given in the column labelled $\texpect{k}$, 
		the number of weakly connected components is given in column labelled ``No.\ Weak Comp.'',
		the number of nodes in the largest weakly connected components is given in column labelled ``Size of LWCC'',
		and the fraction of nodes in the DAG that are in the largest weakly connected components is given in column labelled ``Fraction in LWCC''. The single source node and single weakly connected component for each vaccine DAG is by construction as these networks were constructed following references back from a single document giving the formal approval for each vaccine.
	\\
	\tsecomment{Would be nice to have the unit length of the longest unit path, the SPC and SPE paths in each case.}
	}
	\label{at:otherDAGproperties}
\end{table}


%
%

\newpage
\renewcommand{\thesection}{\Roman{section}}
\setcounter{equation}{0}
\setcounter{section}{0}
\tsecomment{Material in following sections, numbered with Roman numerals, is not in the official appendix. Move material if needed.}

\tsecomment{NOT READ IN}



%
%



\bibliographystyle{dcu}


%

\end{document}